\documentclass[12pt]{report}
    
\usepackage{amsfonts,epsfig,epic,verbatim,amsmath}

\def\d{\mbox{d}}
\def\D{\mbox{D}}
\def\pa{\partial}
\def\ha{\frac{1}{2}}
\def\id{1\!{\rm l}}
\def\>{\rightarrow}

\def\tr{\mbox{Tr}\,}

\def\ep{\epsilon}
\def\mn{_{\mu\nu}}

\def\for{\qquad\mbox{for}\:}
\def\with{\qquad\mbox{with}\:\:}

\def\ca{\mbox{case (a):}\quad}
\def\cb{\mbox{case (b):}\quad}

\begin{document}

\title{Monopoles, Instantons and Confinement}

\author{Gerard 't Hooft, University Utrecht}

\date{Saalburg, September 1999\vspace{5cm}\\
notes written by Falk Bruckmann, University Jena\vspace{0.5cm}\\
(version \today)}

\maketitle

\pagenumbering{Roman}
\tableofcontents

\pagenumbering{arabic}

\chapter{Solitons in 1+1 Dimensions}

As an introduction we consider in this chapter the easiest field
theoretic examples for solitons. These are real scalar field theories
in $1+1$ dimensions with a quartic and a sine-Gordon potential,
respectively. We will concentrate on physical aspects which are
relevant also in higher dimensions and more complicated theories like
QCD.

\section{Definition of the Models}

We investigate the theory of a single real scalar field $\phi(t,x)$
in one time and one space dimension. The usual Lagrangian (density) is,
\begin{eqnarray}
\label{Lagrangian}
\mathcal{L}=\ha\dot{\phi}^2-\ha\left(\pa_x \phi\right)^2-V(\phi)
\end{eqnarray}
We consider two cases for the potential. Case (a) refers to the
`Mexican-hat' potential, well-known from spontaneous symmetry breaking,
\begin{eqnarray}
\label{mex}
\ca V(\phi)=\frac{\lambda}{4!}\left(\phi^2-F^2\right)^2
\end{eqnarray}
while case (b) is the sine-Gordon model,
\begin{eqnarray}
\label{sine}
\cb V(\phi)=A\left(1-\cos\frac{2\pi\phi}{F}\right)
\end{eqnarray}
which is an exactly solvable system.
The important point about these models is  that the vacuum is
degenerate. In case (a) it is two-fold degenerate,
\begin{eqnarray*}
\phi=\pm F
\end{eqnarray*} 
while in case (b) we have an infinite number of vacua,
\begin{eqnarray*}
\phi=nF,\quad\quad n\in \mathbb{Z} 
\end{eqnarray*}

\begin{figure}[t]

\begin{minipage}{0.5\linewidth}
\begin{picture}(190,145)
\small
\put(0,-40){\includegraphics{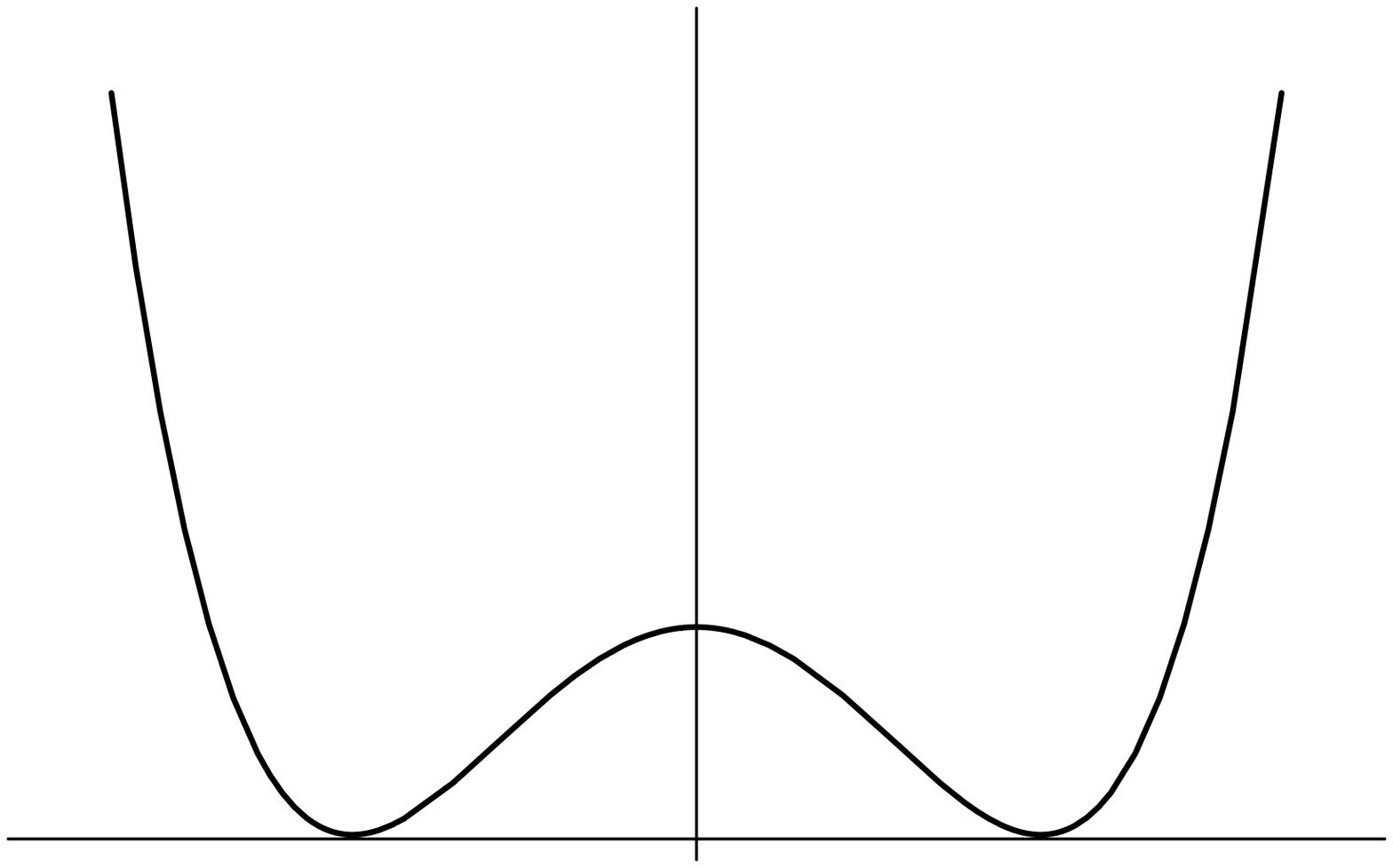}}
\put(167,35){$\phi$}
\put(45,25){$-F$}
\put(123,25){$F$}
\put(82,128){$V(\phi)$}
\put(85,0){(a)}
\normalsize
\end{picture}

\end{minipage}
\begin{minipage}{0.5\linewidth}
\begin{picture}(190,145)
\small
\put(0,-40){\includegraphics{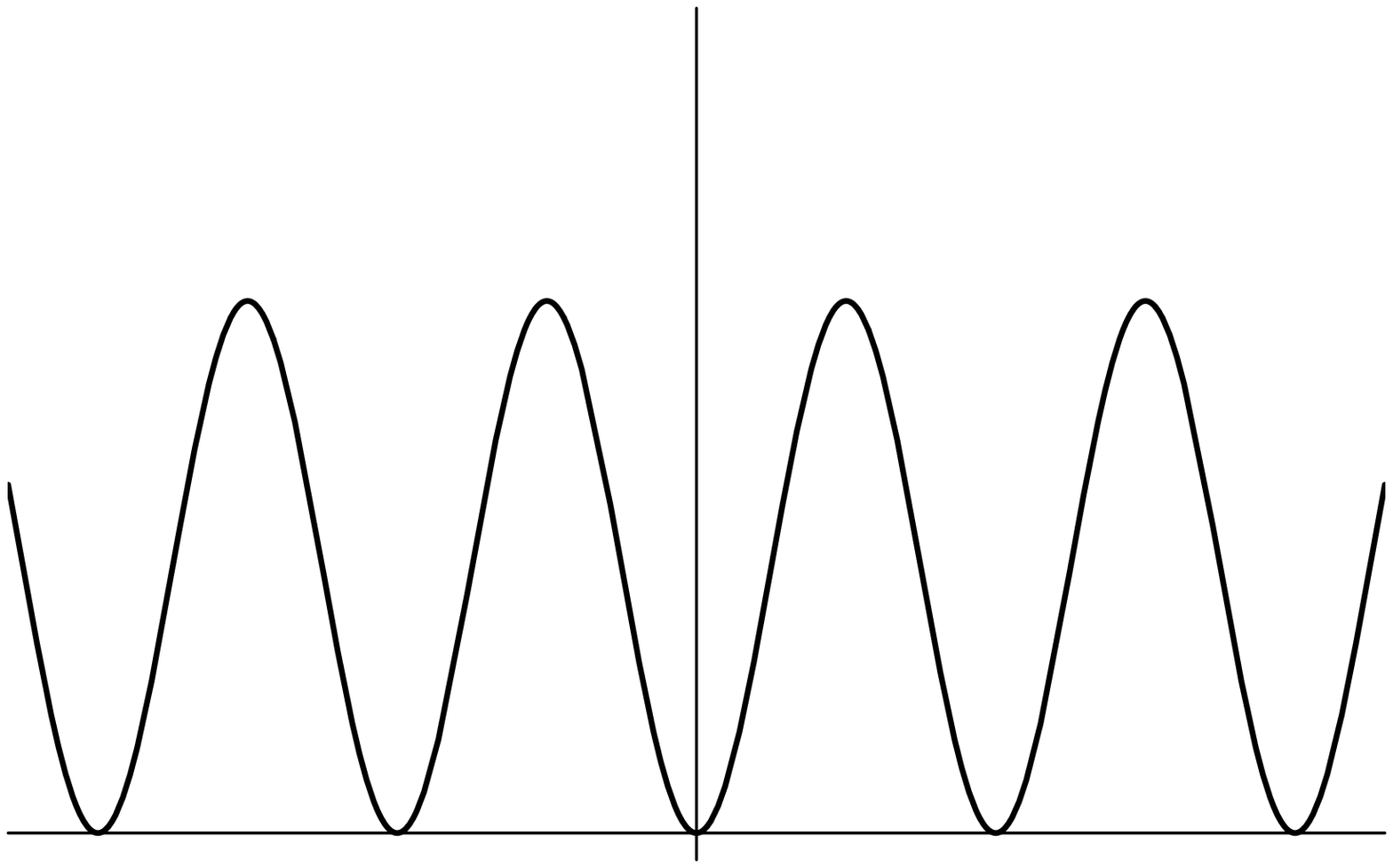}}
\put(167,35){$\phi$}
\put(48,25){$-F$}
\put(89,25){$0$}
\put(117,25){$F$}
\put(82,128){$V(\phi)$}
\put(85,0){(b)}
\normalsize      
\end{picture}
\end{minipage}

\caption{\small{Potentials under consideration: the Mexican-hat (case (a), see
 (\ref{mex})) and the sine-Gordon model (case (b), see
 (\ref{sine})). Both have degenerate vacua and allow for non-trivial
 solutions. }}
\end{figure}

In standard perturbation theory one considers small fluctuations $\eta$ around
the vacua,
\begin{eqnarray*}
\ca \phi&=&F+\eta\\
\cb \phi&=&0+\eta,\quad (\eta\equiv\phi)
\end{eqnarray*}
and expands the potential. Here we get the usual mass term together
with three and four point interactions,
\begin{eqnarray*}
\ca V(\eta)&=&\ha m^2\eta^2
+\frac{g}{3!}\eta^3+\frac{\lambda}{4!}\eta^4\\
&&m^2\equiv\lambda F^2/3,\quad g\equiv\lambda F\\
\cb V(\eta)&=&\ha m^2\eta^2-\frac{\lambda}{4!}\eta^4+\ldots\\
&&\left.\begin{array}{l}
m\equiv 2\pi\sqrt{A}/F\\
\lambda\equiv 16\pi^4 A/F^4
\end{array}\right\}
\left\{\begin{array}{l}
A\equiv m^4/\lambda\\
F\equiv 2\pi m/\sqrt{\lambda}
\end{array}\right.
\end{eqnarray*}
$m$ is the mass of the particles of the theory (we use $\hbar=1$) and
$g$ and $\lambda$ are defined such that the three and four point vertices
are proportional to them. Usually one has small $\lambda$ and large
$F$ such that the mass $\sqrt{\lambda}F$ is fixed.
Notice that the mass square would be negative on the maxima $0$ and
$n+\ha$, respectively,
these so-called tachyons would render the theory unstable.

\section{Soliton Solutions}

The degeneracy of the vacuum results in the fact that these models possess
non-trivial static solutions, which interpolate (in space) between the vacua.
We call them \textit{kinks} or \textit{solitons}.
 Their existence and
shape are given by the Euler-Lagrange equation\footnote{
For time-independent solutions we could also work in the Hamiltonian
formalism. Moreover, any of the following static solutions can be
transformed into a steadily moving one by a Lorentz transfomation.}
derived from $\mathcal{L}$ in (\ref{Lagrangian}),
\begin{eqnarray*}
\ddot{\phi}=\pa_x^2\phi-\frac{\pa V}{\pa \phi}=0
\end{eqnarray*}
If we think of $\phi(x)$ as $x(t)$, this is the equation of motion of
a non-relativistic particle, $\ddot{x}=-\frac{\pa V}{\pa x}$, but in a
potential $-V$. Like the energy in ordinary mechanics,
we find a first integral,
\begin{eqnarray}
\label{firstintegral}
\frac{\d}{\d x}\left(\ha(\pa_x \phi)^2-V(\phi)\right)&=&
\pa_x \phi\left(\pa_x^2\phi-\frac{\pa V}{\pa \phi}\right)=0\nonumber\\
\ha(\pa_x \phi)^2-V(\phi)&=&\mbox{const.}
\end{eqnarray}
Since we want solutions with finite energy, we have to demand that the
energy density,
\begin{eqnarray}
\label{endensity}
E=\int_{-\infty}^{\infty}\!\!\left[\ha(\pa_x
\phi)^2+V(\phi)\right]\d x
\end{eqnarray}
 vanishes at spatial infinity,
\begin{eqnarray*}
|x|\>\infty:\quad\pa_x\phi\>0,\,V(\phi)\>0
\end{eqnarray*}
i.e. the above constant is zero.
The remaining first order differential equation can easily be
solved,
\begin{eqnarray}
\label{solution}
x(\phi)=\int \!\!\frac{\d\phi}{\sqrt{2 V(\phi)}}
\end{eqnarray}
In our models we can write down the solutions exactly,
\begin{eqnarray*}
\ca \phi(x)&=&F\, \mbox{tanh}\:\ha m(x-x_0)\\
\cb \phi(x)&=&\frac{2F}{\pi}\,\mbox{arctan}\left(e^{m(x-x_0)}\right)
\end{eqnarray*}
$x_0$ is an arbitrary constant (of integration) due to translational
invariance.
%\begin{comment}
%---------------------------
\begin{figure}[t]

\begin{minipage}{0.5\linewidth}
\begin{picture}(0,145)
\small
\put(0,-40){\includegraphics{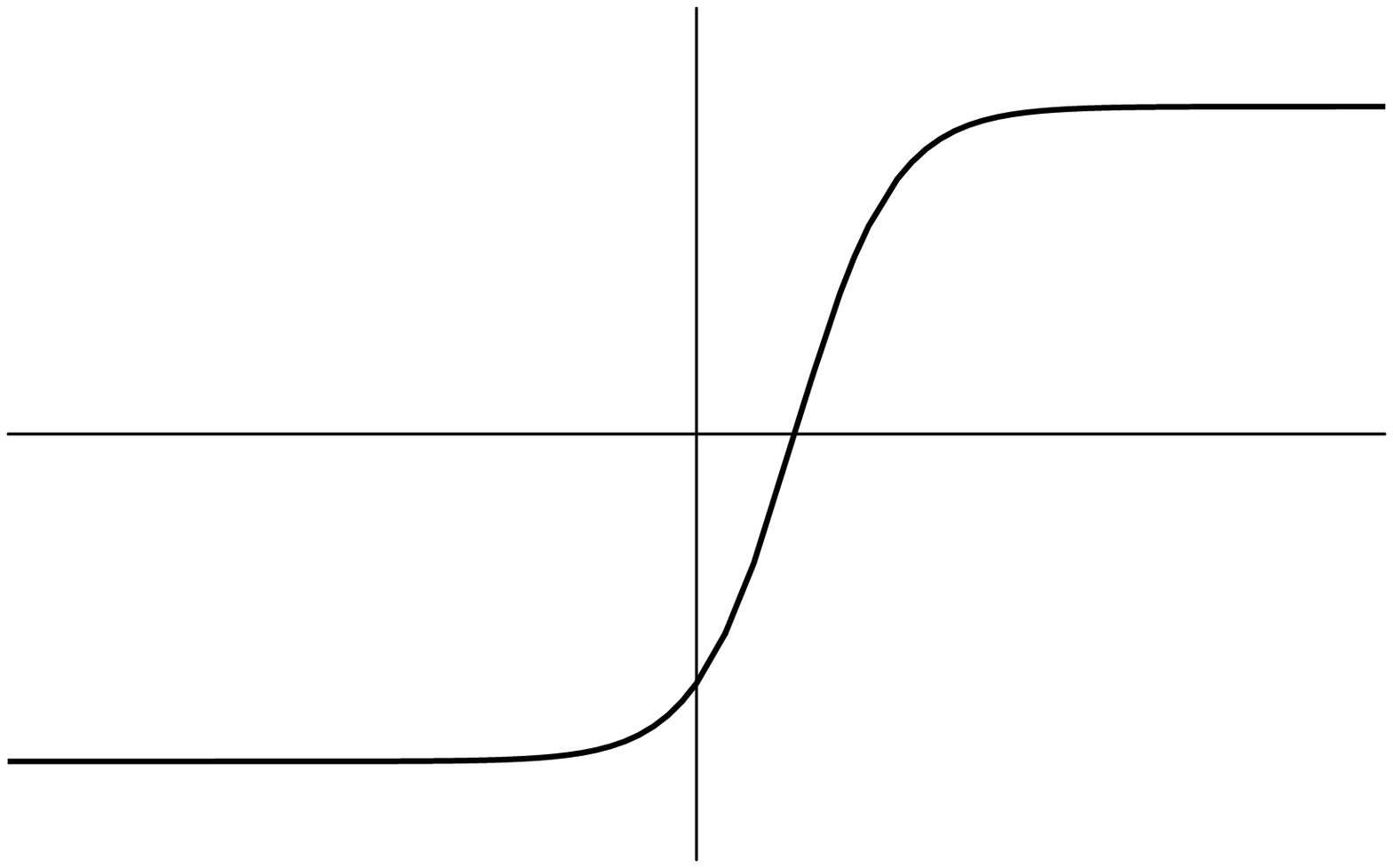}}
\put(167,76){$x$}
\put(102,70){$x_0$}
\put(97,43){$-F$}
\put(77,107){$F$}
\put(82,128){$\phi(x)$}
\put(122,99){$\Gamma\propto 1/m$}
\put(85,0){(a)}
\normalsize
\end{picture}

\end{minipage}
\begin{minipage}{0.5\linewidth}
\begin{picture}(0,145)
\small
\put(0,-40){\includegraphics{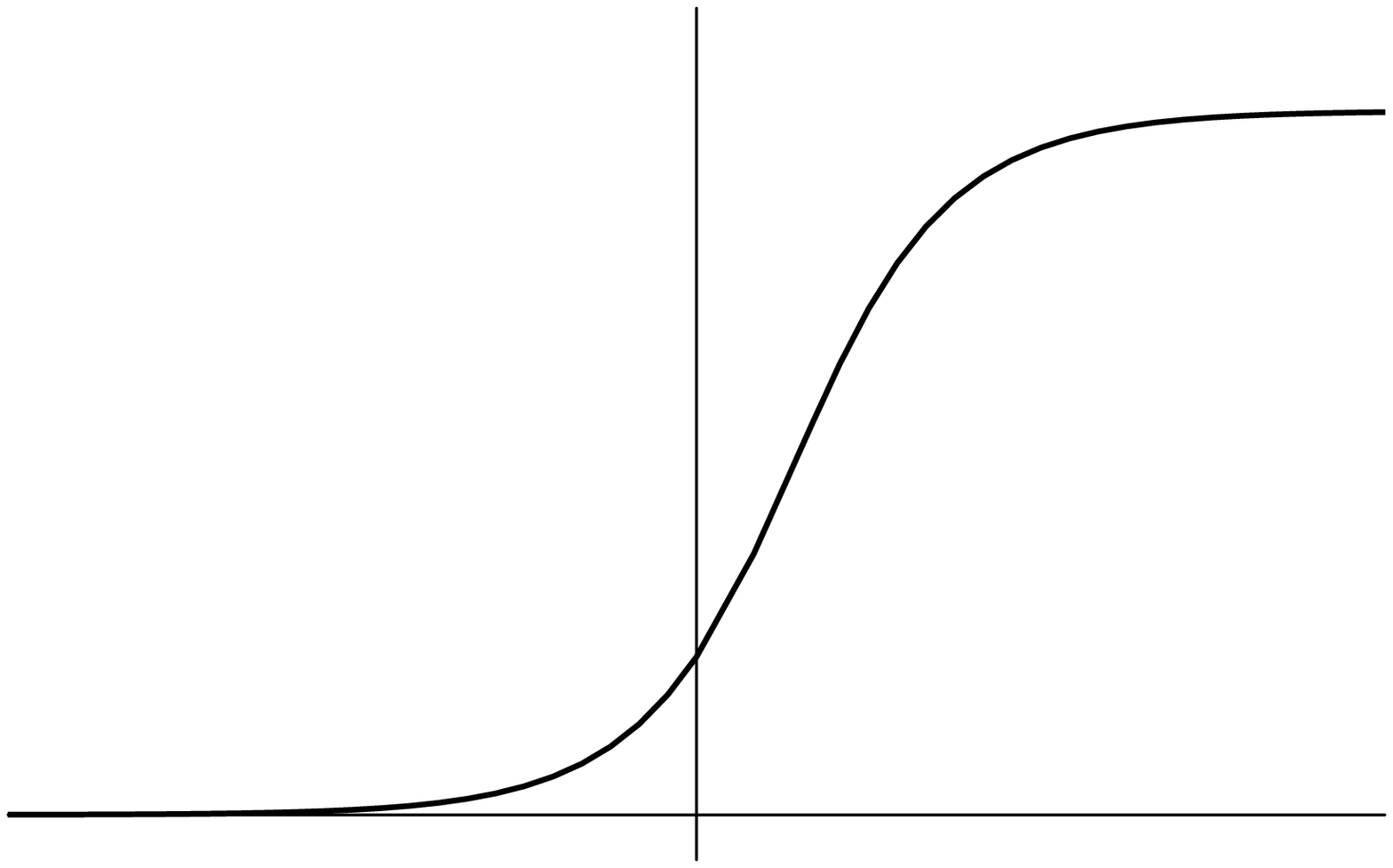}}
\put(167,37){$x$}
\dashline{2}(92,78)(103,78)
\dashline{2}(103,78)(103,42)
\put(102,31){$x_0$}
\put(67,76){$F/2$}
\put(77,107){$F$}
\put(82,128){$\phi(x)$}
\put(130,98){$\Gamma\propto 1/m$}
\put(85,0){(b)}
\normalsize       
\end{picture}
\end{minipage}

\caption{\small{Solitonic solutions in both potentials have nearly
identical shape: The transition from one vacuum value to the next one 
takes place around an arbitrary position $x_0$. It decays with a rate that
is proportional to the inverse mass of the (light) particles of the
theory.}}
\protect{\label{kinks}}
\end{figure}
%---------------------------------
%\end{comment}
Their shapes are very similiar (cf Fig. \ref{kinks})
and show the typical behaviour:

\noindent(i) Solitons interpolate between two neighbouring vacua:
\begin{eqnarray*}
\ca \phi(x\>\pm\infty)&=&\pm F\\
\cb \phi(x\>-\infty)&=&0,\\
\phi(x\>+\infty)&=&F
\end{eqnarray*}

\noindent (ii) The solutions are (nearly) identical to a vacuum
value everywhere except a transition region around an arbitrary point
$x_0$. The shape of the solution there is given by the shape of the
potential between the vacua. Near the vacua the solution is decaying
exponentially with a rate $\Gamma\propto 1/m$. Thus $\phi$ can be
approximated by  a step function, for instance in case (a),
\begin{eqnarray*}
\phi(x)=F\,\mbox{sgn}\:(x-x_0) \for |x-x_0|\gg 1/m
\end{eqnarray*}

In terms of mechanics one could think of a particle which passes the
bottom of a valley at some time $t_0$. It has just the energy to climb
up the hill and stay there. Actually this will take infinitely long,
but after a short time it has already reached a position very near the
top. Of course, the particle must have been on top of the opposite
hill in the infinite past. Whenever both tops have the same height such a
solution exists, no matter which form $V$ has inbetween.

\begin{figure}[b]

\begin{minipage}{0.5\linewidth}
\begin{picture}(0,145)
\small
\put(0,-40){\includegraphics{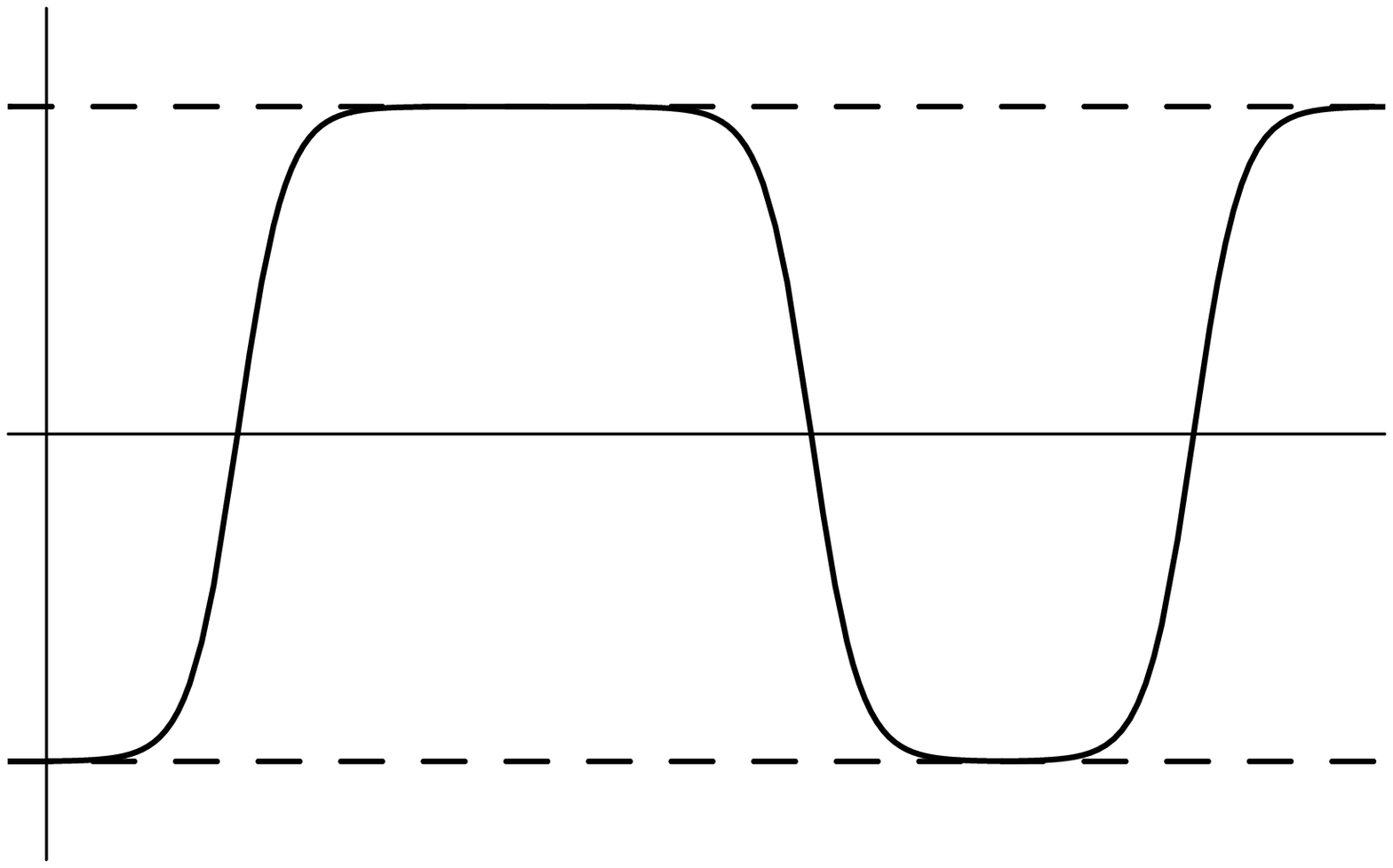}}
\put(19,128){$\phi(x)$}
\put(9,109){$F$}
\put(2,44){$-F$}
\put(167,77){$x$}
\put(47,70){$x_{0,1}$}
\put(108,70){$x_{0,2}$}
\put(145,70){$x_{0,3}$}
\put(85,0){(a)}
\normalsize
\end{picture}

\end{minipage}
\begin{minipage}{0.5\linewidth}
\begin{picture}(0,145)
\small
\put(0,-40){\includegraphics{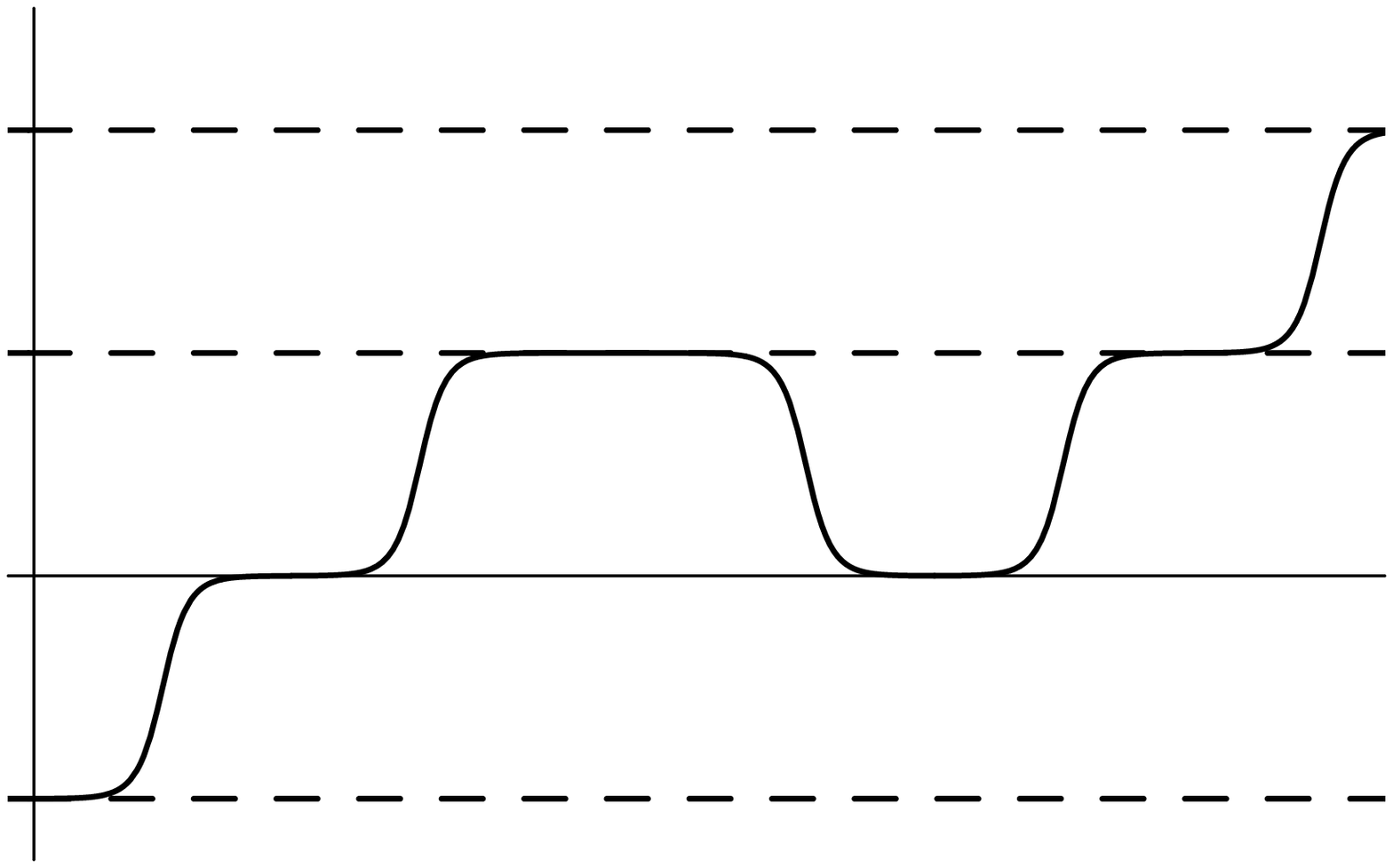}}
\put(19,128){$\phi(x)$}
\put(6,107){$2F$}
\put(11,85){$F$}
\put(12,62){$0$}
\put(2,40){$-F$}
\put(167,62){$x$}
\put(85,0){(b)}
\normalsize
\end{picture}
\end{minipage}

\caption{\small{Multi-Solitons are approximate solutions built of
solitons and anti-solitons at arbitrary positions $x_{0,i}$. The
sequences are strongly constrained in case (a), while in case (b) they
are arbitrary.}}
\protect{\label{sequences}}
\end{figure}

The energy of the solution (its mass) can be computed
from (\ref{endensity}) and (\ref{firstintegral}),
\begin{eqnarray*}
E&=&2\int_{-\infty}^{\infty}\!\!V(\phi)\,\d x
=\int_{\phi_{-\infty}}^{\phi_\infty}\!\!\pa_x\phi\: \d \phi\\
&=&\left\{\begin{array}{l}
2m^3/\lambda \for \mbox{case (a)},\\
8m^3/\lambda \for \mbox{case (b)}.
\end{array}\right.
\end{eqnarray*}
Alternatively we can use the saturation of the Bogomol'nyi bound (cf
Exercise (i)),
\begin{eqnarray}
\label{bogobound}
E=\int\!\!\left[(0)^2+\mbox{total derivative}\right]
=W(\phi)|_{\phi_{-\infty}}^{\phi_\infty}
\end{eqnarray}
In both cases the mass of the soliton is given by the cube of the mass
of the elementary particles divided by $\lambda$ (which has dimension (mass$)^2$).
The soliton is very massive in the perturbative limit. That
means we are dealing with a theory which describes both, light fluctuations,
the elementary particles, and heavy solutions, the solitons. This mass
gap supports the validity of perturbation theory (for example for
tunneling).

Up to now we have only considered one-soliton solutions which
interpolate between neighbouring vacua. Due to the ambiguity of the
square root in (\ref{solution}) there are also solutions interpolating
backwards, anti-solitons. Now one could immediately imagine solutions
consisting of whole sequences of solitons and anti-solitons (see
Fig. \ref{sequences}). As discussed in Exercise (ii), these approximate
solutions are only valid for widely separated objects
$|x_{0,1}-x_{0,2}|\gg 1/m$, such that we have a 'dilute gas'.

Here the Mexican-hat and the sine-Gordon model
differ slightly. Solitons and anti-solitons have to alternate in the
first model. From a particle point of view the anti-soliton is
really the anti-particle of the soliton. If the potential is
symmetric, we cannot distinguish between them. The situation is like in
a real scalar field theory.

In the latter model we can arrive at any vacuum by choosing
the right difference between the number of solitons and
anti-solitons. Now solitons and anti-solitons are distinguishable and
the analog is a complex scalar field theory.

\section{Chiral Fermions}

Now we investigate the (still 1+1 dimensional) system,
\begin{eqnarray*}
\mathcal{L}_\psi=-\bar{\psi}\left(\gamma^\mu\pa_\mu+g\phi(x)\right)\psi
\end{eqnarray*}
where $\psi$ is a Dirac field, $\gamma_\mu$ are the (Euclidean) Dirac
matrices, which we can choose to be the Pauli matrices,
\begin{eqnarray*}
\gamma^1=\sigma_1,\quad \gamma^4=\sigma_3,\quad \gamma_5=\sigma_2
\end{eqnarray*}
and $\phi$ is a solution from above\footnote{
In particular we take $\phi$ from case (a), since with the sine-Gordon
model we would be forced to use a cos-interaction which is not
normalisable in 4 dimensions.}. We could say we put fermions in a
soliton background or we study the consequences to the solution if we
couple fermions to it.

The interaction is provided by the usual Yukawa coupling. The field
$\phi$ acts like a (space-dependent) mass. If $\phi$ takes the vacuum
value $F$ everywhere, then we simply have a theory with massive fermions,
\begin{eqnarray*}
\mathcal{L}_\psi=-\bar{\psi}\left(\gamma^\mu\pa_\mu+m_\psi\right)\psi,
\quad\quad m_\psi=gF 
\end{eqnarray*}
The energies, i.e. the eigenvalues of the (hermitean)
Hamilton-Operator,\footnote{We use the complex notation $\pa_4=-i\pa_t$.}
\begin{eqnarray*}
\mathcal{H}=-i\pa_t=i\sigma_2\pa_x+\sigma_3m_\psi
\end{eqnarray*}
come in pairs $\pm E$ with $|E|\geq m$. The fields with the opposite
energies are generated by $\gamma^1=\sigma_1$,
\begin{eqnarray*}
\{\sigma_1,\sigma_2\}=\{\sigma_1,\sigma_3\}=0\Rightarrow
\mathcal{H}\psi=E\psi\rightleftharpoons\mathcal{H}(\gamma^1\psi)
=-E(\gamma^1\psi)
\end{eqnarray*}

%\begin{comment}
%------------------------------
\begin{figure}[t]

\begin{minipage}{0.5\linewidth}
\begin{picture}(0,145)
\small
\put(0,-40){\includegraphics{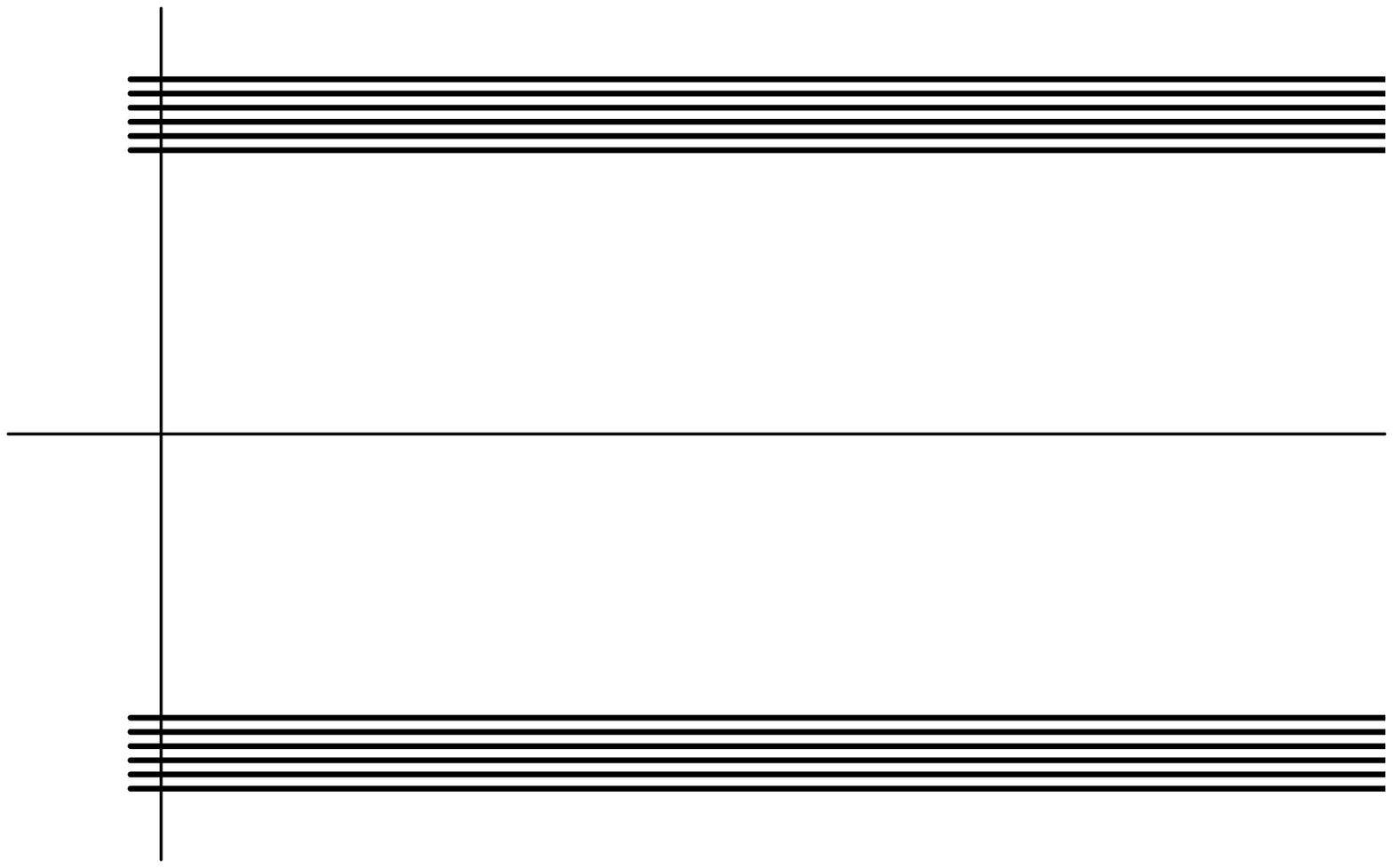}}
\put(16,105){$m_\psi$}
\put(7,48){$-m_\psi$}
\put(32,128){$E$}
\put(85,0){(a)}
\normalsize
\end{picture}

\end{minipage}
\begin{minipage}{0.5\linewidth}
\begin{picture}(0,145)
\small
\put(0,-40){\includegraphics{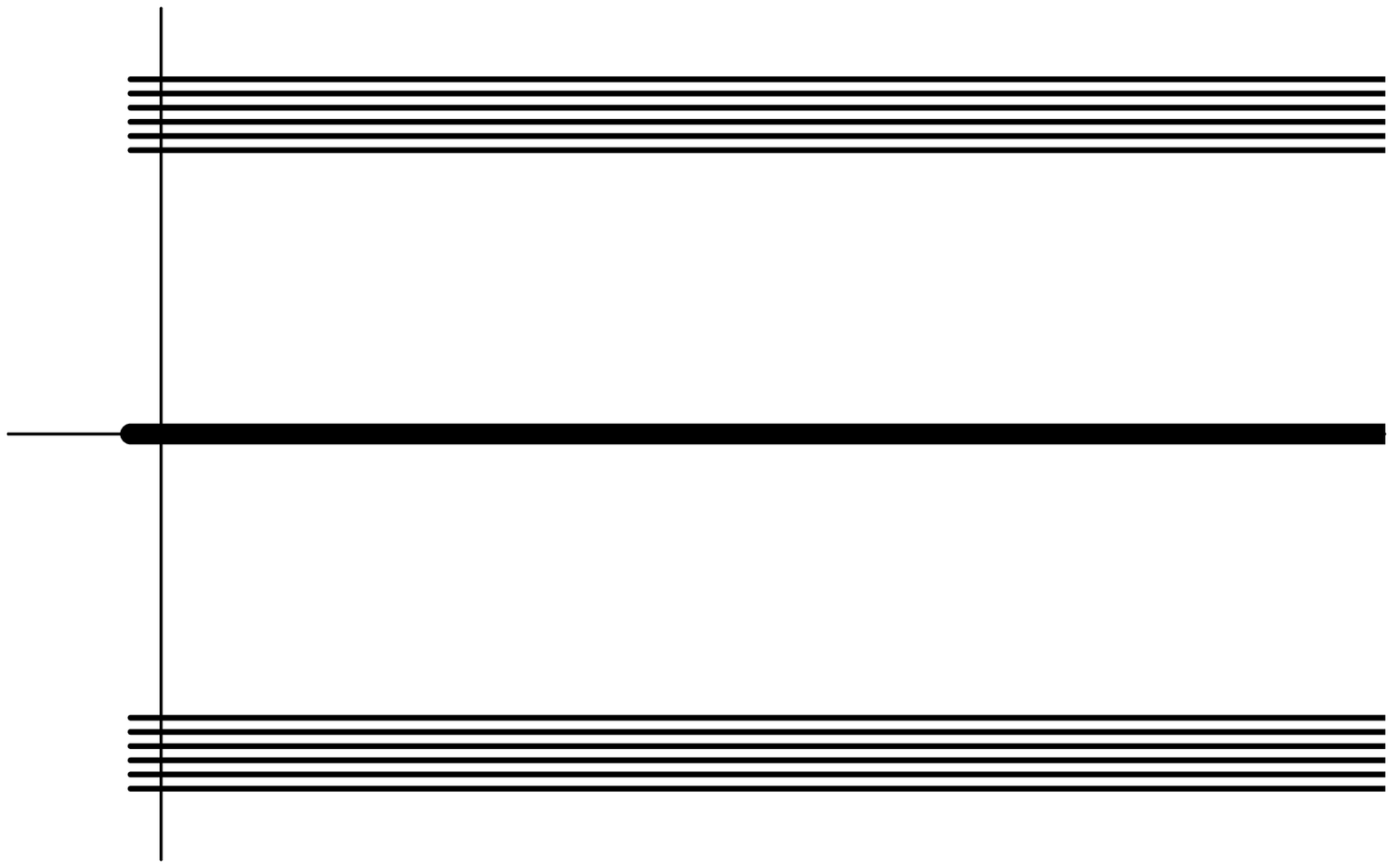}}
\put(32,128){$E$}
\put(85,0){(b)}
\normalsize        
\end{picture}
\end{minipage}

\caption{\small{The usual spectrum of massive fermions (a) is produced
by a constant vacuum solution $\phi\equiv F$. For a soliton there is
\textit{one} additional zero mode (b).}}
\protect{\label{dirac}}
\end{figure}
%----------------------------
%\end{comment}

In the spirit of Dirac we define the vacuum to be filled with
negative energy states, the particles to be excitations with $E>0$ and
the anti-particles to be holes in $E<0$ (cf Fig.
\ref{dirac}). $\gamma^1$ is the fermion number conjugation.

As well-known the chiral
symmetry generated by $\gamma_5$,
\begin{eqnarray*}
\psi\>\gamma_5\psi,\quad
\bar{\psi}\>-\bar{\psi}\gamma_5
\end{eqnarray*}
is broken by the mass term.
But as an interaction this term can be made invariant by
\begin{eqnarray*}
\phi\>-\phi
\end{eqnarray*}
which is again a solution. That means $\gamma_5$ generates a state
with the same energy but with $\phi$ in the opposite vacuum.

Now we really want to insert a non-trivial $\phi$ and look for its
spectrum. Still $\sigma_1$ generates opposite energy solutions. It
comes out that again they are  displaced by the soliton. What
about the special case of solutions of zero energy? Acting on them,
$\mathcal{H}$ and $\sigma_1$ commute, and we choose the zero modes
$\psi$ to be eigenfunctions of $\sigma_1$,
\begin{eqnarray*}
\sigma_1\psi_{\pm}=\pm\psi_{\pm},\quad
\psi_+=\frac{\psi_1(x)}{\sqrt{2}}{1 \choose 1},\quad
\psi_-=\frac{\psi_2(x)}{\sqrt{2}}{1 \choose -1}
\end{eqnarray*}
Now we have to solve
\begin{eqnarray*}
(\sigma_1\pa_x+g\phi)\psi_{\pm}=\sigma_3 E\psi_{\pm}=0
\end{eqnarray*}
which becomes
\begin{eqnarray}
\label{log}
\pa_x\psi_{1,2}=\mp g\phi\psi_{1,2},\qquad
\ln \psi_{1,2}=\mp g\int\!\phi\,\d x
\end{eqnarray}

%\begin{comment}
%---------------------------
\begin{figure}[t]

\begin{minipage}{0.5\linewidth}
\begin{picture}(0,145)
\small
\put(0,-40){\includegraphics{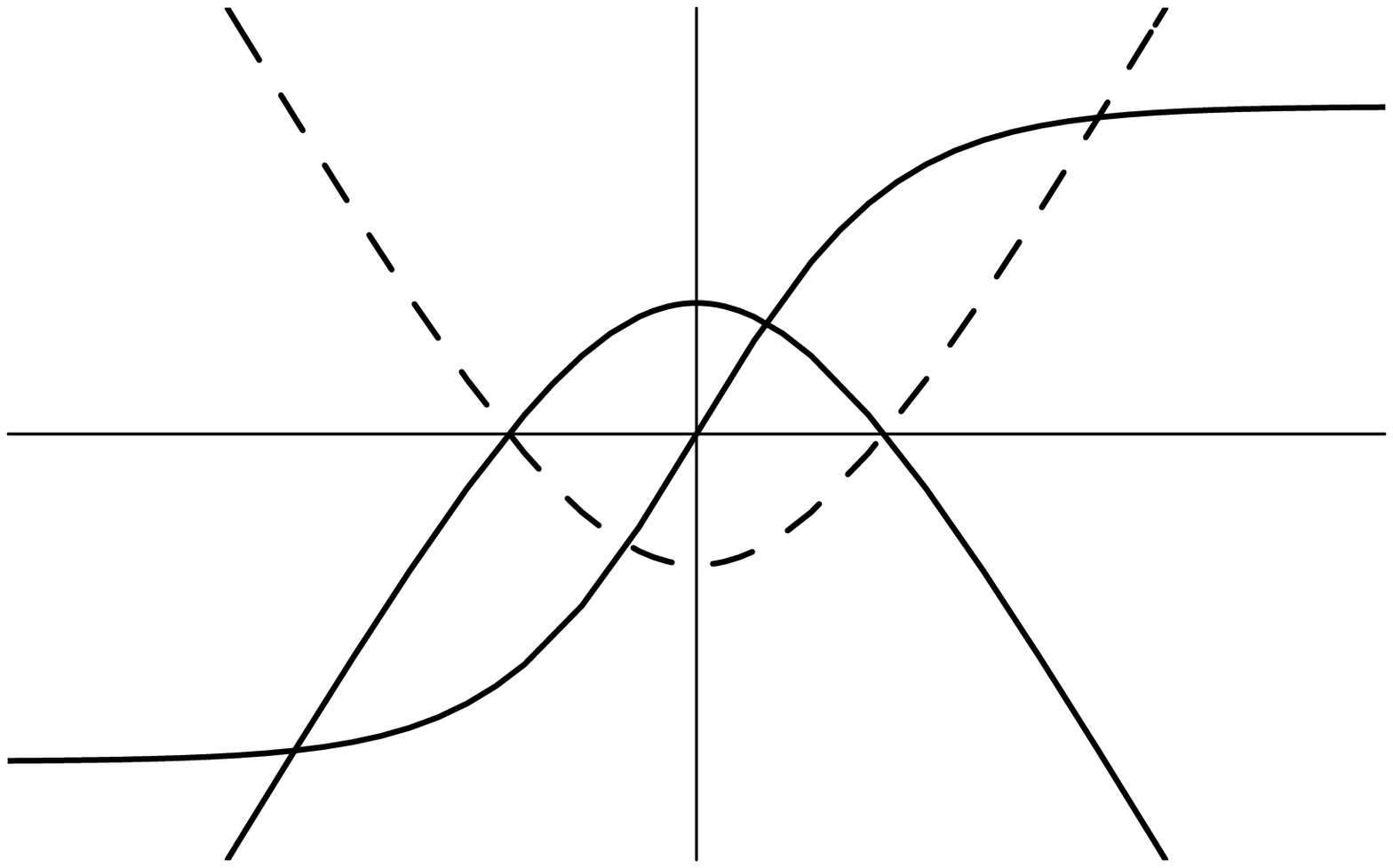}}
\put(167,76){$x$}
\put(132,128){$\ln\psi_2$}
\put(167,110){$\phi$}
\put(132,23){$\ln\psi_1$}
\put(85,0){(a)}
\normalsize
\end{picture}

\end{minipage}
\begin{minipage}{0.5\linewidth}
\begin{picture}(0,145)
\small
\put(0,-40){\includegraphics{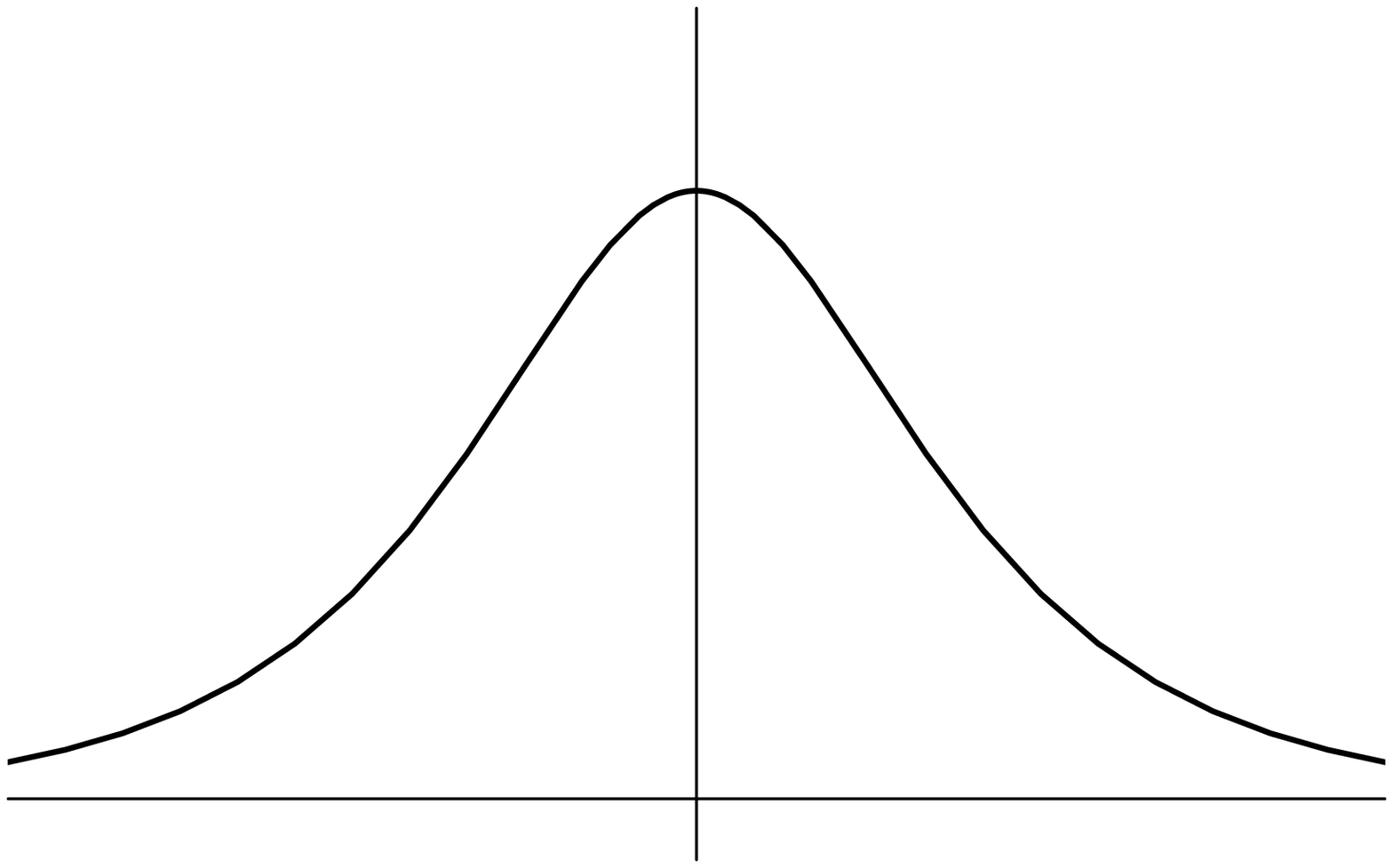}}
\put(167,37){$x$}
\put(82,128){$\psi_1(x)$}
\put(85,0){(b)}
\normalsize       
\end{picture}
\end{minipage}

\caption{\small{Zero energy solutions in a soliton background:
$\ln\psi_{1,2}\propto\pm\int\!\phi\d x$ (cf (\ref{log})).
 Only one of the solutions is
normalisable, the one with the full line in (a). It is localised
at the soliton position which we have chosen to be 0 here (b).
A similiar picture applies for the anti-soliton.}}
\protect{\label{zero}}
\end{figure}
%---------------------------------
%\end{comment}

Knowing the general shape of the soliton $\phi$ we see immediately
that the solution with the lower sign is non-normalisable, while the
one which the upper sign fulfills every decent boundary condition,
since it drops exponentially. Notice that $\psi_1\psi_2=\mbox{const.}$
as a general property. Thus, if we add the continuum,
 there is an `odd' (but infinite) number of
solutions.

For completeness we give the formula for the  normalisable zero mode,
\begin{eqnarray*}
\psi_+=\mbox{const}\left(\cosh\frac{m}{2}(x-x_0)\right)^{-2m/m_\psi}
{1 \choose 1}
\end{eqnarray*}
It is strongly localized at the
position $x_0$ of the soliton (cf Fig. \ref{zero}).
For the anti-soliton $-\phi$ the solution with the lower sign is
normalisable,
\begin{eqnarray*}
\psi_-=\mbox{const}\left(\cosh\frac{m}{2}(x-x_0)\right)^{-2m/m_\psi}
{1 \choose -1}
\end{eqnarray*}
This of course agrees with the chiral transformed $\psi_+$,
\begin{eqnarray*}
\psi_+&\>&\gamma_5\psi_+=
\mbox{const}\left(\cosh\frac{m}{2}(x-x_0)\right)^{-2m/m_\psi}
\sigma_2{1 \choose 1}\\
&&=\mbox{const}\left(\cosh\frac{m}{2}(x-x_0)\right)^{-2m/m_\psi}
{i \choose -i}\propto\psi_-
\end{eqnarray*}

These zero modes are called Jackiw-Rebbi modes \cite{JaRe}. We view
them as soliton-fermion bound states indistinguishable from the
original true soliton\footnote{The counting of states depends on this
interpretation (cf \cite{JaRe}).}.

When we quantise the theory, $\psi$ becomes an operator with
Fermi-Dirac statistics/anti-commutation relations, especially
$\hat{\psi}^2=0$. $\hat{\psi}_+$ and $\hat{\psi}_-$ commute with the
Hamiltonian: $[\hat{\psi}_{\pm},\hat{H}]=0$. In general this commutator
involves the energy, but here we have $E=0$. Whether these states are
filled or empty has no effect on the energy, they are `somewhere
inbetween fermions and anti-fermions'. As we have seen they are
related by $\gamma_5$. In fact Jackiw and Rebbi \cite{JaRe} have
shown that the soliton has two states with
\begin{eqnarray*}
\framebox{fermion number: $n=\pm1/2$,
no spin, no Fermi-Dirac statistics!}
\end{eqnarray*}
$n$ is the expectation value of the conserved charge
$\hat{Q}_0=\int\!\d x :\hat{\bar{\psi}}\gamma^0\hat{\psi}:$
in these states.

\section{Outlook to Higher Dimensions}

In higher dimensions the field will still like to sit in a vacuum for
most of the space-time. The kinks will now be substituted by (moving)
\textit{domain walls}, i.e. transition regions between
domains with different $\phi$-values. Their shape will depend on the
model. In any case passing domain walls will have huge physical
consequences. For example in case (a) passing from $\phi$ to $-\phi$
means transforming (by chiral symmetry) matter into anti-matter.

The domains themselves are related through a discrete global symmetry,
namely $Z_2$ in case (a) and $\mathbb{Z}$ in case (b). In other words
we have two sorts of domains in case (a) and infinitely many in case
(b), respectively.

%\begin{comment}
%------------------------------
\begin{figure}[t]

\begin{picture}(380,145)
\small
\put(110,0){\includegraphics{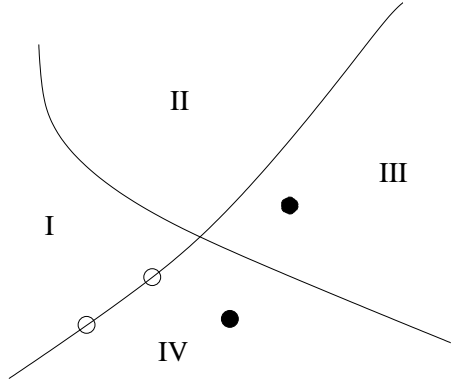}}
\normalsize
\end{picture}

\caption{\small{The picture we expect in higher dimensions: In the
domains I to IV $\phi$ sits in different vacua, which are related by a
discrete symmetry. The transition takes place on domain walls. Inside
the domains fermions are massive ($\bullet$), on the walls they are
massless ($\circ$) and localised perpendicular to the walls.}}
\end{figure}
%----------------------------
%\end{comment}

Concerning the chiral fermions, they are massive inside the domains
and massless on the walls. The latter are localised in the direction
perpendicular to the domain wall. Along the domain wall we can view
them as lower-dimensional Dirac fermions.
This scenario has become helpful for studying fermions on the lattice.

\chapter{The Abrikosov-Nielsen-Olesen-Zumino~Vortex}

\section{Approach to the Vortex Solution}

\begin{figure}[t]
\begin{picture}(380,145)
\small
\put(110,-50){\includegraphics{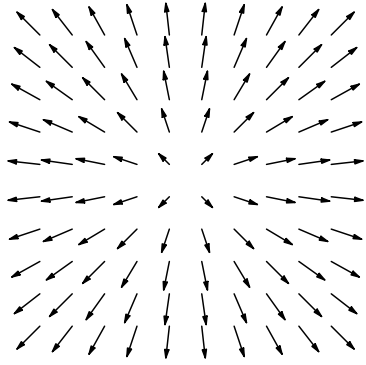}}
\normalsize
\end{picture}
\caption{\small{A typical vortex
 solution with isospace vectors $\vec{\phi}$
depicted in (2 dimensional) coordinate space $\vec{x}$. The field
`winds around once' at spatial infinity as a general feature. 
Angle and length of $\vec{\phi}$ inside, especially the position of
the zero, are still arbitrary. All these configurations are specified
by the winding number 1.}}
\protect{\label{hedge}}
\end{figure}

We try to find the analogue of domain walls in 2+1 dimensions. They
will come out as \textit{vortices} or \textit{strings}.
We have to take a complex (or two component real) scalar
field,
\begin{eqnarray}
\phi=\phi_1+i\phi_2,\:\:\vec{\phi}={\phi_1 \choose \phi_2}
\end{eqnarray}
We use the generalization of (\ref{Lagrangian}) with a global U(1)
invariance,
\begin{eqnarray}
\label{general}
\mathcal{L}=-\pa_\mu\phi^\ast\pa_\mu\phi
-\frac{\lambda}{2}\left(\phi^\ast\phi-F^2\right)^2
\end{eqnarray}
as our starting point. Notice that the vacuum manifold is now a
circle $|\phi|=F$. For the desired soliton solution we
combine it with the directions in space at spatial infinity,
\begin{eqnarray}
\label{asymp}
|x|\>\infty:\,\,\vec{\phi}\>F\frac{\vec{x}}{|x|},\:\:\phi\>Fe^{i\varphi}
\end{eqnarray}
where $\varphi$ is the polar angle in coordinate space.
One such solution
%\footnote{
%For `higher windings' we can take the naive generalization
%$\phi\> Fe^{in\varphi}$ with $n\in\mathbb{Z}$.} 
is depicted in Fig.~\ref{hedge}.
The solution could also be a deformation of this, but should go a
full circle around the boundary. Since the map on the boundary is
non-trivial, $\phi$ must have a zero inside.

But this non-trivial map at spatial infinity has the effect that the
energy,
%(per unit length in the time-like direction, i.e. the string
%tension):
\begin{eqnarray}
\label{stringtension}
E=\int\!\d^2 x\left(\vec{\pa}\phi^\ast\vec{\pa}\phi
+V(\phi,\phi^\ast)\right)
\end{eqnarray} 
is divergent, since the rotation of $\phi$ enters the
kinetic energy,
%\footnote{The contribution of the potential energy
%depends on how $|\phi|$ approaches its asymptotic value $F$.}.
%We can easily compute it from (\ref{asymp}):
\begin{eqnarray*}
|x|\>\infty:\quad
\pa_i\phi_j &\>& \frac{F}{|x|}\left(\delta_{ij}
-\frac{x_i x_j}{|x|^2}\right)\\
\sum_{i,j=1}^2 (\pa_i\phi_j)^2 &\>& \frac{F^2}{|x|^2}
(2-2+1)=\frac{F^2}{|x|^2}\\ 
%\end{eqnarray*}
%Inserting this into (\ref{stringtension}) gives a logarithmic
%divergence:
%\begin{eqnarray*}
\int\!\d^2x\,\vec{\pa}\phi^\ast\vec{\pa}\phi&\>&
2\pi\!\int_0^\infty\!\!\!\d|x|\, \frac{F^2}{|x|}
\ldots\mbox{log. divergent} 
\end{eqnarray*}
Thus in a theory with \textit{global} $U(1)$ invariance, there exists
a vortex, but its energy (per time unit in three dimension) is
logarithmically divergent!
 
Derrick's Theorem \cite{Derrick} states
that this divergence is unavoidable for
time-in\-de\-pen\-dent solutions in $d\geq 2$.
Since it is only a mild divergence, the solution still plays a role
in phase transitions in statistical mechanics.
%Nevertheless, we will
%cure it in the next section.

\section{Introduction of the Gauge Field}

%For what follows now, it is crucial to note that (\ref{general})
%is a theory with a \textit{global} $U(1)$ invariance.
Now we cure the above divergence by making the U(1)
invariance \textit{local} in the standard manner. We add a %$U(1)$
gauge field $A_\mu$ and replace the partial derivative in
(\ref{general}) by the covariant one,
\begin{eqnarray}
\label{covariant}
\pa_\mu\phi\>\D_\mu\phi=\left(\pa_\mu-ieA_\mu\right)\phi
\end{eqnarray}
This gives $\vec{\D} \phi$ the chance to converge better than
$\vec{\pa}\phi$ (we still deal with static solutions). In other words
the divergence is absorbed in $\vec{A}$. Since asymptotically $\phi$
depends only on the angle $\varphi$, $\vec{A}$ will only have a
component in this direction
%orthogonal to the field $\phi$.

Asymptotically, $\phi$ is real at the $x$-axis,
\begin{eqnarray*}
\phi\>Fe^{i\varphi}|_{\varphi=0}=F
\end{eqnarray*}
and the gradient has only a $y$ component,
\begin{eqnarray*}
\vec{\pa}\phi\>\left.{\pa_x\phi \choose \pa_y\phi}\right|_{\varphi=0}
=\left.{\pa_r\phi \choose \frac{1}{r}\pa_\varphi \phi}\right|_{\varphi=0}
={0 \choose iF/r}
\end{eqnarray*}
We can read off $\vec{A}$ from the demand of
vanishing covariant derivative,
\begin{equation}
\label{puregauge}
\vec{A}\>\frac{1}{ie}\phi^{-1}\vec{\pa}\phi,\qquad
A_x\>0\nonumber,\qquad
A_y\>\frac{1}{er}
\end{equation}
For the general case (at any point $(x,y)$) we perform a trick, namely
we can rotate $\phi$ locally to be real,
\begin{eqnarray*}
\phi\>\Omega F\with \Omega(\vec{x})=e^{i\varphi}
\end{eqnarray*}
thus,
%As in (\ref{puregauge}) we choose $\vec{A}$ to be:
\begin{eqnarray*}
\vec{A}\>-\frac{1}{ie}\Omega \vec{\pa} \Omega^{-1}
\end{eqnarray*}
In fact the covariant derivative vanishes asymptotically,
\begin{eqnarray*}
\vec{\D}\phi\>\left(\vec{\pa}\Omega+
\Omega(\vec{\pa} \Omega^{-1})\Omega\right) F
=\Omega\vec{\pa}\left(\Omega^{-1}\Omega\right) F=0
\end{eqnarray*}
\begin{figure}[t]
\begin{picture}(380,155)
\small
\put(110,0){\includegraphics{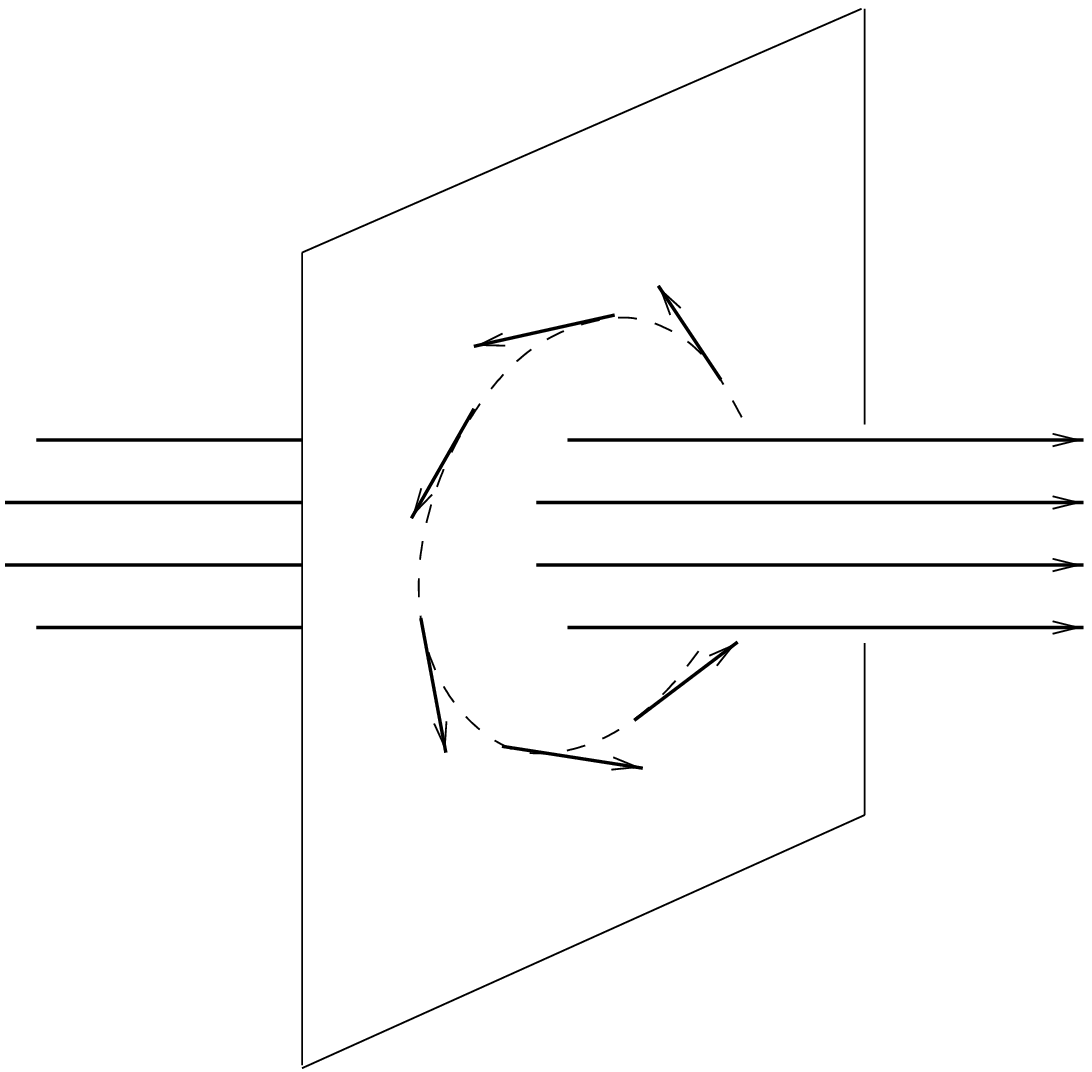}}
\put(210,114){$\vec{A}$}
\put(244,51){$\vec{B}$}
\put(273,74){$\Phi$}         
\normalsize
\end{picture}
\caption{\small{The introduction of a circular gauge field $\vec{A}$
(cf (\ref{A_i})) leads to a
vortex with quantised magnetic flux
$\Phi=n \frac{2\pi}{e}$.}}
\end{figure}
The general form of $\vec{A}$ is,
\begin{eqnarray}
\label{A_i}
A_i\>-\frac{1}{e}\ep_{ij}\frac{x_j}{r^2}
\end{eqnarray}
As we expected it has only a $\varphi$-component,
\begin{eqnarray*}
A_r\>0,\qquad
A_\varphi\>\frac{1}{er}
\end{eqnarray*}
Furthermore it is a pure gauge asymptotically and the field strength
vanishes,
\begin{eqnarray*}
\vec{A}\>\frac{1}{e}\vec{\pa}\varphi,\qquad F_{ij}\>0
\end{eqnarray*}
%keeping (the chance for)
giving a  solution with finite energy per unit length.

It can be shown that the choices for $\phi$ and $A$ are solutions of
the Euler-Lagrange equations asymptotically. If we try to extend them
naively towards the origin, $A$ runs into a singularity. Instead one
 could make an
ansatz for $\phi$ and $A$ and try to solve the remaining
equations numerically \cite{NiOl}. But already from the asymptotic
behaviour we can deduce a \textit{quantised magnetic flux},
\begin{eqnarray*}
\Phi=\int_S\!\vec{B}\d\vec{\sigma}
=\int_{C=\pa S}\!\vec{A}\,\d\vec{x}=g_m,\qquad g_m=\frac{2\pi}{e}
\end{eqnarray*}
For higher windigs we have analogously,
\begin{eqnarray*}
\Phi=n g_m  \with n\in\mathbb{Z}
\end{eqnarray*}

The situaton is very much like in the Ginzburg-Landau theory for the
superconductor. In this theory an electromagnetic
field interacts with a fundamental scalar field  describing 
Cooper pairs. The latter are bound states of two electrons with
opposite momentum and spin. As bosonic objects they can fall into the
same quantum state resulting in \textit{one} scalar field $\phi$.
The potential of the scalar field is of the Mexican-hat form with
tem\-pe\-ra\-ture-de\-pen\-dent coefficients.
 In the low temperature phase the
symmetry is broken and the photons become massive. That means if a
magnetic field enters the superconductor at all, it does so in
flux tubes. Performing an Aharonov-Bohm gedankenexperiment
around such a tube leads to a flux quantum,
\begin{eqnarray*}
\Phi_{\rm SC}=n g_{\rm SC}
\end{eqnarray*}
The only difference to the above model is a factor 2 from the
\textit{pair} of electrons,
\begin{eqnarray*}
q_{\rm SC}=2e,\qquad g_{\rm SC}=\frac{2\pi}{q_{\rm SC}}=\frac{\pi}{e}
\end{eqnarray*} 

\section{Bogomol'nyi Bound for the Energy}

%The two dimensional energy of the soliton is its energy per unit length
%in three dimensions.
Adding the field strength term to (\ref{general})
and (\ref{covariant}), the complete Lagrangian reads,
\begin{eqnarray}
\label{U1theory}
\mathcal{L}=-\D_\mu\phi^\ast\D_\mu\phi
-\frac{\lambda}{2}\left(\phi^\ast\phi-F^2\right)^2
-\frac{1}{4}F_{\mu\nu}F_{\mu\nu}
\end{eqnarray}
The energy integral is now,
\begin{eqnarray*}
E=\int\!\!\d^2x\left[\D_i\phi^\ast\D_i\phi+\ha F_{12}^2+
\frac{\lambda}{2}\left(\phi^\ast\phi-F^2\right)^2\right]
\end{eqnarray*}
In the gauge where $\phi$ is real\footnote{
At the origin this gauge $\Omega^{-1}=e^{-i\varphi}$
becomes singular,
since $\varphi$ is ambiguous at this point.}
 the integrand consists of a sum of squares,
\begin{eqnarray*}
E=\int\!\!\d^2x\left[(\pa_i\phi)^2+
e^2\vec{A}^2\phi^2+
\ha F_{12}^2+
\frac{\lambda}{2}\left(\phi^\ast\phi-F^2\right)^2\right]
\end{eqnarray*}
where the second and the fourth term cannot be zero at the same time.
For the Bogomolnyi bound we reduce the number of squares 
by partial integration as for the soliton (cf (\ref{bogobound}) and
Exercise (i)),
\begin{eqnarray*}
(\pa_i\phi)^2+e^2\vec{A}^2\phi^2&=&
\left(\pa_i\phi\pm e\ep_{ij}A_j\phi\right)^2\pm e\phi^2F_{12}+
\mbox{total der.}\\
\ha F_{12}^2+
\frac{\lambda}{2}\left(\phi^\ast\phi-F^2\right)^2&=&
\ha\left(F_{12}\pm\sqrt{\lambda}(\phi^2-F^2)^2\right)^2
\mp \sqrt{\lambda}(\phi^2-F^2)F_{12} 
\end{eqnarray*}
Notice that the new square in the first equation looks like a
covariant derivative, but it is not. The boundary contributions are
easily to be calculated.

For the special choice\footnote{
This choice corresponds to the type I/type II phase boundary of the
superconductor.}
of
\begin{eqnarray*}
\framebox{$\lambda=e^2$} \qquad m_\phi=m_A=\sqrt{2}eF
\end{eqnarray*}
the energy simplifies further,
\begin{eqnarray*}
E&=&\int\!\!\d^2 x\left[
\left(\pa_i\phi\pm e\ep_{ij}A_j\phi\right)^2+
\ha\left(F_{12}\pm\sqrt{\lambda}(\phi^2-F^2)^2\right)^2
\pm e F^2 F_{12}\right]\\
&\geq&eF^2\left|\int\!\!F_{12}\,\d^2x\right|
\end{eqnarray*}
For the saturation of the bound the first two equations can be solved
numerically, while the rest gives the total magnetic flux,
\begin{eqnarray}
\label{Bbound}
E\geq eF^2\,n\frac{2\pi}{e}=n\frac{\pi m^2}{e^2}
\end{eqnarray}
Again we have found the typical dependence mass$^2/$coupling for
heavy topological objects.

\section{Gauge Topology Description}

For the vortex as well as for the soliton we have seen that the
asymptotic behaviour is important, in the sense that the requirement
of finite energy forces the configurations to fall into disjoint
`classes'. Interpolating between these classes must include
configurations with divergent energy. Now we want to clarify this
topological property.

Along the lines of spontaneous symmetry breaking, (\ref{U1theory}) is
a $U(1)$
gauge theory coupled to a Higgs field $\phi$. The vacuum manifold
$|\phi|=F$ is $U(1)$-invariant, but the special choice $\phi=F$
 breaks the $U(1)$ down to $\id$:
No gauge transformation leaves this special value
invariant. That is, the gauge transformation leading to this gauge
must be a mapping from the boundary of $\mathbb{R}^2$ to 
$U(1)/\id$,
\begin{eqnarray*}
\Omega:S^1\longrightarrow U(1)/\id\equiv U(1)
\end{eqnarray*}
The identity has been (formally) divided out,
since for the general case $\Omega$ need
not come back to the same group element. It is allowed to differ
by another group element belonging to the subgroup which
leaves the vacuum choice invariant. We say the Higgs field $\phi$
transforms under the group $ U(1)/\id$.

\begin{figure}[t]
\begin{picture}(380,155)
\small
\put(110,0){\includegraphics{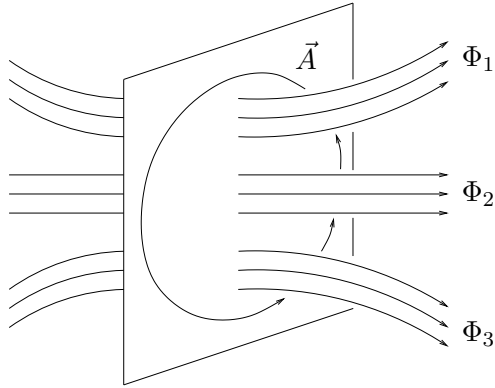}}
\put(219,120){$\vec{A}$}
\put(282,121){$\Phi_1$}
\put(282,70){$\Phi_2$}
\put(282,17){$\Phi_3$}         
\normalsize
\end{picture}
\caption{\small{A non-trivial configuration of vortices carrying total
flux $(n_1+n_2+n_3)\frac{2\pi}{e^2}$. For topological reasons it
cannot be continuously shrinked to the trivial vacuum (unless
$n_1+n_2+n_3=0$). The total flux also results in a lower bound for the
energy (cf (\ref{Bbound})): $E\geq|n_1+n_2+n_3|\frac{\pi m^2}{e^2}$
for $\lambda=e^2$.}}
\end{figure}

The mappings from $S^1$ into a manifold $M$ themselves form a group,
called the \textit{first homotopy group} $\pi_1(M)$.
%Mappings, which
%can be continuously deformed into each other are called homotopic  and
%give the same group element.
$\pi_1$ measures the non-contractibiliy of $M$, i.e. the
`existence of holes'.
For contractible $M$ all mappings are identified and $\pi_1$
is simply the identity.

The Lie group $U(1)$ itself is a circle $S^1$. The first homotopy
group of $S^1$ is well-known to be the group of integers,
%with the sum as the composition law
\begin{eqnarray*}
\pi_1(S^1)=\mathbb{Z}
\end{eqnarray*} 
Notice that this group is Abelian.

Thus from topological arguments each vortex carries a quantum number
\begin{eqnarray*}
Q\in\pi_1(U(1)/\id)=\mathbb{Z}
\end{eqnarray*}
which can be identified with the total flux number $n$. We have found
an abstract reasoning for the quantisation of this physical quantity.
There are infinitely many $U(1)/\id$ vortices and they are
\textit{additively stable}.

\begin{figure}[t]
\begin{picture}(380,125)
\small
\put(110,10){\includegraphics{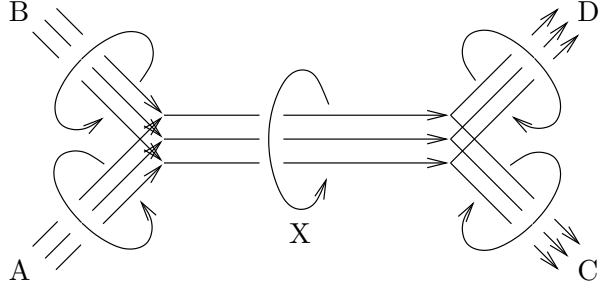}}
\put(102,106){B}
\put(102,8){A} 
\put(208,22){X}
\put(317,106){D}
\put(317,8){C}         
\normalsize
\end{picture}
\caption{\small{In a general Higgs theory $G\>G_1$ each vortex
represents an element of the group $G_2=\pi_1(G/G_1)$. The fusion
rules are governed by this group, which might be non-Abelian. For the
depicted case we have $g_{\rm A}g_{\rm B}=g_{\rm X}=g_{\rm C}g_{\rm
D}$. }}
\protect{\label{fusion}}
\end{figure}

Other situations may occur. Let a group $G$ be spontaneously broken
down to a subgroup $G_1$,
\begin{eqnarray*}
G\stackrel{\mbox{Higgs}}{\longrightarrow}G_1
\end{eqnarray*}
\begin{figure}[t]
\begin{minipage}{0.5\linewidth}
\begin{picture}(180,140)
\small
\put(30,0){\includegraphics{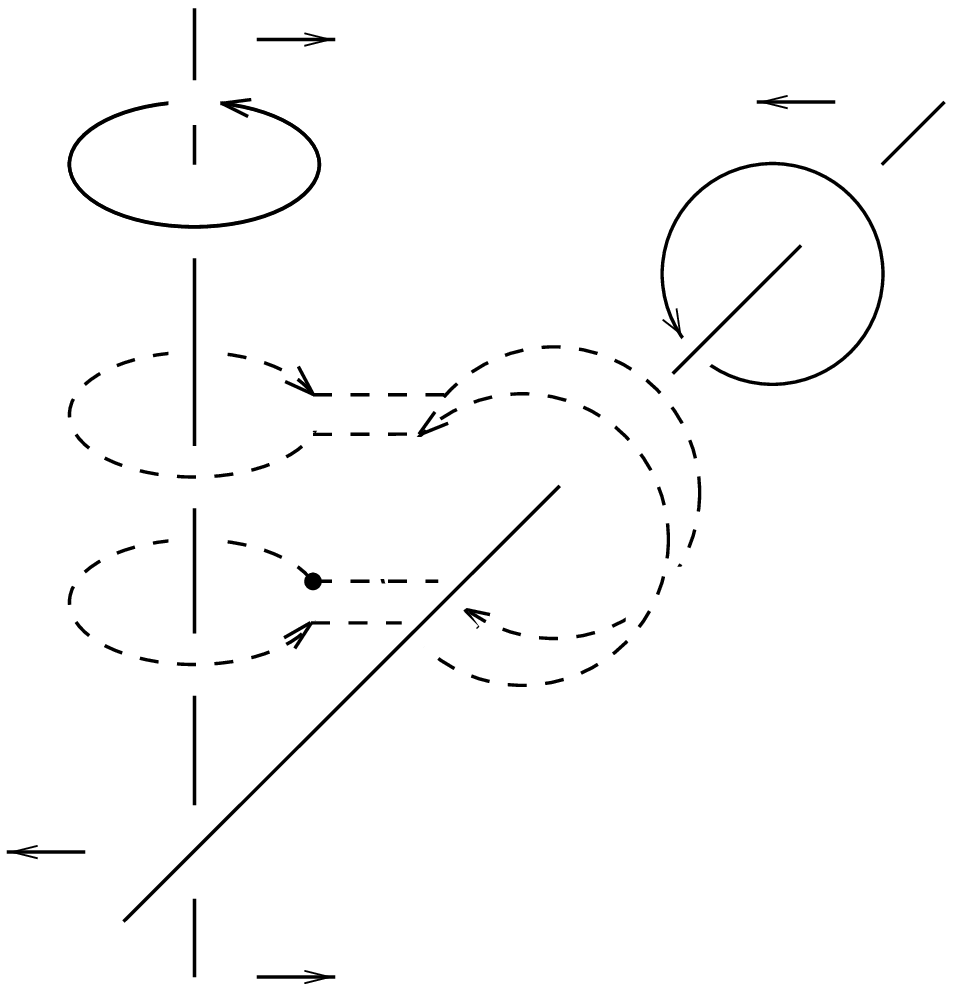}}
\put(80,117){$A$}
\put(114,109){$B$}
\normalsize
\end{picture}
\end{minipage}
\begin{minipage}{0.5\linewidth}
\begin{picture}(180,140)
\small
\put(30,20){\includegraphics{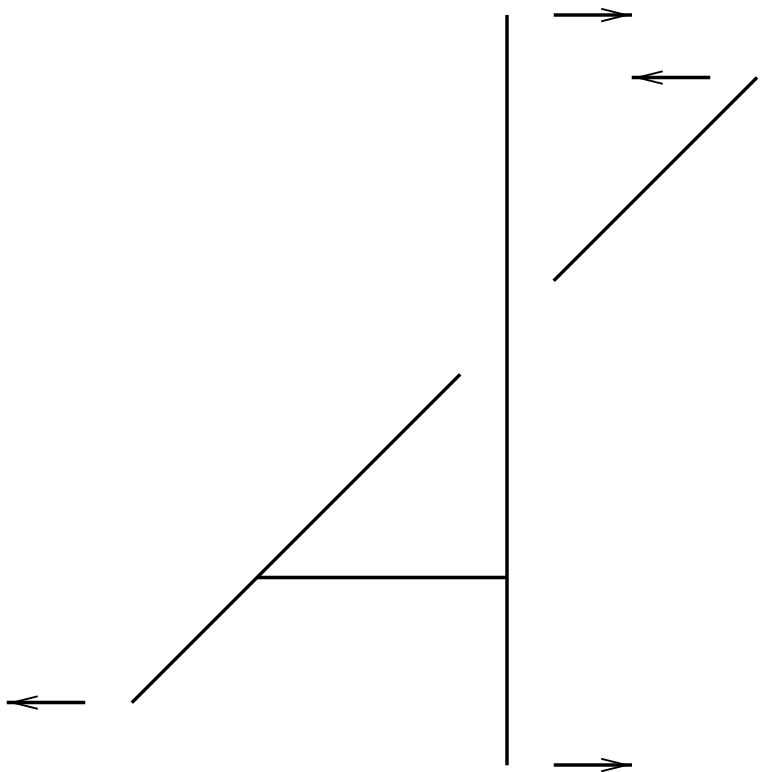}}
\put(98,7){$A$}
\put(38,15){$B$}
\put(80,35){$C$}         
\normalsize       
\end{picture}
\end{minipage}
\caption{\small{Alice strings: When a non-Abelian group $G_2$ is associated to the
vortices, the hitting of two of them ($A$ and $B$) will lead to a
third one $C=ABA^{-1}B^{-1}$ as indicated by the dashed contour.}} 
\protect{\label{Alice}}
\end{figure}
Then in the same spirit
\begin{eqnarray*}
G_2=\pi_1(G/G_1)
\end{eqnarray*}
is the group of vortex quantum numbers. Whenever $G_2$ is non-trivial
$G_2\neq \id$, there are stable vortices. Their fusion rules are given
by the composition law of the group $G_2$, which in general might be
non-Abelian. Then the quantum number of the vortex is not additive,
and one vortex cannot `go through the other one' without leaving a third
vortex (cf Fig. \ref{fusion}, \ref{Alice} and Exercise (iii)).
 This situation plays a role in the
theory of crystal defects, where the vortices go under the name of
`Alice strings'. Generically it does not occur in the Standard Model
of elementary particle physics.

But let us consider a `double Higgs theory',
\begin{eqnarray*}
SU(2)\left.\begin{array}{c}
\mbox{Higgs}\\\longrightarrow\\I=1
\end{array}\right.
U(1)
\left.\begin{array}{c}
\mbox{Higgs}\\\longrightarrow\\I=1
\end{array}\right.
 Z_2
\end{eqnarray*}
The $SU(2)$ theory is broken by a Higgs field in the adjoint
$(I=1)$ representation down to the maximal Abelian subgroup $U(1)$. This 
$U(1)\cong SO(2)$ corresponds to the residual rotations around the 
preferred vacuum direction. Afterwards the theory is broken down
further to $Z_2$ by another adjoint Higgs field. $Z_2$ as the center
of $SU(2)$ is mapped onto the identity in the adjoint 
representation (cf (\ref{center}))
and thus acts trivially on the Higgs field.

\begin{figure}[t]
\begin{picture}(380,175)
\small
\put(110,15){\includegraphics{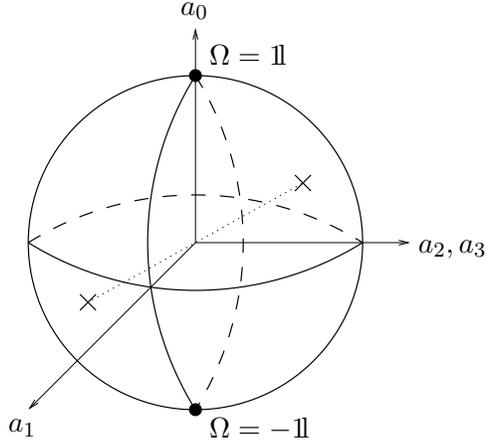}}
\put(168,167){$a_0$}
\put(258,77){$a_2,a_3$}
\put(103,9){$a_1$}         
\put(179,7){$\Omega=-\id$}
\put(179,148){$\Omega=\id$}
\normalsize
\end{picture}
\caption{\small{The group $SU(2)$ parametrised as a three-sphere
$a_0^2+\sum_{i=1}^3 a_i^2=1$. The center $Z_2=\pm\id$ sits on the
poles, and every closed path can be contracted to a point:
$\pi_1(SU(2))=\id$. After identification of opposite points $(\times)$
one arrives at the group $SO(3)$. Every path connecting two
opposite points is now closed, but not contractible:
$\pi_1(SO(3))=Z_2$.}}
\protect{\label{su2sphere}}
\end{figure}

To be explicit we parametrise $SU(2)$ as a three-sphere
%\footnote{
%One should not be confused: The Lie \textit{group} $SU(2)$ can be 
%parametrized by the generators $\sigma_i$ of its
%Lie \textit{algebra} plus the identity.}
(cf Fig. \ref{su2sphere}),
\begin{eqnarray}
\label{para}
SU(2)\ni \Omega=a_0 \id+ia_i\sigma_i,
\quad a_\mu\:\mbox{real},
\quad a_0^2+\vec{a}^2=1
\end{eqnarray}
The center $Z_2$ sits on the poles $a_0=\pm 1,\:\vec{a}=0,\:\Omega=\pm\id$.
Clearly it sends an $I=1$ field $\phi$ back to itself,
\begin{eqnarray}
\label{center}
\phi \>^\Omega\!\!\phi=\Omega^\dagger\phi\,\Omega=(\pm)^2\id\phi\id=\phi 
\end{eqnarray}
It is just the identification of opposite points that leads to the 
group $SO(3)$,
\begin{eqnarray*}
 SU(2)/Z_2 \cong SO(3)
\end{eqnarray*}
In addition non-contractible closed paths are created, namely those 
which connect two opposite points. The first homotopy group of $SO(3)$
is non-trivial,
%\footnote{$SO(3)$ is \textit{doubly connected}.}
\begin{eqnarray*}
\pi_1(SO(3))=Z_2
\end{eqnarray*}
%with multiplication as the composition law.
Therefore, a $SU(2)/Z_2$
vortex carries a \textit{multiplicative} quantum number $\pm 1$. $+1$
stands for the contractible situation, which is homotopic to the
trivial vacuum. Unlike the case above there is only a finite
number of different vortices, namely two.

\begin{figure}[t]
\begin{minipage}{0.5\linewidth}
\begin{picture}(0,70)
\small
\put(30,0){\includegraphics{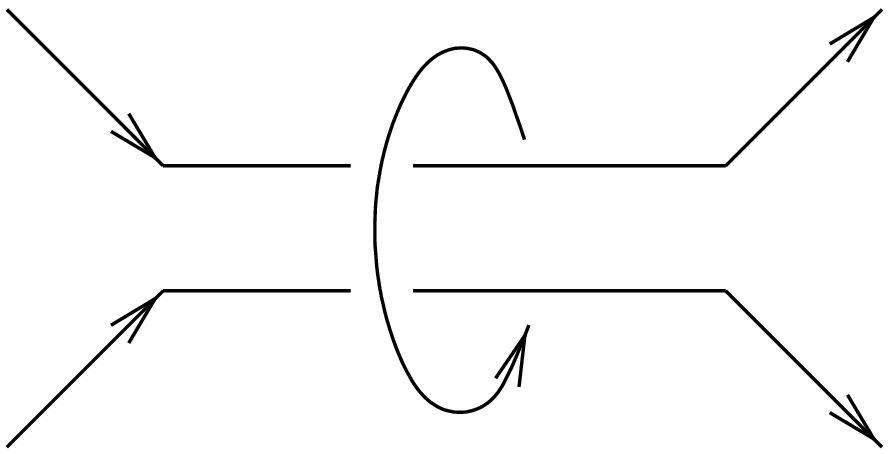}}
\put(45,55){$-$}
\put(45,7){$-$}
\put(100,-2){$+$}
\normalsize
\end{picture}
\end{minipage}
\begin{minipage}{0.5\linewidth}
\begin{picture}(0,70)
\small
\put(30,0){\includegraphics{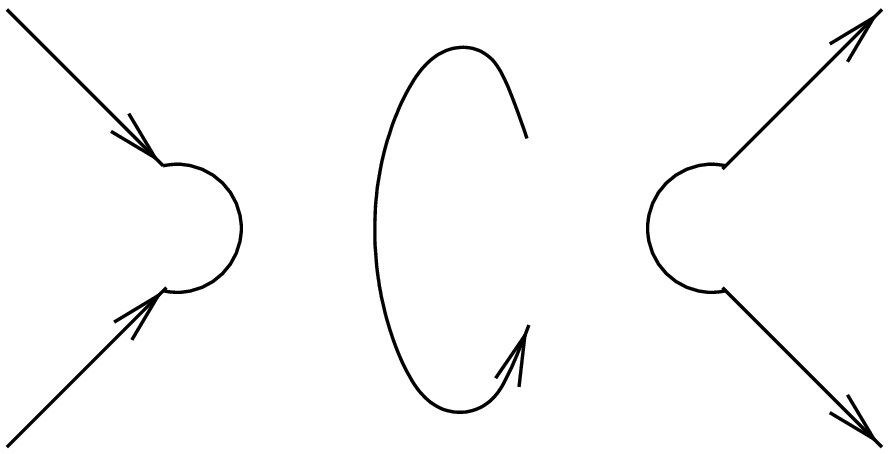}}
\put(100,-2){$+$}
\normalsize       
\end{picture}
\end{minipage}
\caption{\small{A graphical proof that the $SU(2)/Z_2$ vortex is
non-orientable: Two incoming vortices with quantum number $-1$ produce
a vortex with quantum number $+1$. It is equivalent to the vacuum, and
the arrows become inconsistent.}}
\protect{\label{orientable}}
\end{figure}

Another significant difference to the $U(1)/\id$ case is the
\textit{orientability}: One could try to label the quantum numbers of
the vortices by arrows. But as Fig. \ref{orientable} indicates,
these arrows are unstable in the $SU(2)/Z_2$ case.

\begin{figure}[h]
\begin{minipage}{0.5\linewidth}
\begin{picture}(0,90)
\small
\put(15,20){\includegraphics{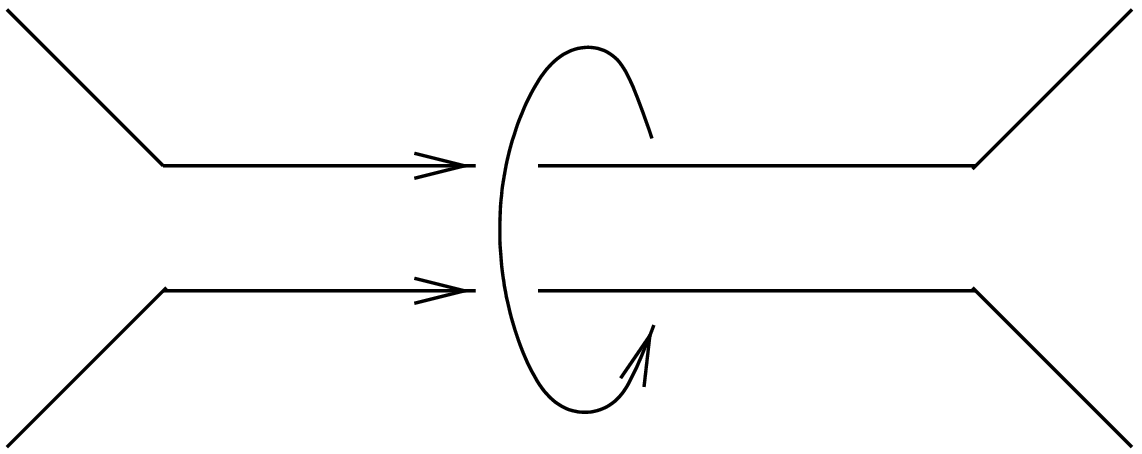}}
\put(75,0){$U(1)\>Z_2$}
\normalsize
\end{picture}
\end{minipage}
\begin{minipage}{0.5\linewidth}
\begin{picture}(0,90)
\small
\put(15,20){\includegraphics{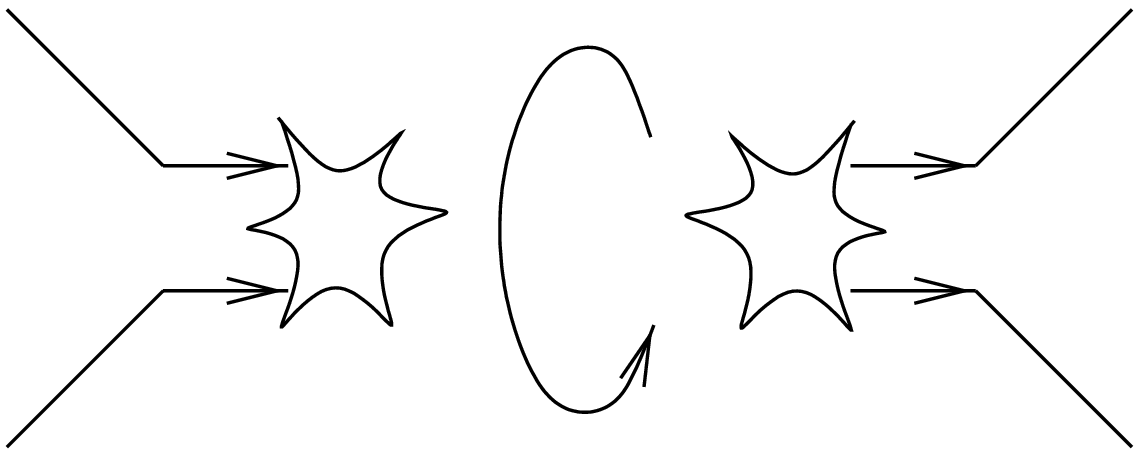}}
\put(50,0){$SU(2)\>U(1)\>Z_2$}
\normalsize       
\end{picture}
\end{minipage}
\caption{\small{The snapping of vortices (see text).}}
\protect{\label{snapping}}
\end{figure}

Altogether we have that
\begin{eqnarray*}
\framebox{$\left.\begin{array}{c}
\mbox{the } U(1)\>\id
\mbox{ vortex  has an additive quantum number}\\
n\in \mathbb{Z} \mbox{ and is orientable.}
\end{array}\right.$}
\end{eqnarray*}
while
\begin{eqnarray*}
\framebox{$\left.\begin{array}{c}
\mbox{the }SU(2)\>U(1)\>Z_2
\mbox{ vortex  has a multiplicative quantum number}\\
\pm 1\in Z_2 \mbox{ and is non-orientable.}
\end{array}\right.$}
\end{eqnarray*} 
This statement has a very interesting physical consequence
(cf Fig. \ref{snapping}). Imagine
two $U(1)\>\id$ vortices with flux $2\pi/e$, respectively. The
total flux is $4\pi/e$. But seen as
$SU(2)\>U(1)\>Z_2$ vortices the intermediate vortex  is equivalent to
the vacuum with flux zero. The vortices have snapped creating a
pair of something that carries magnetic charge.
 We conclude there must be \textit{magnetic monopoles}
with magnetic charge $4\pi/e$ (or an integer multiple of
it). We will analyse these magnetic monopoles in the next chapter.

\chapter{Magnetic Monopoles}

\section{Electric and Magnetic Charges and the Dirac Condition}

Studying the vortices of Chapter 2 automatically revealed the
existence of pure magnetic charges in non-Abelian gauge theories
$G\>U(1)$.
As worked out,
\begin{eqnarray*}
SU(2)\stackrel{I=1}{\>}
U(1)
\end{eqnarray*}
produces magnetic monopoles with magnetic flux $\pm4\pi/e=g_m$.
The minimally allowed electric charge is
$q=e/2$
for $I=1/2$ doublets. Indeed the \textit{Dirac condition},
\begin{eqnarray*}
\framebox{$q g_m=2\pi n$}\qquad n\in \mathbb{Z}
\end{eqnarray*}
is exactly obeyed.

We remind the reader of its origin. In Maxwell's theory isolated
magnetic sources are excluded and the magnetic field is 
the curl of a smooth gauge field. Thus for a monopole the Maxwell
field has to be singular at the so-called Dirac string. This is a
curve which extends from the monopole to infinity\footnote{
It could also extend to a second monopole with inverse charge such that
the net flux through a surface including both vanishes.}
and carries a magnetic flux $g_m$. Physically the magnetic monopole is the
endpoint of a tight magnetic solenoid which is too thin to detect. The
string itself can be moved to a different position by a singular gauge
transformation.

\begin{figure}[t]
\begin{picture}(380,120)
\small
\put(140,0){\includegraphics{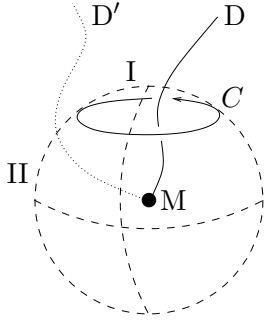}}
\put(212,110){D}
\put(162,110){D$^\prime$}
\put(175,89){I}
\put(211,79){$C$}
\put(130,50){II}
\put(188,40){M}         
\normalsize
\end{picture}
\caption{\small{For introducing a magnetic monopole (M) into an
Abelian theory, a Dirac string (D) is needed. When moving around the
string on a circle $C$ the wave function picks up a phase. This phase
is proportional to the magnetic flux carried by the string, and the Dirac
condition follows. The Dirac string can be put on a different position
(D$^\prime$) by a singular gauge transformation.
In the fibre bundle construction the circle $C$ is
the overlap region of two patches (I,II).}}
\end{figure}

Now we consider a matter field in the presence of that string. The
vector potential enters the Schr\"odinger equation via the conjugate
momentum,
\begin{eqnarray*}
H=H(\vec{p}-q\vec{A},V)
\end{eqnarray*}
When we go around the string the wave function picks up a phase,
\begin{eqnarray*}
q\int_C\!\!\vec{A}\,\d\vec{r}=qg_m
\end{eqnarray*}
In the Aharonov-Bohm effect the same consideration leads to a phase
shift of two electron beams. Since the wave function has
to be single-valued and
no AB effect shall take place, we have the restriction that $qg_m=2\pi n$.
The existence of one monopole quantises all electric charges.\footnote{
Introducing the magnetic charge $q_{\rm mag}=g_m/4\pi$ the Dirac condition
reads $qq_{\rm mag}=n/2$.}

One can avoid the singularities by a fibre bundle construction:
Every two-sphere around the monopole consists of two patches on which
the gauge fields are regular, respectively. The patches overlap on some
circle $C$ around the string. There a gauge transformation (`transition
function') $\Omega=e^{ie\Lambda}$ connects the fields,
\begin{eqnarray*}
\vec{A}^{\mbox{(II)}}=\vec{A}^{\mbox{(I)}}+\vec{\nabla}\Lambda,\qquad
\psi^{\mbox{(II)}}=\psi^{\mbox{(I)}}e^{ie\Lambda}
\end{eqnarray*}
For $\Omega$ to be single-valued $\Lambda$ has to fulfil,
\begin{eqnarray*}
q[\Lambda(\varphi=2\pi)-\Lambda(\varphi=0)]=2\pi n
\end{eqnarray*}
The functions $\Lambda$ fall into disjoint classes, the simplest
representatives of which are just proportional to the angle $\varphi$
around the string,
\begin{eqnarray*}
\Lambda=\frac{n}{q}\varphi
\end{eqnarray*}
On the other hand the magnetic flux is given by the $A$-integral on
the boun\-da\-ry. Here it is the $\vec{\nabla}\Lambda$-integral on that
circle,
\begin{eqnarray}
\label{Diracwinding}
g_m=\int_C\!\!\vec{\nabla}\Lambda\d\vec{r}=\Lambda|^{2\pi}_0=2\pi n/q
\end{eqnarray}
The Dirac condition has a topological meaning: The transition function
$\Omega:S^1\>U(1)$ has a winding number and (\ref{Diracwinding}) is
how to compute it.

\section{Construction of  Monopole Solutions}
\label{construct}

After the excursions through lower dimensions we present in this
section a 3+1 dimensional theory. No surprise, it is a non-Abelian
gauge theory with gauge group $SU(2)$ and a Higgs field $\phi$ in the
$I=1$ representation,
\begin{eqnarray}
\label{Lmonopole}
\mathcal{L}=-\ha\left(\D_\mu\phi_a\right)^2
-\frac{\lambda}{8}\left(\phi_a^2-F^2\right)^2
-\frac{1}{4}F_{\mu\nu}^a F_{\mu\nu}^a
\end{eqnarray}
Both $\phi$ and $A$ are elements of the Lie algebra $su(2)\cong
\mathbb{R}^3$,
\begin{eqnarray*}
\phi=\phi^a\tau_a,\qquad A_\mu=A_\mu^a\tau_a,\qquad \tau_a=\sigma_a/2
\end{eqnarray*}
and the non-Abelian definition of the covariant derivative
and the field strength includes commutator terms,
\begin{eqnarray*}
\D_\mu\phi_a=\pa_\mu\phi_a+\ep_{abc}A_\mu^b\phi^c,\qquad
F_{\mu\nu}^a=\pa_\mu A_\nu-\pa_\nu A_\mu+\ep_{abc} A_\mu^b A_\nu^c
\end{eqnarray*}

\begin{figure}[t]
\begin{picture}(380,125)
\small
\put(105,-60){\includegraphics{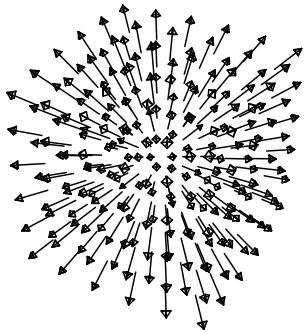}}
\normalsize
\end{picture}
\caption{\small{The Higgs field of a monopole configuration shows a
`hedgehog' behaviour. It points in the same direction (in isospace)
like its argument (in coordinate space) and has winding number 1.}}
\protect{\label{hedgehog}}
\end{figure}

We look for static solutions of the field equations. Repeating the
arguments from the previous chapter we expect $\phi$ to live on a
sphere with radius $F$ asymptotically:
$\phi^a\phi^a=F^2$. Topologically it is a mapping from $S^2$ (as the
boundary of the coordinate space) to another $S^2$ (of algebra
elements with fixed length). The degree of this mapping is an
integer. Alternatively one can see immediately that $\phi$ transforms
under $(SU(2)_{I=1}\equiv SO(3))/U(1)$. Its second homotopy
group\footnote{
Analogously to $\pi_1$, the second homotopy group $\pi_2$
is the group of mappings from $S^2$ into the given manifold.}
is the group of (even) integers. The one to one mapping,
\begin{eqnarray*}
\phi^a(x)\>F\hat{x}^a,\qquad \hat{x}^i=x^i/|\vec{x}|,
\,|\vec{x}|=\sqrt{x^ix^i},
\:\:i=1,2,3
\end{eqnarray*}
is the first non-trivial mapping. On the boundary the same things
happen as before, $\phi$ `winds around once'. Notice that the isospace
structure (indices $a$) is mixed with the space-time structure (indices
$i$). Inside, $\phi$ is of the same form,
\begin{eqnarray}
\label{reg_mon_1}
\phi^a(x)=\phi(|\vec{x}|)\hat{x}^a
\end{eqnarray}
with a regular function $\phi(|\vec{x}|)$.
The solution is depicted in Fig. \ref{hedgehog}. It is called
a `hedgehog' and has a zero inside.

Corresponding to our previous discussions we make the natural ansatz:
\begin{eqnarray}
\label{reg_mon_2}
A_0=0,\qquad A_i^a=\ep_{iaj}\hat{x}_jA(|\vec{x}|)
\end{eqnarray}
The first condition means that there is no electric field, 
the second one is the analogue of the circular gauge
field in (\ref{A_i}). Again it exploits the mixing of isospace and
coordinate space indices. The magnetic field at spatial infinity looks like if
there were a magnetic charge inside: $B_i\propto x_i/|\vec{x}|^3$.

What happens after spontaneous symmetry breaking? To extract the
physical content one usually makes use of the local gauge symmetry. We
diagonalise $\phi$, i.e. force it to have only a third
component.
The corresponding gauge is called `unitary gauge',
\begin{eqnarray}
\label{unitary}
\phi\>\left(\begin{array}{c}0\\0\\F
\end{array}\right)+
\left(\begin{array}{c}0\\0\\ \eta
\end{array}\right)
\end{eqnarray}
$F$ and $\eta$ are the vacuum expectation value and the fluctuations
of the Higgs field, respectively. The third component of the
Higgs field gets a mass,
\begin{eqnarray*}
M_\eta\equiv M_H=F\sqrt{\lambda}
\end{eqnarray*}
For the gauge fields it is the other way round. $A_\mu^1$ and $A_\mu^2$ are
massive vector bosons, $A_\mu^3$ is the massless photon referring to the
unbroken $U(1)$ in the third direction,
\begin{eqnarray*}
M_{A^{1,2}}\equiv M_{W^\pm}=eF,\qquad M_{A^3}=0
\end{eqnarray*}
The gauge field coupling is denoted by `$e$', since this is also the
charge unit with respect to the residual Maxwell potential $A_\mu^3$.

It can be shown that under spontaneous symmetry breaking the hedgehog
configuration turns into a Dirac monopole. It resides in the origin,
while the Dirac string is placed along the negative $z$-axis. The last
point is not difficult to explain (see also Exercise (iv)): The gauge
transformation has to rotate $\phi$ onto the positive $z$-axis in the
algebra. It can be written in terms of the spherical coordinates
$\theta$ and $\varphi$. The latter becomes ambiguous on the
$z$-axis. This does not matter at its positive part, since $\phi$ is
already of the desired form there. But on the negative part it points
just in the opposite direction and there are a lot of rotation
matrices.
Whatever direction we choose for the spontaneous symmetry breaking in
(\ref{unitary}), a singularity occurs: the unitary gauge changes
the asymptotic behaviour of $\phi$ from the hedgehog to the trivial one.
Like in the explicit case the singularity is always situated on the opposite
 part of the chosen axis. 
We have found again that the existence of the Dirac string is
gauge invariant, its position is gauge dependent. 

\section{Existence of Monopoles}

What is the general feature of theories which allow for monopole
solutions? The $U(1)_{em}$ must be embedded as a
subgroup\footnote{
It is also possible to embed a product of $U(1)$'s into $G$ which will
be the case for the Abelian Projection of $SU(3)$ and higher groups in
chapter 5.} in a larger non-Abelian group $G$, and
\begin{eqnarray*}
\pi_1(G)<\mathbb{Z}
\end{eqnarray*} 
For the winding number to be finite, $G$ must have a \textit{compact}
covering group\footnote{The covering group of a given Lie group is
constructed from the same Lie algebra, but is simply connected.}. This
is the topological reason for the statement, that there are
\textbf{no magnetic monopoles in the electroweak sector of the
Standard Model}
\begin{eqnarray*}
SU(2)_I\times U(1)_Y\>U(1)_{em}
\end{eqnarray*}
$U(1)$ has a non-compact covering group, namely
$\mathbb{R}^+$. Thus $\pi_1(SU(2)_I\times U(1)_Y)$ is still
$\mathbb{Z}$ and vortices refuse to snap.

In Grand Unified Theories (GUT) the Standard Model is embedded in a
larger group like $SU(5)$. Then monopole solutions become possible
again and have magnetic flux $2\pi/e$. As we will see in the next
section, its mass is bigger than the mass of the massive vector bosons
$W$ of the theory. The GUT scale is $10^{16}{\rm GeV}$ and the monopole
mass is of the order of $m_{\rm Planck}$.

So far no experiment has detected magnetic monopoles. Perhaps they
exist somewhere in the universe. Not only GUT's but also cosmological
models predict their existence.

The generalization to monopoles with added \textit{electric} charge
was introduced by Julia and Zee \cite{JuZe}. These particles are
called `dyons'. It is easy to imagine that one can add multiples of
$A_\pm$ to a monopole,
\begin{eqnarray*}
g_m=\frac{4\pi}{e},\qquad q=ne
\end{eqnarray*}

\section{Bogomol'nyi Bound and BPS States}

For estimating the monopole mass we again use the Bogomol'nyi trick,
\begin{eqnarray*}
E=\int\!\!\d^3x\left(\ha(\vec{\D}\phi_a)^2+
\frac{\lambda}{8}(\phi_a^2-F^2)^2+
\ha \vec{B}_a^2\right)
\end{eqnarray*} 
Note that the homogeneous field equation for non-Abelian theories
read,
\begin{eqnarray*}
\D_iB_i^a=\ha \ep_{ijk}\D_iF_{jk}^a=0
\end{eqnarray*}
It is the usual Bianchi identity that allows for the introduction of
the $A$-field. We use it to reduce the number of squares,
\begin{eqnarray*}
(\vec{\D}\phi_a)^2+\vec{B}_a^2=
(\vec{\D}\phi_a \pm \vec{B}_a)^2 \mp 2 \vec{B}_a\vec{\D}\phi_a 
\end{eqnarray*}
We rewrite  the last term in a total derivative,
\begin{eqnarray*}
\vec{B}_a\vec{\D}\phi_a =\vec{\pa}(\vec{B}_a\phi_a)
\end{eqnarray*}
Its contribution to the energy is gauge invariant, and we compute it
in the unitary gauge,
\begin{eqnarray*}
\int\!\!\d^3x\vec{\pa}(\vec{B}_a\phi_a)
=F\int_{S^2_\infty}\vec{B}_3\vec{n}
=\frac{4\pi}{e}F
\end{eqnarray*}
Thus the energy of the monopole is bounded from below by the mass of the
$W$-boson,
\begin{eqnarray}
\label{monmass}
E=\int\!\!\d^3x\left[\ha(\vec{\D}\phi_a \pm \vec{B}_a)^2
+\frac{\lambda}{8}(\phi_a^2-F^2)^2\right]
+\frac{4\pi}{e^2}M_W
\end{eqnarray}
The bound is saturated for vanishing potential, $\lambda=0$. The exact
solution to the remaining equations,
\begin{eqnarray*}
\vec{\D}\phi_a \pm \vec{B}_a=0,\qquad |\phi|\>F
\end{eqnarray*}
was given by Sommerfield and Prasad. These so-called BPS states have a
mass,
\begin{eqnarray*}
M_{\rm mon}=\frac{4\pi}{e^2}M_W
\end{eqnarray*}
and are important for supersymmetric theories.

\section{Orbital Angular Momentum for $\mathbf{qg}$ Bound States}

In this section we look for electrically charged particles bound to
the monopole. All particles with $U(1)$ charge originate from $SU(2)$
representations,
\begin{eqnarray*}
I=\mbox{integer} &\longrightarrow&q_{U(1)}=ne\\
I=\mbox{integer}+\ha &\longrightarrow&q_{U(1)}=(n+\ha)e
\end{eqnarray*}
consistent with the Dirac condition,
\begin{eqnarray*}
g_{\rm mon}=\frac{4\pi}{e}\longrightarrow qg_{\rm mon}=2\pi n
\end{eqnarray*}
Let us take a minimal charge $q=\ha e$, i.e. a field $\psi$ in the
defining representation of $SU(2)$,
\begin{eqnarray*}
I=\ha:\quad\psi={\psi_1 \choose \psi_2}
\end{eqnarray*}
and consider $\psi$ near a monopole. The wave equation reads:
\begin{eqnarray*}
\D^2\psi+\mu^2\psi\>0\qquad
\mbox{or}\qquad\left(\gamma_\nu\D_\nu+\mu\right)\psi\>0
\end{eqnarray*}
$\mu$ plays the role of a binding potential.
In the regular description the monopole solution has
$\phi^a(x)=\phi(|x|)\hat{x}^a$. Its rotational symmetry
%\footnote{
%which corresponds to the mixing of isospace and coordinate space in
%$\ep_{iaj}$ for the gauge field.} 
can only be exploited if we rotate
$\phi^a$ together with $\vec{x}$. That is spacial $SO(3)$ rotations
must be \textit{coupled} to isospin $SU(2)$ rotations,
\begin{eqnarray*}
SU(2)_{\rm space}\times SU(2)_{\rm isospin}\longrightarrow
SU(2)_{\rm diag} 
\end{eqnarray*}
where $SU(2)_{\rm diag}$ is the invariance group of the monopole. The
representation of our $\psi$ in this $SU(2)_{\rm diag}$ is,
\begin{eqnarray*}
L_{\rm tot}&=&L_{\rm space}+L_{\rm isospin}\\
l_{\rm tot}&=&l_{\rm space}\pm \ha
\end{eqnarray*}
$\psi$ may be a scalar under spacial rotations but carries now half
spin! Similar things happen, when we give $\psi$ an ordinary spin
$\pm\ha$: the angular momentum becomes an integer! We have found that
\cite{tHooftHasenfratz}
\begin{eqnarray*}
\framebox{$\left.\begin{array}{c}
q=\ha e\: 
\mbox{particles will bind to a magnetic monopole }\\
\mbox{with}\: g_m=\frac{4 \pi}{e}=\frac{2\pi}{q}\:
 \mbox{in such a way that}\\
 \mbox{ the orbital angular momentum is
  integer}+\ha.
\end{array}\right.$}
\end{eqnarray*}
The spin becomes half-odd integer, although the monopole is a spin 0
object. The \textit{anomalous spin addition theorem} for $qg$ bound
states with $q\cdot g=2\pi$ reads:
\begin{eqnarray*}
\framebox{integer $+$ integer $\longrightarrow$ integer$+\frac{1}{2}$}
\end{eqnarray*}
etc. Something like this was never seen in quantum field theory before.

\begin{figure}[t]

\begin{minipage}{0.32\linewidth}
\begin{picture}(100,115)
\small
\put(10,15){\includegraphics{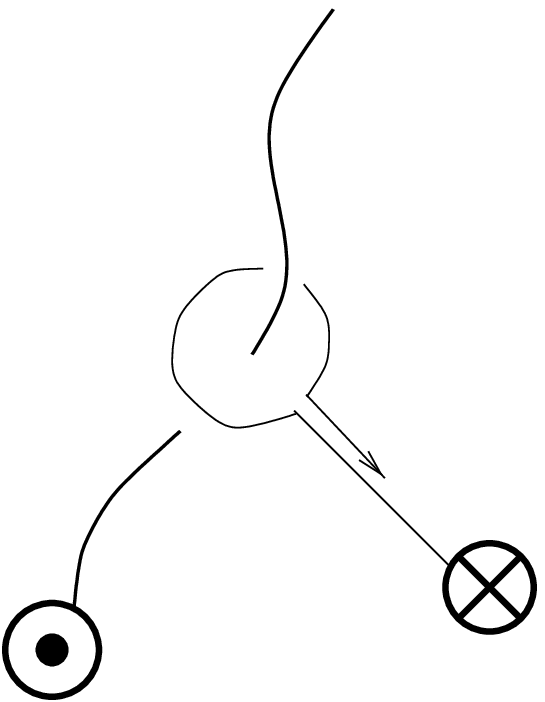}}
\put(59,105){\mbox{D}}
\put(28,20){\mbox{mon}}
\put(91,29){\mbox{el}}
\put(45,3){\mbox{(a)}}
\normalsize
\end{picture}
\end{minipage}
\begin{minipage}{0.32\linewidth}
\begin{picture}(100,115)
\small
\put(10,15){\includegraphics{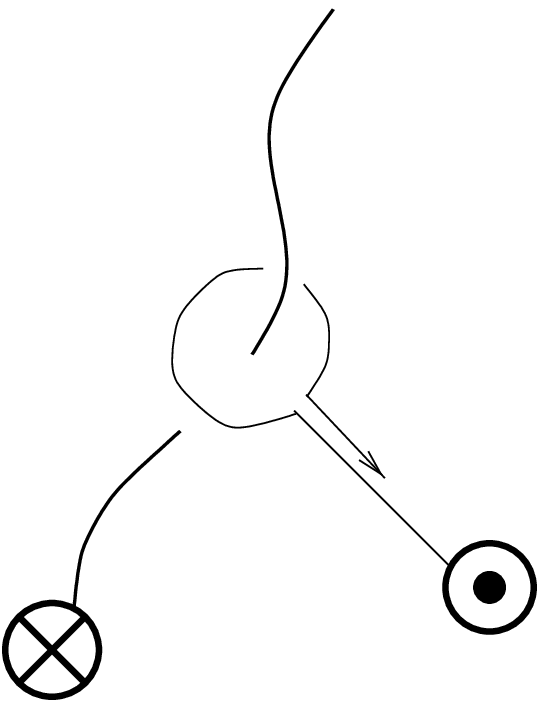}}
\put(45,3){\mbox{(b)}}
\normalsize      
\end{picture}
\end{minipage}
\begin{minipage}{0.32\linewidth}
\begin{picture}(100,115)
\small
\put(10,15){\includegraphics{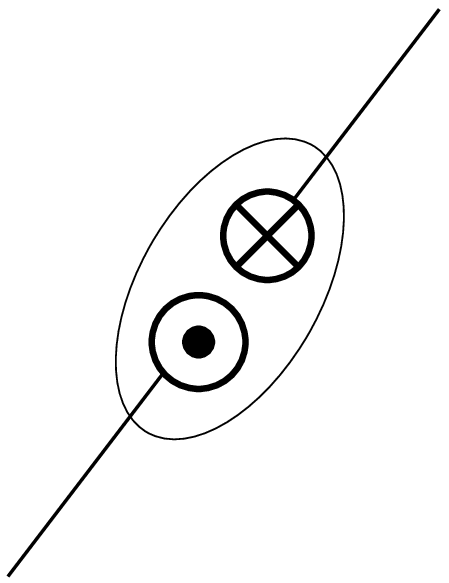}}
\put(45,3){\mbox{(c)}}
\normalsize      
\end{picture}
\end{minipage}
\caption{\small{An electrically charged particle (el) feels the Dirac
string (D) of a magnetic monopole (mon). When moving around the string,
$\psi_{\rm el}$ picks up a phase factor (a). This is equivalent to
move the monopole around the electric charge with its electric string
(b). For bound states (c) we choose both strings to point in opposite
directions. }}
\protect{\label{objects}}
\end{figure}

The reasoning heavily relies on the existence of Dirac strings. Imagine
an electric charge in the fundamental representation and a monopole
like in Fig. \ref{objects}. The wave function of the electric charge
$\psi_{\rm el}$ feels the string coming from the monopole. The Maxwell
equations allow us to interprete the resulting phase shift also after
interchanging electric and magnetic charges. Then $\psi_{\rm mon}$
feels the string coming from the electric charge.
Accordingly, the $eg$ bound state has two strings. When they are
oppositely oriented, the bound state looks like if it has a string
running from $-\infty$ to $\infty$. We remind the reader that one part
of the string is only felt by the magnetic monopole wave function,
while the other part only by the electric charge wave function.

\begin{figure}[t]
\begin{minipage}{0.32\linewidth}
\begin{picture}(110,115)
\small
\put(0,25){\includegraphics{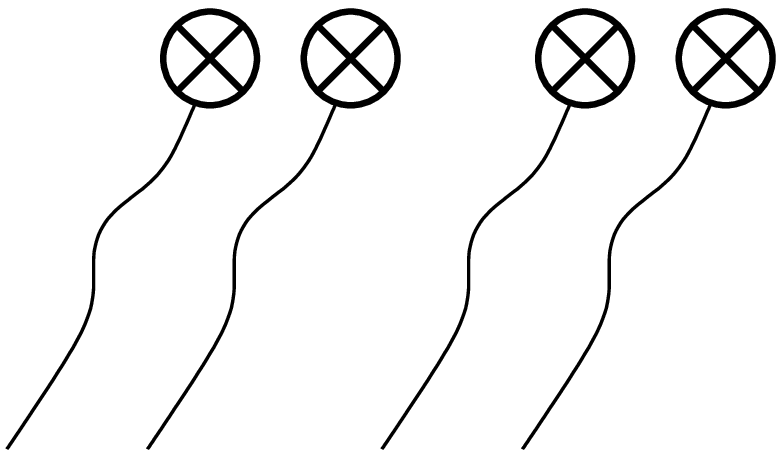}}
\put(28,96){\mbox{1}}
\put(48,96){\mbox{2}}
\put(63,79){$=$}
\put(83,96){\mbox{2}}
\put(103,96){\mbox{1}}
\put(45,3){\mbox{(a)}}
\normalsize
\end{picture}
\end{minipage}
\begin{minipage}{0.32\linewidth}
\begin{picture}(110,115)
\small
\put(0,40){\includegraphics{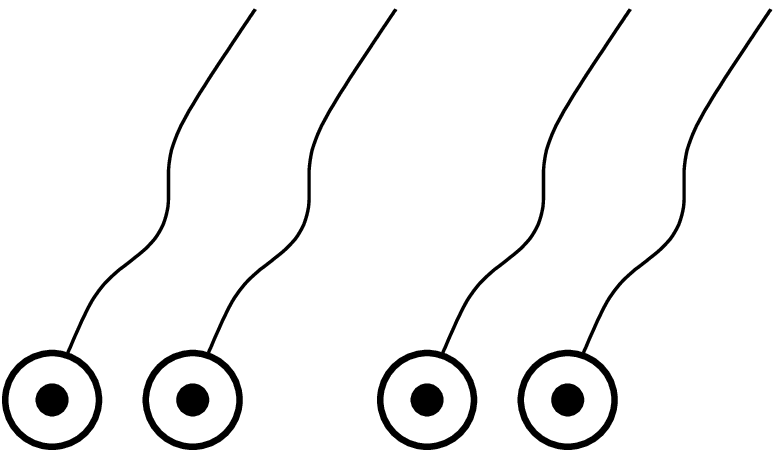}}
\put(5,27){\mbox{1}}
\put(25,27){\mbox{2}}
\put(41,45){$=$}
\put(60,27){\mbox{2}}
\put(80,27){\mbox{1}}
\put(45,3){\mbox{(b)}}
\normalsize      
\end{picture}
\end{minipage}
\begin{minipage}{0.32\linewidth}
\begin{picture}(110,115)
\small
\put(5,30){\includegraphics{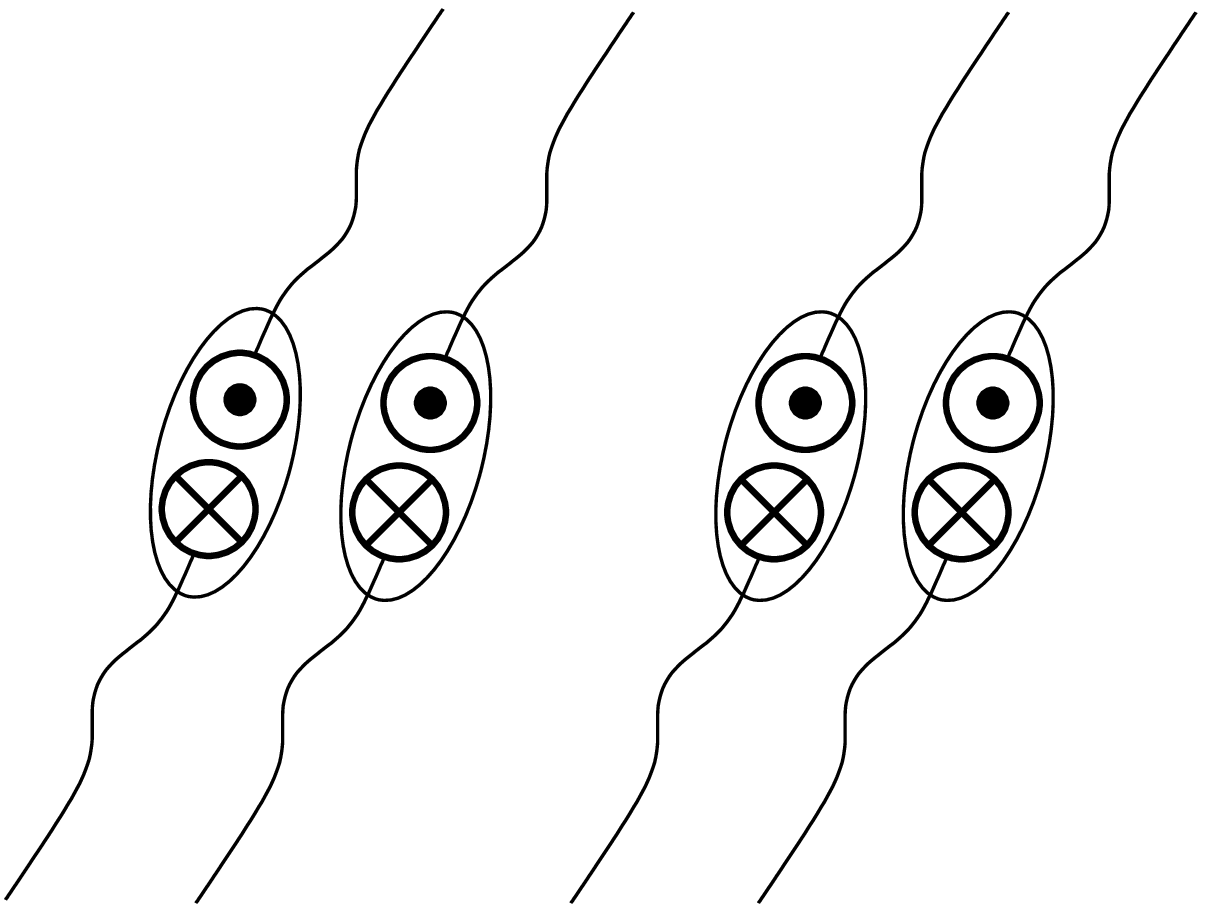}}
\put(3,19){\mbox{1}}
\put(21,19){\mbox{2}}
\put(51,66){$\stackrel{(?)}{=}$}
\put(52,19){\mbox{2}}
\put(70,19){\mbox{1}}
\put(45,3){\mbox{(c)}}
\normalsize      
\end{picture}
\end{minipage}
\caption{\small{The two-particle-wave function of monopoles (a) and
electric charges (b) is symmetric due to the fact that the charges do
not feel string of their own kind. What will happen for bound states
(c) is discussed in the text. }}
\protect{\label{pair}}
\end{figure}

Now consider two identical \textit{bosonic} monopoles and two
identical \textit{bosonic} electric charges. Since the charges do not
feel strings of its own kind, they can be moved around freely
(cf Fig.~\ref{pair}(a) and \ref{pair}(b)). 
In the same way combine the charges into \textit{identical} bound
states. What is their statistics? What happens to the wave function
when we interchange two of these?

 They are the states of Fig.~\ref{pair}(c),
 and, considering the wave functions of these objects,
\textit{with all strings attached}, there will be no anomalous sign
switch if we interchange the two objects.

However, we may now observe that, as long as the objects remain
tightly bound, each as a whole feels a string that runs from $-\infty$
to $+\infty$: since they carry both electric and magnetic charge, they
each feel the combination of the strings from Fig.~\ref{pair}(a) and
\ref{pair}(b). To be precise: if $\vec{r}_1$ is the center of mass of
bound state 1 and $\vec{r}_2$ is the center of mass of bound state 2,
the wave function is,
\begin{eqnarray*}
\psi_{12}(\vec{r}_1,\vec{r}_2)=\psi_{\rm
cm}(\frac{\vec{r}_1+\vec{r}_2}{2})
\psi_{\rm rel}(\vec{r}_1-\vec{r}_2)
\end{eqnarray*}
and it is $\psi_{\rm rel}(\vec{r}_1-\vec{r}_2)$ that feels a Dirac
string running through the origin from $z=-\infty$ to $z=+\infty$.

The point is now that we may \textit{remove} this Dirac string by
multiplying $\psi_{\rm rel}$ with
\begin{eqnarray*}
e^{i\varphi(\vec{r}_1-\vec{r}_2)}
\end{eqnarray*}  
This produces a minus sign under the interchange
$\vec{r}_1\leftrightarrow\vec{r}_2$.
The bound states obey Fermi-Dirac statistics \cite{Goldhaber}!
After the Dirac string is removed, the system of two identical bound
states is treated as an ordinary system of particles such as molecules.

\section{Jackiw-Rebbi States at a Magnetic Mo\-no\-pole}

In the first chapter we have seen that there are chiral fermions in
the background of a kink. We briefly discuss this effect for the
monopole.
We introduce fermions $\psi$ transforming  under some representation
of the gauge group $SU(2)$ given by the generators $(T^a)_{ij}$,
\begin{eqnarray*} 
\mathcal{L}=\mathcal{L}_{\rm mon}-
\bar{\psi}\gamma\D\psi-G\bar{\psi}_i\phi_aT^a_{ij}\psi_j
\end{eqnarray*}
The first part $\mathcal{L}_{\rm mon}$ is the theory
(\ref{Lmonopole}) we have discussed so far. Since the fermion couples
to the Higgs field it gets a mass,
\begin{eqnarray*}
\mbox{unitary gauge:}\:\langle\phi_a\rangle\>
\left(\begin{array}{c}0\\0\\F
\end{array}\right)\!:\:\:
m_\psi=C\cdot GF
\end{eqnarray*}
where $C$ is a coefficient depending on the representation.
%, the usual square of
%the angular momentum for $SU(2)$.

The energies are again the eigenvalues of the
Hamilton operator,
\begin{eqnarray*}
\gamma_4\frac{\pa}{i\pa t}\>\gamma_4 E
\end{eqnarray*}
We use an off-diagonal representation for the matrices $\vec{\alpha}$
and $\beta$,
\begin{eqnarray*}
\gamma_4\vec{\gamma}=-i\vec{\alpha},
\quad \vec{\alpha}=\left(\begin{array}{cc}
0&\vec{\sigma}\\~\vec{\sigma}&0
\end{array}\right),
\quad \gamma_4=\beta,
\quad \beta=-i\left(\begin{array}{cc}
0&\id\\-\id&0
\end{array}\right)
\end{eqnarray*}
The energy equation reads
\begin{eqnarray*}
\left[\vec{\alpha}(\vec{p}+gT^a\vec{A}_a)+
\beta GT^a\phi_a\right]\psi=E\psi
\end{eqnarray*}
We split $\psi$ into its chirality components $\psi={\chi^+ \choose
\chi^-}$ and insert the magnetic monopole in its regular form
((\ref{reg_mon_1}),(\ref{reg_mon_2})),
\begin{eqnarray*}
\left[\vec{\sigma}(\vec{p}+gA(|\vec{x}|)T^a\left(\vec{\sigma}\wedge\vec{r})_a)
\pm iG\phi(|\vec{x}|\right)T^a\hat{x}_a\right]\chi^{\pm}=E\chi^{\mp}
\end{eqnarray*}
which for $E=0$ separates into equations for $\chi^+$ and $\chi^-$,
respectively.

\begin{figure}[b]
\begin{minipage}{0.5\linewidth}
\begin{picture}(380,135)
\put(30,-100){\includegraphics{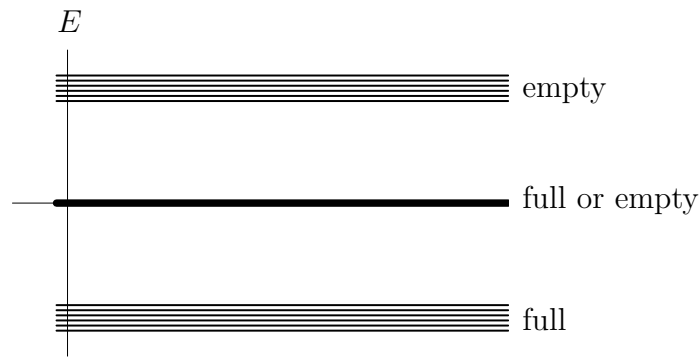}}
\put(76,124){$E$}
\put(252,99){\mbox{empty}}
\put(252,57){\mbox{full or empty}}
\put(252,11){\mbox{full}}        
\end{picture}
\end{minipage}
\caption{\small{The spectrum of fermions in the background of a
magnetic monopole. There are $E=0$ Jackiw-Rebbi states, the degeneracy 
of which depends on the total angular momentum $j$.}}
\protect{\label{full_or_empty}}
\end{figure}

We have already discussed the symmetry of this equation. Invariant
rotations are generated by the total angular momentum $\vec{J}$,
\begin{eqnarray*}
\vec{J}=\vec{L}+\vec{S}+\vec{T}
\end{eqnarray*}
where $\vec{L}$, $\vec{S}$ and $\vec{T}$ are the ordinary angular
momentum, the spin and the isospin, respectively. Let us work in the
defining representation $t=1/2$ ($q=\pm e/2$)
and look for the simplest solutions
with $\vec{J}=\vec{L}=0$,
\begin{eqnarray*}
\vec{S}+\vec{T}=0
%,\qquad (\vec{\sigma}+\vec{T})\psi=0
\end{eqnarray*}
For this case Jackiw and Rebbi found \textit{one} solution,
\begin{eqnarray*}
E=0, \qquad(j=l=0).
\end{eqnarray*}
Note that $j=0$ inspite of $s=1/2$.

As for the kink, the Jackiw-Rebbi state lies inbetween the fermion and 
anti-fermion eigenstates. Whether it is full or empty does not change
the energy of the system (Fig. \ref{full_or_empty}). In most cases
the baryon number is just a conserved charge, the monopole
(anti-monopole) contributes to since it has:
\begin{eqnarray*}
\mbox{baryon number}\:&=&\pm\, 1/2\\
\mbox{electric charge}\:&=&\pm\, e/4
\end{eqnarray*}
Much more is to be said about the electric charges. Here we only state 
that the bound states behave like particles or anti-particles under
$U(1)_{\rm charge}$.

For the adjoint representation $t=1$ we have $j=1/2$. There are now
two solutions with $E=0$ and $j_z=\pm\, 1/2$ and we get a $2^2$-fold
degeneracy.

Notice that the Jackiw-Rebbi solution is a \textit{chiral} wave
function: $\chi^+$ and $\chi^-$ are eigenstates of $\gamma_5$, which
is block-diagonal in the chosen representation.
\chapter{Instantons}

New topological objects, the so-called instantons, arise in pure
non-Abelian gauge (Yang Mills) theories in four
dimensions. We approach the topic by investigating the structure of
gauge transformations.

\section{Topological Gauge Transformations}

Let us work in the Weyl (=temporal) gauge $A_0=0$, where the theory reduces to
\begin{eqnarray*}
F_{0i}^a=\pa_0 A_i^a,\qquad
\mathcal{L}=\ha(\pa_0 A_i^a)^2-\frac{1}{4}F_{ij}^a F_{ij}^a
=\ha\left(\vec{E}_a^2-\vec{B}_a^2\right). 
\end{eqnarray*}
The Lagrangian density is nothing but
the difference of kinetic and potential energy in a Yang Mills 
sense. The action of a gauge transformation $\Omega$ on a gauge field
$A$ is,
\begin{eqnarray*}
A_\mu\>\Omega(x)(\frac{1}{ie}\pa_\mu+A_\mu)\Omega^{-1}(\vec{x})
\end{eqnarray*}
Obviously the surviving invariance of the gauge $A_0=0$ consists of
time-independent gauge transformations,
\begin{eqnarray*}
\pa_t\Omega=0\Rightarrow \Omega(\vec{x},t)=\Omega(\vec{x})
\end{eqnarray*}
just like a global symmetry in time. The Hamiltonian of the theory, 
\begin{eqnarray*}
H=\int\!\!\d^3\vec{x}(E_i^a\pa_0 A_i^a-\mathcal{L})
=\ha\int\!\!\d^3\vec{x}(\vec{E}_a^2+\vec{B}_a^2)
\end{eqnarray*}
is the sum of the kinetic and potential energy, and commutes with
these gauge transformations,
\begin{eqnarray*}
[H,\Omega(\vec{x})]=0
\end{eqnarray*}
We can diagonalise both operators simultaneously,
\begin{eqnarray*}
H|\psi\rangle=E|\psi\rangle,\qquad
\Omega(\vec{x})|\psi\rangle=\omega(\vec{x})|\psi\rangle
\end{eqnarray*} 
The eigenvalues $\omega$ are constants of motion. Now 
infinitesimal gauge transformations give rise to eigenvalues
$\lambda$,
\begin{eqnarray*}
\Omega(\vec{x})=\id+i\ep\Lambda(\vec{x}),\qquad
\Lambda(\vec{x})|\psi\rangle=\lambda(\vec{x})|\psi\rangle
\end{eqnarray*}
The only values of $\lambda$ consistent with the unbroken
spacial
%\footnote{
%Translational invariance demands constant $\lambda$, but non-vanishing 
%constants lead to conflicts with rotational invariance.}
Lorentz transformations is
\begin{eqnarray*}
\framebox{$\lambda(\vec{x})=0$}
\end{eqnarray*}

\begin{figure}[t]
\begin{picture}(380,110)
\small
\put(80,0){\includegraphics{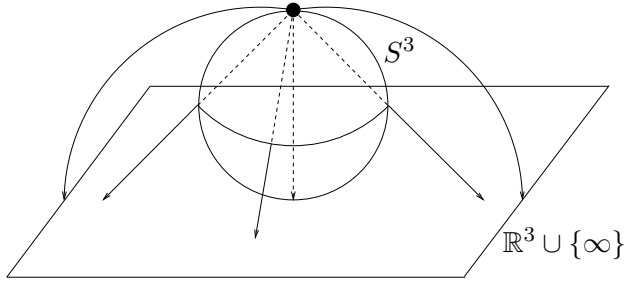}}
\put(223,82){$S^3$}
\put(268,10){$\mathbb{R}^3\cup\{\infty\}$}
\normalsize
\end{picture}
\caption{\small{The stereographic projection identifies the three-sphere
with the three-space compactified at spacial infinity, which is the image
of the north pole.}}
\protect{\label{stereo}}
\end{figure}

However, \textit{a class of $\Omega(\vec{x})$ exists that cannot be obtained
from infinitesimal gauge rotations $\Lambda(\vec{x})$}. We remind the
reader of the stereographic projection, which identifies the three-space
$\mathbb{R}^3$ compactified at spacial infinity with the three-sphere
$S^3$ (Fig. \ref{stereo}). If $\Omega$ has the same limit when going
to spacial infinity in any direction, it can be regarded as a function 
on $\mathbb{R}^3\cup\{\infty\}\cong S^3$. Since $SU(2)$ is again a
three-sphere we have,
\begin{eqnarray*}
\Omega: \: S^3\>S^3
\end{eqnarray*}
Similiarly to vortices and monopoles, these mappings are classified by 
the third homotopy group, which for $SU(2)$ is an integer,
\begin{eqnarray*}
\pi_3(SU(2))=\pi_3(S^3)=\mathbb{Z}
\end{eqnarray*}
Again the one to one mapping $\Omega_1$ is distinct from the trivial
mapping $\Omega_0(\vec{x})\equiv \id$ and has winding number
one. Representatives of higher windings are delivered by raising this
function to the $n$th power,
\begin{eqnarray*}
\Omega_n(\vec{x})=\left(\Omega_1(\vec{x})\right)^n
\end{eqnarray*}
Still these operators can be diagonalised together with the
Hamiltonian. Since they are unitary, their contants of motion are
characterised by an angle $\theta$,
\begin{eqnarray}
\label{intheta}
\Omega_1(\vec{x})|\psi\rangle=e^{i\theta}|\psi\rangle,\qquad
\Omega_n(\vec{x})|\psi\rangle=e^{i n\theta}|\psi\rangle,\qquad
\theta\in[0,2\pi)
\end{eqnarray}
$\theta$ is a Lorentz invariant. It is called the \textit{instanton
angle}. It is a fundamental parameter of the theory, which could be
measured in principle\footnote{The experimental evidence that there is 
little CP violation in QCD indicates that $\theta$ must be very small
or zero.}.

\begin{figure}[t]

\begin{minipage}{0.5\linewidth}
\begin{picture}(190,138)
\small
\put(40,40){\includegraphics{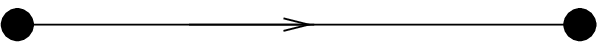}}
\put(18,25){$\lambda=0:\,A$}
\put(96,30){$\lambda$}
\put(125,25){$\lambda=1:\,^{\Omega_1}\!A$}
\put(90,0){(a)}
\normalsize
\end{picture}
\end{minipage}
\begin{minipage}{0.5\linewidth}
\begin{picture}(190,138)
\small
\put(20,30){\includegraphics{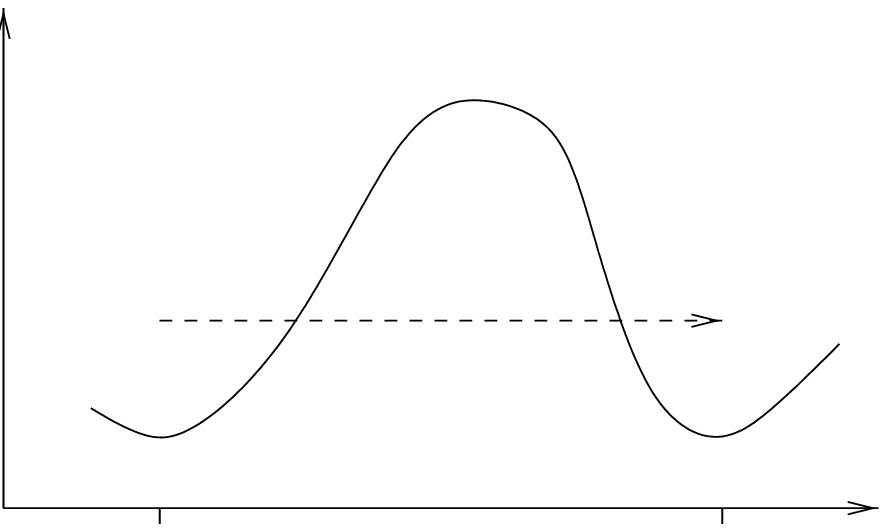}}
\put(10,128){$E[A_i(\lambda,\vec{x})]$}
\put(176,30){$\lambda$}
\put(45,20){$0$}
\put(142,20){$1$}
\put(93,52){$\Omega_1$}       
\put(90,0){(b)}
\normalsize      
\end{picture}
\end{minipage}
\caption{\small{A continuous line of gauge fields connects two gauge
equivalent configurations (a). Since the intermediate points are
physically different, they have different energies (b) and tunnelling
is expected.}}
\protect{\label{line}}
\end{figure}

Although $\Omega_n(\vec{x})$ form topologically distinct gauge
transformations, they act on the space $\{A_i(\vec{x})\}$ which is
topologically trivial.
%\footnote{
%The same consideration holds for other representations of the local
%gauge symmetry like Higgs or fermion fields.}. 
Consider a continuous
line of gauge fields connecting two gauge equivalent $A$'s (Fig.
\ref{line}(a)), 
\begin{eqnarray*}
A_i(\vec{x})\>A_i(\lambda,\vec{x}),\qquad
A_i(1,\vec{x})=\,^{\Omega_1}\!\!A_i(0,\vec{x})
\end{eqnarray*}
But at fractional $\lambda$ this is \textit{not} a gauge
transformation. These gauge fields lie on different orbits, i.e. are
physically different! So do their energies, i.e. the expectation value 
of $H$ in these configurations,
\begin{eqnarray*}
E[A_i(\lambda,\vec{x})]=\langle\{A_i(\lambda,\vec{x})\}|H|
\{A_i(\lambda,\vec{x})\}\rangle
\end{eqnarray*}
If $\lambda=0$ and $\lambda=1$ are vacua, the energy 
is higher inbetween as drawn in Fig. \ref{line}(b). The system may
\textit{tunnel} through the gauge transformation $\Omega_1$.
 How one actually computes the tunnelling rate and how the action
enters this calculation will be explained in the next section.

\section{Semiclassical Approximation for Tunnelling}

For the eigenfunctions $\psi$ of the Hamiltonian $H$ of an ordinary one
dimensional quantum mechanical system
\begin{eqnarray*}
H\psi=E\psi,\qquad H=\ha p^2+V(x)\qquad(\hbar=m=1)
\end{eqnarray*}
we write formally,
\begin{eqnarray*}
p\psi=-i\frac{\pa}{\pa x}\psi=\sqrt{2(E-V(x))}\psi
\end{eqnarray*}
Thus
\begin{eqnarray*}
\psi\propto \exp(i\!\int\!\sqrt{2(E-V(x))}\,\d x)
\end{eqnarray*}
is an approximate solution, i.e. describes the leading effects (in
$\hbar$). In the classically allowed regions $E>V(x)$ the wave
function just oscillates, while in the forbidden regions there is an
exponential suppression,
\begin{eqnarray*}
E<V(x):\quad \psi\propto \exp(-\!\int\!\sqrt{2(V(x)-E)}\,\d x)
\end{eqnarray*}
We deduce that the following quantity approximates the tunnelling
amplitude,
\begin{eqnarray*}
\exp(-\!\int_A^B\!\!\sqrt{2(V(x)-E)}\,\d x)
\end{eqnarray*}
where $A$ and $B$ are the boundary points of the forbidden region
$V(A)=V(B)=E$. 

\begin{figure}[t]
\begin{picture}(380,138)
\small
\put(120,18){\includegraphics{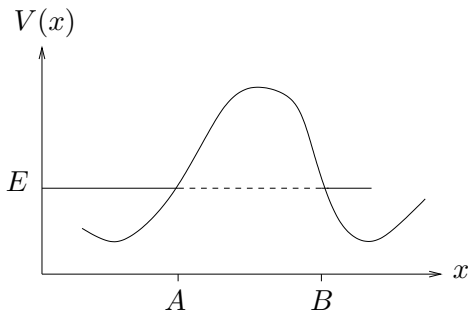}}
\put(110,114){$V(x)$}
\put(276,20){$x$}
\put(167,8){$A$}
\put(222,8){$B$}
\put(107,53){$E$}
\normalsize
\end{picture}
\caption{\small{The semiclassical situation for tunnelling through a 
potential barriere (see text).}}
\protect{\label{tunnelling}}
\end{figure}

The sign switch $V-E\>E-V$ is equivalent to $p\>ip,\,p^2\>-p^2$ or to,
\begin{eqnarray*}
t\>it=\tau,\quad E\>iE,\quad V\>iV
\end{eqnarray*}
The first replacement means that we can interchange the meaning of
`allowed' and `forbidden' by going to an imaginary time. For field
theories one passes from Minkowski to \textit{Euclidean} space,
accordingly.

Moreover, the integral can be rewritten as the action for imaginary
times,
\begin{eqnarray*}
\int_A^B\!\!\sqrt{2(V(x)-E)}\,\d x
=\int_{t_A}^{t_B}\!\!p\,\dot{x}\,\d t
=\int_{\tau_A}^{\tau_B}\!\!\mathcal{L}(\tau)\,\d \tau
=S_{\rm tot}\qquad(\mbox{if } E=0)
\end{eqnarray*}
Thus the dominant contribution to a tunnelling transition is obtained
by computing the \textit{action} of a classical motion in 
\textit{Eulidean} space, and write,
\begin{eqnarray}
\label{rate}
e^{-|S_{\rm tot}|}
\end{eqnarray}
For tunnelling in the space of gauge fields we are automatically
driven to the following topic.

\section{Action for a Topological Transition, Explicit Instanton Solutions}

\begin{figure}[t]
\begin{picture}(380,115)
\small
\put(120,3){\includegraphics{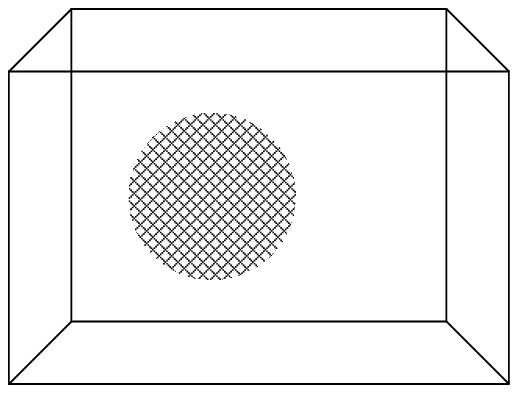}}
\put(144,101){$\vec{A}=\Omega_1(\vec{x})\frac{1}{ig}\vec{\pa}\,
\Omega_1^{-1}(\vec{x})$}
\put(270,55){$A=0$}
\put(180,10){$A=0$}
\put(206,55){$\vec{A}(\lambda,\vec{x})$}
\put(88,55){$A=0$}
\put(70,2){$x_4\>-\infty$}
\put(70,92){$x_4\>+\infty$}
\normalsize
\end{picture}
\caption{\small{A tunnelling process in Yang Mills theory. A trivial
vacuum at $x_4\>-\infty$ evolves into a vacuum with winding number 1 at 
$x_4\>+\infty.$}}
\protect{\label{box}}
\end{figure}

Let us seek for a tunnelling configuration along the lines of
Fig. \ref{box}. In the
infinite (Euclidean) past the gauge field is trivial $A=0$. Then it
evolves somehow and arrives at the first non-trivial vacuum
$\vec{A}=\Omega_1(\vec{x})\frac{1}{ig}\vec{\pa}\,
\Omega_1^{-1}(\vec{x})$ in the
infinite future. During the whole process $A$ should vanish at the
spacial boundary. For $x_4\>+\infty$ we already know this, since
$\Omega_1(\vec{x})$ becomes constant there. But now we can write $A$
as a pure gauge on the whole boundary of $\mathbb{R}^4$,
\begin{eqnarray*}
\vec{A}\>\Omega_1(x)\frac{1}{ig}\vec{\pa}\,\Omega_1^{-1}(x)
\quad\mbox{with}\:\:\Omega_1(x)=
\left\{\begin{array}{ll}
\Omega_1(\vec{x}) & \mbox{at}\:x_4\>+\infty,\\ 
\mbox{const.}     & \mbox{elsewhere}. 
\end{array}\right.
\end{eqnarray*}

A gauge equivalent (now we leave $A_4=0$), but more symmetric way is
to choose
\begin{eqnarray}
\label{new1}
A_\mu\>\Omega_1(x)\frac{1}{ig}\pa_\mu\,\Omega_1^{-1}(x)
\end{eqnarray}
with
\begin{eqnarray*}
\Omega_1(x)\> \frac{x_4 \id+i x_i \tau_i}{|x|},\qquad
|x|=\sqrt{x_\mu x_\mu} 
\end{eqnarray*}
Notice that $\Omega_1$ lives on the boundary of
$\mathbb{R}^4$ which is a three-sphere.
%\footnote{
%In the gauge above we could identify this three-sphere as the compactified 
%3-space at $x_4\>+\infty$.}
 It has the same degree as discussed above
and mixes coordinate space and isospace.

The last point will be crucial for finding explicit instanton 
solutions. The problem becomes simpler due to the higher symmetry. The 
action of $\Omega_1$ on a \textit{fundamental spinor} is,
\begin{eqnarray}
\label{new2}
\Omega_1(x){1 \choose 0}={x_4+ix_3 \choose
-x_2+ix_1}\frac{1}{|x|}
\end{eqnarray}
and covers the whole sphere. The symmetry is such that an $SO(4)$
rotation in Euclidean space is linked to isospin $SU(2)$ rotations,
\begin{eqnarray*}
\left.\begin{array}{ccccc}
SO(4)&\cong&SU(2)_{\rm L}&\otimes&SU(2)_{\rm R}\\
SU(2)&  \> &  SU(2)  &\otimes& \id 
\end{array}\right.
\end{eqnarray*}
%Actually the first relation\footnote{
%comparison to Lorentz decomp.} is rigorous only locally, i.e. on the
%algebra level we have $so(4)\cong su(2)_{\rm L}\oplus su(2)_{\rm R}$.
Obviously the Lie algebra
$so(4)$ is 6=3+3 dimensional. For a matrix $\alpha\in
so(4)$,
\begin{eqnarray*}
\alpha_{\mu\nu}\in\mathbb{R},\quad \alpha\mn=-\alpha_{\nu\mu}
\end{eqnarray*}
we define the `dual transform' $\tilde{\alpha}$ as,
\begin{eqnarray*}
\tilde{\alpha}_{\mu\nu}:=\ha\ep_{\mu\nu\rho\sigma}\alpha_{\rho\sigma}
\end{eqnarray*}
The six degrees of freedom can be divided as follows,
\begin{eqnarray*}
\left.\begin{array}{ccccc}
\alpha_{\mu\nu}&=&\ha(\alpha+\tilde{\alpha})\mn&+&
\ha(\alpha-\tilde{\alpha})\mn\\
6&=&3&+&3
\end{array}\right.
\end{eqnarray*}
The terms on the right handside are selfdual and anti-selfdual
($\tilde{\tilde{\alpha}}\equiv \alpha$) and correspond to
representations of $su(2)_L$ and $su(2)_R$, respectively.

A field $\psi^a$ transforming as an $I=1$ representation under
$SU(2)_{\rm L}$ can be written as,
\begin{eqnarray*}
\psi^a=\eta^a\mn a\mn
\end{eqnarray*}
The coefficients are denoted by the tensor $\eta$. It is selfdual,
\begin{eqnarray*}
\eta^a\mn=\tilde{\eta}^a\mn
\end{eqnarray*}
and of course anti-symmetric in $(\mu,\nu)$, as easily seen from the
explicit representation,
\begin{eqnarray*}
\framebox{$\eta^a_{ij}=\ep^{aij},\quad\eta^a_{i4}=\delta^a_i,
\quad\eta^a_{4i}=-\delta^a_i$}
\end{eqnarray*}
As $\ep_{iaj}$ in three dimensions it provides the mixing of coordinate
space and isospace.

The $\eta$-tensor can now be used to describe 
the \textit{vector} field $A_\mu^a$ in the adjoint representation.
One finds from (\ref{new1}) and (\ref{new2}) that, asymptotically,
\begin{eqnarray*}
A_\mu\>\Omega_1\frac{1}{ig}\pa_\mu\,\Omega_1^{-1}
\equiv 2 \eta^a\mn \frac{x_\nu}{|x|^2} \tau_a
\end{eqnarray*}
It becomes singular when approaching the origin. Which
\textit{smoothened} connection near the origin \textit{minimises} the
action?
With our knowledge we try,
\begin{eqnarray*}
A_\mu^a=\eta^a\mn x_\nu A(|x|)
\end{eqnarray*}
Indeed, the profile,
\begin{eqnarray}
\label{profile}
A(|x|)=\frac{2}{|x|^2+\rho^2}
\end{eqnarray}
makes the action minimal,
\begin{eqnarray}
\label{action1}
S=\frac{1}{4}\int F^a\mn F^a\mn=-\frac{8\pi^2}{g^2}
\end{eqnarray}
%The term 'minimum' is understood w.r.t. the topological constraints we put
%on $A_\mu$, as discussed below. 
This number has to be exponentiated in
(\ref{rate}). If $|g|$ is small, the resulting rate is very very small.
 Furthermore since its expansion
for small $g$ gives zero in all orders, tunnelling
processes will not be seen in pertubation theory.

The new length $\rho$ is the width of the profile. Since these
configurations are local events in space \textit{and} time, they are
called \textit{instantons} or \textit{pseudo-particles}. The action
is independent of $\rho$, i.e. we have found a whole manifold of
instanton solutions. This so-called moduli space also contains the
position $z_\mu$ of the center of the instanton which was chosen to be
at the origin in above.
% SCALE INVARIANCE

\section{Bogomol'nyi Bound and Selfdual Fields}

The instanton fulfills a Bogomol'nyi bound. We write
\begin{eqnarray}
\label{action2}
-S=\frac{1}{8}\int(F^a\mn-\tilde{F}^a\mn)^2+
\frac{1}{4}\int F^a\mn \tilde{F}^a\mn
\end{eqnarray}
The number of squares has reduced from $3 \cdot 6$ in
(\ref{action1}) to $3 \cdot 3$. To see that the second term is a total
derivative needs some effort,
\begin{eqnarray*}
\frac{1}{4} F^a\mn \tilde{F}^a\mn
&=&\frac{1}{8}\ep_{\mu\nu\rho\sigma} F^a\mn F^a_{\rho\sigma}\\
&=&\ha\ep_{\mu\nu\rho\sigma}(\pa_\mu A_\nu^a+\frac{g}{2}\ep^{abc}
A_\mu^b A_\nu^c)(\pa_\rho A_\sigma^a+\frac{g}{2}\ep^{ade}
A_\rho^d A_\sigma^e)\\
&=&\ha\ep_{\mu\nu\rho\sigma}(\pa_\mu A_\nu^a\pa_\rho A_\sigma^a+
g\ep^{ade}\pa_\mu A_\nu^a A_\rho^d A_\sigma^e+
\frac{g^2}{4}\ep^{abc}\ep^{ade}A_\mu^b A_\nu^c A_\rho^d A_\sigma^e)
\end{eqnarray*}
The $g^2$ term vanishes because of the symmetry of the $\delta$'s in
$(b,c,d,e)$ together with the anti-symmetry of $\ep$ in
$(\mu,\nu,\rho,\sigma)$. Similar symmetry arguments give the following
result\footnote{
For readers familiar with differential forms we give the following
equivalent equation: $\mbox{tr}F\wedge F\propto\d \,\mbox{tr}(A\wedge\d A-
\frac{2ig}{3}A\wedge A\wedge A)$ with a proper definition of the wedge
product for algebra elements},
\begin{eqnarray}
\label{top/surf}
\frac{1}{4} F^a\mn \tilde{F}^a\mn=
\frac{8\pi^2}{g^2}\pa_\mu K_\mu 
\end{eqnarray}
with the Chern-Simons current
\begin{eqnarray}
\label{CS}
K_\mu=\frac{g^2}{16\pi^2}\ep_{\mu\nu\rho\sigma}
(A_\nu^a\pa_\rho A_\sigma^a+\frac{g}{3}\ep^{abc} A_\nu^a A_\rho^b
A_\sigma^e)
\end{eqnarray}
being a gauge \textit{variant} quantity. 
The asymptotic
behaviour of this current,
\begin{eqnarray*}
|x|\>\infty:\quad K_\mu\>\frac{1}{2\pi^2}\frac{x_\mu}{|x|^4}
\end{eqnarray*}
gives the following surface integral
\begin{eqnarray}
\label{8pig}
-S=\frac{8\pi^2}{g^2}\int_{S^3_\infty}\d^3\sigma K_\perp
=\frac{8\pi^2}{g^2}|x|^3\mbox{area}(S^3_1)\frac{1}{2\pi^2}\frac{1}{|x|^3}
=\frac{8\pi^2}{g^2}
\end{eqnarray}
The vanishing of the square in (\ref{action2}) means that the field
strength is selfdual. From (\ref{profile}) we compute
\begin{eqnarray*}
 F^a\mn=\tilde{F}^a\mn=-\frac{4}{g}\eta^a\mn\frac{\rho^2}{(|x|^2+\rho^2)^2}
\end{eqnarray*}
and indeed,
\begin{eqnarray*}
\D_\mu F\mn(\equiv\D_\mu\tilde{F}\mn)=0.
\end{eqnarray*}

In general, the Bogomol'nyi bound is a useful tool to solve the
Yang-Mills equations. After having introduced the $A$-field, one needs
to solve $\D_\mu F\mn=0$. This equation corresponds to the
inhomogeneous Maxwell equation and therefore is \textit{second
order} in $A$. The demand for selfdual fields $ F\mn=\tilde{F}\mn$ is
only \textit{first order} in $A$. Now the Yang-Mills equation is
automatically fulfilled because of the Bianchi identity
$\D_\mu\tilde{F}\mn=0$.

For \textit{all} configurations the second term in (\ref{action2}) is
a multiple of $-\frac{8\pi^2}{g^2}$. It is a topological quantity,
called the \textit{Pontryagin index}. Since the integral can be
reduced to the surface, it corresponds to the winding number
$\Omega_1:S^3\>S^3$ discussed above. 

\section{Intermezzo: Massless Fermions in a Gauge Theory}

\begin{figure}[b]
\begin{picture}(380,100)
\small
\put(120,12){\includegraphics{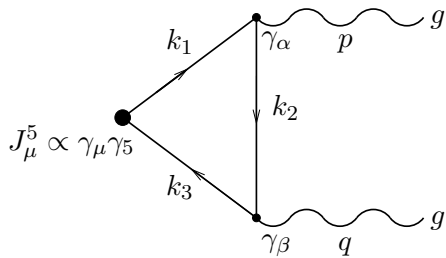}}
\put(80,42){$J_\mu^5\propto\gamma_\mu\gamma_5$}
\put(176,5){$\gamma_\beta$}
\put(176,82){$\gamma_\alpha$}
\put(140,25){$k_3$}
\put(140,80){$k_1$}
\put(180,55){$k_2$}
\put(205,81){$p$}
\put(205,4){$q$}
\put(240,90){$g$}
\put(240,14){$g$}
\normalsize
\end{picture}
\caption{\small{The lowest order Feynman graph leading to the chiral anomaly.}}
\protect{\label{Feynman}}
\end{figure}

The coupling of fermions to the gauge field is done in a standard way
by the \textit{vector current},
\begin{eqnarray*}
J_\mu=\bar{\psi}\gamma_\mu\psi
\end{eqnarray*}
where we dropped the isospace structure. For massless fermions,
the \textit{axial current} is conserved, too,
\begin{eqnarray*}
J_\mu^5=\bar{\psi}\gamma_\mu\gamma_5\psi,
\qquad \pa_\mu J_\mu=\pa_\mu J_\mu^5=0 
\end{eqnarray*}
Denoting by $J_\mu^{\rm L}$ and $J_\mu^{\rm R}$ the projections onto
$\gamma_5$ eigenstates, we can write,
\begin{eqnarray*}
J_\mu=J_\mu^{\rm L}+J_\mu^{\rm R},\qquad
J_\mu^5=J_\mu^{\rm L}-J_\mu^{\rm R}
\end{eqnarray*}
Thus the total number of fermions as well as the difference of
left-handed and right-handed fermions are classically conserved.

In order to look whether these statements survive the quantisation of
the theory, consider the matrix element
\begin{eqnarray*}
\langle 0 |J_\mu^5|gg\rangle
\end{eqnarray*}
$g$ are the gauge photons (gluons) which couple to $J_\mu$, not to
$J_\mu^5$. The corresponding lowest order Feynman diagram  is a
one-loop graph depicted in Fig. (\ref{Feynman}). We do not want to
go into the details of the calculation, but rather sketch the Dirac
matrix structure,
\begin{eqnarray}
\label{gamma}
\Gamma_{\mu\alpha\beta}(k,p,q)\propto\tr
\gamma_\mu\gamma_5\frac{(\gamma,k_1)}{k_1^2}
\gamma_\alpha\frac{(\gamma,k_2)}{k_2^2}
\gamma_\beta\frac{(\gamma,k_3)}{k_3^2}
\end{eqnarray}
$k_i$ and $(p,q)$ are the momenta of the fermions and the gauge
photons, respectively ($k+p+q=0$). The diagram is totally symmetric in
the sense that in (\ref{gamma}) we can put $\gamma_5$ also after
$\gamma_\alpha$ or $\gamma_\beta$ due to the anti-commutation
relations. But the diagram is linearly divergent, and the infinity must
be regularised. We prefer the introduction of Pauli-Villars mass
terms,
but the
result will be independent of the regularisation method,
\begin{eqnarray*}
\Gamma_{\mu\alpha\beta}^{\rm PV}(k,p,q)\propto\tr
\gamma_\mu\gamma_5\frac{M-i(\gamma,k_1)}{k_1^2+M^2}
\gamma_\alpha\frac{M-i(\gamma,k_2)}{k_2^2+M^2}
\gamma_\beta\frac{M-i(\gamma,k_3)}{k_3^2+M^2}
\end{eqnarray*}
The symmetry is lost by renormalisation, namely \textit{the finite
part of the diagram will depend on where one puts} $\gamma_5$. The
ambiguity in $\gamma_5$ is removed by the following choice,
\begin{eqnarray*}
p_\alpha\Gamma_{\mu\alpha\beta}(k,p,q)=
q_\beta\Gamma_{\mu\alpha\beta}(k,p,q)&=&0\\
k_\mu\Gamma_{\mu\alpha\beta}(k,p,q)\propto
\ep_{\alpha\beta\gamma\delta}p_\gamma q_\delta&\neq& 0
\end{eqnarray*}
The gauge invariance due to the two gauge photons has survived, but
$J_\mu^5$ is \textit{not} conserved anymore,
\begin{eqnarray*}
\pa_\mu\langle 0|J_\mu(x)|gg\rangle&=&0\\
\pa_\mu\langle 0|J_\mu^5(x)|gg\rangle&=&\frac{g^2}{16\pi^2}
\langle 0|F^a\mn\tilde{F}^a\mn|gg\rangle
\end{eqnarray*}
The last identity is the non-Abelian version of the Adler-Bell-Jackiw
 anomaly\footnote{
The Adler-Bardeen theorem guaranteees that there are no effects in higher
order pertubation theory.}. The topological density enters here,
remember that $\int\!\d^4 x F^a\mn\tilde{F}^a\mn=\frac{32\pi^2}{g^2}$
for an instanton. It effects the charges
$Q_5=\int\!\d^3 x J_0^5(x)$ in the way that the charge `after the
instanton' (at $x_4\>-\infty$) differs by two from the charge `before the
instanton' (at $x_4\>+\infty$),
\begin{eqnarray*}
\int\!\d^4 x\,\pa_\mu J_\mu^5=Q_5^{\rm after}-Q_5^{\rm before}=2
\end{eqnarray*}
One fermion has flipped its helicity from right to left. In other
words, the instanton adds a left-handed particle and removes a
right-handed anti-particle (the other way round for right-handed
particles).

\section{Jackiw-Rebbi States at an Instanton}

\begin{figure}[t]
\begin{picture}(380,90)
\small
\put(120,5){\includegraphics{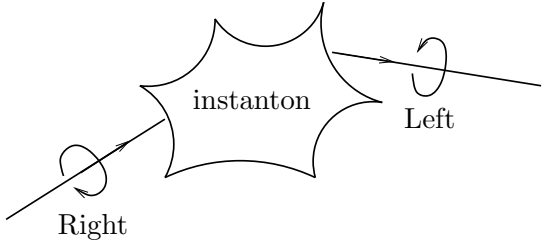}}
\put(141,1){\mbox{Right}}
\put(192,48){\mbox{instanton}}
\put(272,41){\mbox{Left}}
\normalsize
\end{picture}
\caption{\small{The (interaction with the) instanton flips the
helicity of the fermion from right to left as shown in the text.}}
\protect{\label{flip}}
\end{figure}

How to understand the fact, that the interaction
with an instanton flips the helicity of the fermion (Fig.
\ref{flip})?

Let us investigate the gauge group $SU(2)$ with fundamental fermions
($I=1/2$). As we know the spinorial group $SU(2)_{\rm L}\otimes
SU(2)_{\rm R}$ couples to the gauge group $SU(2)_{\rm L}$,
\begin{eqnarray*}
SU(2)_{\rm L}\otimes\left(SU(2)_{\rm L}\otimes
SU(2)_{\rm R}\right)
\end{eqnarray*}
For left-handed and right-handed fermions we have,
\begin{eqnarray*}
2_{\rm L}\times(2_{\rm L}\times 1_{\rm R})=3_{\rm L}+1_{\rm L},
\quad
2_{\rm L}\times(1_{\rm L}\times 2_{\rm R})=2_{\rm L}\times 2_{\rm R}
\end{eqnarray*}
respectively. There is one state with $j_{\rm L}=j_{\rm R}=0$ which
indeed has a normalisable solution in four-space,
\begin{eqnarray*}
\psi=\frac{\rm const.}{(|x|^2+\rho^2)^{3/2}}
\end{eqnarray*}
This Jackiw-Rebbi state is a chiral eigenstate and fulfils the
(Euclidean) Dirac equation
\begin{eqnarray*}
\gamma\D \psi=0
\end{eqnarray*}

\begin{figure}[t]
\begin{picture}(380,120)
\small
\put(100,-60){\includegraphics{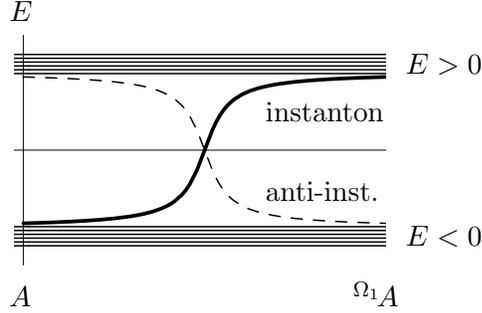}}
\put(120,0){$A$}
\put(250,0){$^{\Omega_1}\!A$}
\put(270,88){$E>0$}
\put(270,23){$E<0$}
\put(217,70){\mbox{instanton}}
\put(217,40){\mbox{anti-inst.}}
\put(120,108){$E$}
\normalsize
\end{picture}
\caption{\small{The instanton not only changes the winding number by
1, but also creates a (left-handed) particle and removes a
(right-handed) anti-particle. For the anti-instanton it is vice
versa. Infact the picture is delicate: In the original theory there is
no mass and no gap, but we could add a small mass and the
considerations still hold.}}
\protect{\label{trans}}
\end{figure}

Let us come back to the line $\vec{A}(\lambda,\vec{x})$ connecting two
vacua in the gauge $A_4=0$ (Fig. \ref{line}(a)) and choose just
$x_4$ as the parameter of the configuration,
\begin{eqnarray*}
\lambda\equiv x_4
\end{eqnarray*}
The operator $\frac{\pa}{\pa \lambda}\equiv\frac{\pa}{\pa x_4}$ enters
the Dirac equation and extracts the energy,
\begin{eqnarray*}
(\gamma_4\pa_4+\vec{\gamma}\vec{\D})\psi=0=
(-\gamma_4 E +\vec{\gamma}\vec{\D})\psi
\end{eqnarray*}
We represent its action on $\psi$ by two functions of $\lambda$,
\begin{eqnarray*}
\frac{\pa}{\pa \lambda}\psi&=&+\alpha(\lambda)\psi\quad
\lambda\>-\infty\\
\frac{\pa}{\pa \lambda}\psi&=&-\beta(\lambda)\psi\quad
\lambda\>+\infty
\end{eqnarray*}
The signs follow from the general shape of normalisable modes. For the
case at hand it has a power law
behaviour: $\psi(\lambda,\vec{x})\propto\frac{1}{\lambda^3}$. If we
approximate it by an exponential law, $\alpha$ and $\beta$ become
constants and we arrive at the qualitative spectrum shown in Fig.
\ref{trans}: The instanton provides a transition from $A$ to
$^{\Omega_1}\!A$
during which the number of left-handed particles increases by 1,
while the number of right-handed anti-particles drops by 1,
\begin{eqnarray*}
\triangle Q^{\rm L}=-\triangle Q^{\rm R}=1
\end{eqnarray*}
accordingly -1 for anti-instantons.

\section{Estimate of the Flip Amplitude}

How to calculate the amplitude for a process with $\triangle Q_5=2$?
As a device for such a process we add the following term to the
Lagrangian,
\begin{eqnarray*}
\mathcal{L}=-\frac{1}{4}F^a\mn\tilde{F}^a\mn
-\bar\psi\gamma\D\psi
-J\bar{\psi}^{\rm R}(x)\psi^{\rm L}(x^\prime)
\end{eqnarray*}
It simulates the flip from right to left resulting from an interaction
with an instanton. We wrote $x$ and $x^\prime$ allowing $J$ to be
non-local. The vacuum-to-vacuum amplitude is given by the usual path
integral,
\begin{eqnarray*}
\int\D A_\mu\D\bar{\psi}\D\psi\exp[\,i\!\int\!\mathcal{L}_A-
\bar{\psi}(\gamma\D+J)\psi]
\end{eqnarray*}
The fermionic part gives the determinant of the operator $\gamma\D+J$,
therefore we have to solve,
\begin{eqnarray*}
\gamma\D\psi(x^\prime)+J\psi(x)=\lambda\psi(x^\prime),
\qquad \lambda\neq 0
\end{eqnarray*}
The Jackiw-Rebbi mode has $\gamma\D\psi=0$ and thus $\lambda\propto J$.

Not only $\gamma\D$ but also the fluctuation operator of $A_\mu$ has
zero modes. We expand around the instanton field,
\begin{eqnarray*}
A_\mu=A_\mu^{\rm inst}+\delta A_\mu
\end{eqnarray*}
Since $A_\mu^{\rm inst}$ is a classical solution the change in the
action is second order,
\begin{eqnarray*}
\delta\int\!\mathcal{L}=\int\delta A_\mu M\mn\delta A_\nu
\end{eqnarray*}
Zero modes of $M$ are connected with the collective coordinates for
the instanton. We already mentioned five of them, namely the width $\rho$
and the position $z_\mu$,
\begin{eqnarray*}
A^{\rm inst}(z_\mu+\delta z_\mu,\rho+\delta \rho)=
A^{\rm inst}+\delta A
\Rightarrow \delta\int\!\mathcal{L}_A=0
\end{eqnarray*}
There are also three gauge-collective coordinates for $SU(2)$, so in
total we have 5+3=8 zero modes.

All other eigenstates of $\gamma\D$ and $M$ have non-vanishing
eigenvalues $\lambda_i$. The path integral (in a semiclassical
approximation for the gauge field) is the product,
\begin{eqnarray*}
\int\D\delta A_\mu\D\bar{\psi}\D\psi\propto\prod_i \lambda_i
\end{eqnarray*}
%This is only a formal expression since the determinant gives
%infinity. The regularisation is provided by det$^{-1}$. The large
%eigenvalues cancel and we are left with $\prod_i \lambda_i\propto
%J$. 
If $J=0$ then the amplitude vanishes, it would have been an amplitude
with $\triangle Q_5=0$. The term linear in $J$ has
\begin{eqnarray*}
\triangle Q_5=\pm 2
\end{eqnarray*}
and this amplitude is not equal to zero.

We now ask: which \textit{effective} Lagrangian would mimic the
instanton effects? It must be a Lagrangian which (a) gives an
interaction with $\triangle Q_{\rm L}=-\triangle Q_{\rm
R}=\frac{1}{2}\triangle Q_5=N_f$ (the number of flavours) and (b) 
obeys the flavour symmetry $SU(N_f)^{\rm L}\times SU(N_f)^{\rm
R}\times U(1)^{\rm vector}$ and violates $U(1)^{\rm axial}$.

For one flavour the instanton induces an effective mass term,
\begin{eqnarray*}
\triangle\mathcal{L}_{\rm eff}\propto e^{-8\pi^2/g^2}\bar{\psi}\psi,
\qquad
\bar{\psi}\psi\propto
\bar{\psi}^{\rm R}\psi^{\rm L}+\bar{\psi}^{\rm L}\psi^{\rm R}
\end{eqnarray*}
For $N_f$ flavours
$\pa_\mu J_\mu^{5\,ij}$ is diagonal in the flavour indices $i$ and $j$,
\begin{eqnarray*}
\pa_\mu J_\mu^{5\,ij}\propto
F^a\mn\tilde{F}^a\mn \delta^{ij},\quad i,j=1..N_f
\end{eqnarray*}
but the effective instanton Lagrangian may contain Dirac indices.
The original gauge Lagrangian has
no mass and thus the following symmetry,
\begin{eqnarray*}
U(N_f)^{\rm L}\times U(N_f)^{\rm R}=SU(N_f)^{\rm L}\times SU(N_f)^{\rm
R}\times \underbrace{U(1)^{\rm L}\times U(1)^{\rm R}}_
{U(1)^{\rm V}\times U(1)^{\rm A}}
\end{eqnarray*}
The instanton contribution violates $U(1)^{\rm A}$ only
and the first approximation
\begin{eqnarray*}
e^{-8\pi^2/g^2}(\bar{\psi}_1\psi_1)\cdot\ldots\cdot(\bar{\psi}_{N_f}\psi_{N_f})
\end{eqnarray*}
should better be replaced with
\begin{eqnarray}
\label{det}
e^{-8\pi^2/g^2}\underset{ij}{\det}(\bar{\psi}_i\psi_j)
\end{eqnarray}
Even more precisely the effective Lagrangian is \cite{tHooftU1}
\begin{eqnarray*}
e^{-8\pi^2/g^2}\sum_{a_i,b_i,j_i}
R(a_i,b_i)\ep^{j_1\ldots j_n}
(\bar{\psi}_1^{a_1}[1+\gamma_5]\psi_{j_1}^{b_1})\cdot\ldots\cdot
(\bar{\psi}_{N_f}^{a_{N_f}}[1+\gamma_5]\psi_{j_{N_f}}^{b_{N_f}})
\end{eqnarray*}

\section{Influence of the Instanton Angle}

If we allow for $\theta$, the
effective Lagrangian includes a multiplication with $e^{i\theta}$ for
the instanton ($e^{-i\theta}$ for the anti-instanton),
\begin{eqnarray*}
e^{-8\pi^2/g^2+i\theta}\det(\bar{\psi}_{\rm R}\psi_{\rm L})
\end{eqnarray*}
To get the same effect on the level of the original Lagrangian, we
have to add the topological/surface term from (\ref{top/surf}),
\begin{eqnarray*}
\mathcal{L}=-\frac{1}{4}F^a\mn F^a\mn+
\frac{g^2}{4\cdot8\pi^2}i\theta F^a\mn \tilde{F}^a\mn
\end{eqnarray*}
Then the instanton action becomes
$\exp(-S)\>\exp(-S+i\theta\frac{g^2}{8\pi^2}\int\frac{1}{4}F\tilde{F})$
with $\int\frac{1}{4}F\tilde{F}=\frac{8\pi^2}{g^2}$ for an instanton
(cf. (\ref{8pig})). In the preferred notation
$g A_\mu=:\mathcal{A}_\mu$, $g F\mn=:\mathcal{F}\mn$, where $D_\mu$ is
free of $g$, we have,
\begin{eqnarray*}
\mathcal{L}=-\frac{1}{g^2}\cdot
\frac{1}{4}\mathcal{F}^a\mn \mathcal{F}^a\mn+
\frac{i\theta}{8\pi^2}\cdot
\frac{1}{4}\mathcal{F}^a\mn \tilde{\mathcal{F}}^a\mn
\end{eqnarray*}
Pure gauge theory has two constants of nature, $g$ and $\theta$, both
of which are in principle observable. Especially in SUSY theories
they are combined as,
\begin{eqnarray*}
z=\frac{1}{g^2}+\frac{i\theta}{8\pi^2}
\end{eqnarray*}

How to observe $\theta$? Let us study the effect of $\theta$ on the
\textit{electric} charge of a magnetic monopole,
\begin{eqnarray*}
\mathcal{L}&=&-\frac{1}{4}F^a\mn F^a\mn+
\frac{i\theta g^2}{8\pi^2}F^a\mn \tilde{F}^a\mn
=\ha (\vec{E}_a^2-\vec{B}_a^2)-
\frac{\theta g^2}{8\pi^2}\vec{E}_a\cdot\vec{B}_a\\
&=&\ha (\vec{E}_a-\frac{\theta g^2}{8\pi^2}\vec{B}_a)^2-
\ha  (1+\frac{\theta^2 g^4}{(8\pi^2)^2})\vec{B}_a^2
=\ha\vec{E}_a^{\prime\,2}-\ha \vec{B}_a^{\prime\,2} 
\end{eqnarray*}
$\theta$ \textit{shifts} the electric field
and changes the energy of the magnetic field (slightly).
This can be interpreted as the energy of a background electric field,
which gives small corrections to the mass.

Integrating the $E^\prime$ equation around a monopole gives a relation for
the charges,
\begin{eqnarray*}
q_e^\prime=q_e-\frac{\theta g^2}{8\pi^2}g_m,
\quad q_e=0,
\quad g_m=\frac{4\pi}{g}
\end{eqnarray*}
The monopole behaves like it has a \textit{fractional electric}
charge,
\begin{eqnarray*}
\framebox{$q_e=\frac{\theta}{2\pi}g$}
\end{eqnarray*}
Notice that we were not allowed to shift the $B$ field, since this
would give magnetic charges to electric objects.

\begin{figure}[t]

\begin{minipage}{0.5\linewidth}
\begin{picture}(190,155)
\small
\put(7,20){\includegraphics{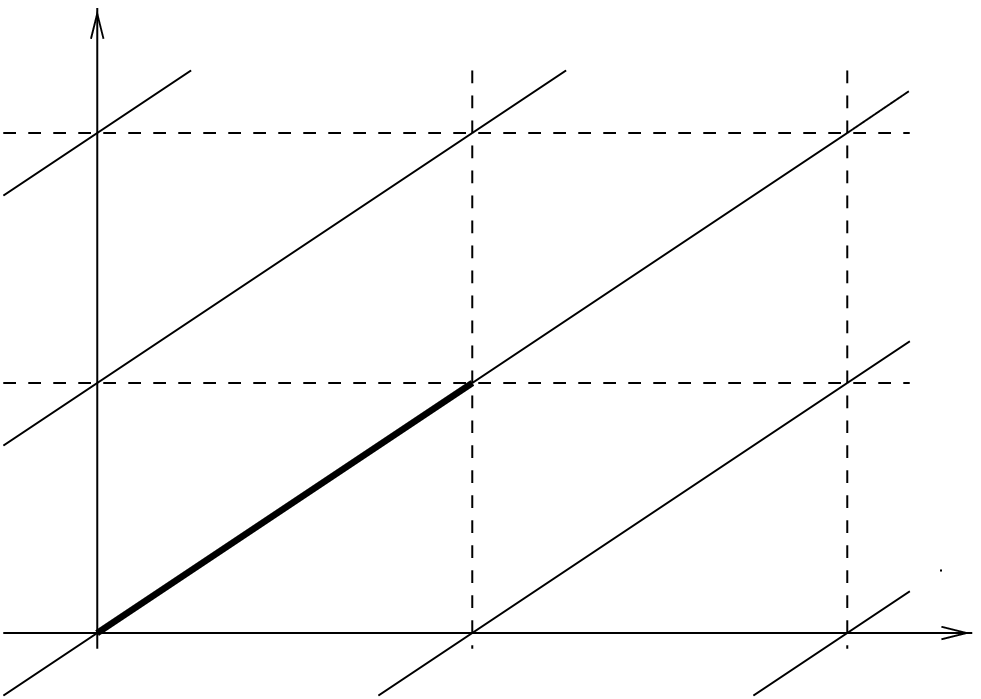}}
\put(20,145){$q_e$}
\put(22,20){$0$}
\put(86,19){$2\pi$}
\put(148,20){$4\pi$}
\put(178,28){$\theta$}
\put(16,79){$g$}
\put(11,122){$2g$}
\put(90,0){(a)}
\normalsize
\end{picture}
\end{minipage}
\begin{minipage}{0.5\linewidth}
\begin{picture}(190,155)
\small
\put(20,20){\includegraphics{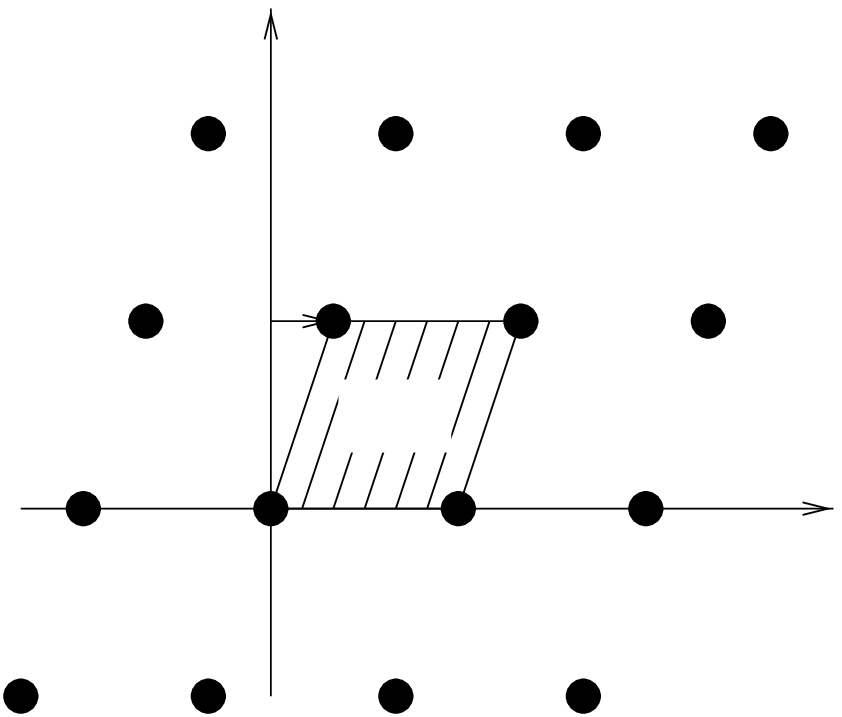}}
\put(62,148){$g_m$}
\put(168,55){$q_e$}
\put(69,95){$\theta/2\pi$}
\put(82,69){$2\pi$}       
\put(90,0){(b)}
\normalsize      
\end{picture}
\end{minipage}
\caption{\small{The figures show (a) the electric charge $q_e$ of a
monopole as a function of the instanton angle $\theta$ and (b) the
possible values of electric and magnetic charges.}}
\protect{\label{diagrams}}
\end{figure}

The function $q_e(\theta)$ for the monopole is plotted in
Fig. \ref{diagrams}(a).
We already know that the monopoles could get an integer
electric charge by binding to an electric particle, i.e. the original
line in the figure has vertically shifted copies. In this
way everything becomes periodic in $\theta$ as it should be (from its
introduction via $e^{i\theta}$, cf (\ref{intheta})). It is helpful to
plot the possible electric and magnetic charges like in Fig.
\ref{diagrams}(b).

Since the monopole now has an electric charge the Dirac condition has
to be modified as well,
\begin{eqnarray*}
\framebox{\rule[-1.8mm]{0mm}{5.8mm}
 $q_e^1 g_m^2- q_e^2 g_m^1=2\pi n_{12}$ }\qquad n_{12}\in \mathbb{Z}
\end{eqnarray*}
In a theory with gauge group $U(1)^{N-1}$ like in the next chapter only
the sum of all $U(1)$ charges is constrained,
\begin{eqnarray*}
\sum_{i=1}^{N-1}
q_e^{1\:(i)} g_m^{2\:(i)}- q_e^{2\:(i)} g_m^{1\:(i)}=2\pi n_{12}
\qquad n_{12}\in \mathbb{Z}
\end{eqnarray*}
Geometrically, the area of the elementary cell in Fig.
\ref{diagrams}(b) is $2\pi$, independent of the value of $\theta$.

\chapter{Permanent Quark Confinement}

The phenomenological observation that quarks cannot be seen in isolation is
called quark \textit{confinement}. Although being a basic issue of the
strong interaction theory, it remained a mystery before the 70's. The
reason is that confinement is not a feature of perturbation
theory. There one has to fix the gauge first, then quantise the theory
and expand in a small coupling constant. Certain phenomenological aspects
are described correctly, but there is no confinement.

To derive this new non-perturbative effect, one has to investigate the
theory more precisely. As in the theory of electrodynamics there occur
gauge fixing ambiguities when fixing the gauge. These so-called
\textit{Gribov ambiguities} may explain confinement.

Here we can give only a qualitative picture of the mechanism. We use a
special partial gauge fixing, the \textit{Abelian projection}, where
the (local) gauge group  is reduced to its (local) Abelian
subgroup. Besides quarks and gauge photons the Abelian theory contains
magnetic monopoles, which via a (dual) Meissner effect should confine
all chromoelectric charges.

\section{The Abelian Projection}

In the following we deal with $SU(N)$ as the prototype of a
non-Abelian gauge group. We do not focus on renormalisation (i.e.~we
do not attempt to account for infinities),
because it does not have much to do with confinement.

The principle of the Abelian projection is to fix the gauge `as
locally as possible' (using the gauge field only). How is it done?

\vspace{0.3cm}
\noindent\large\textbf{\framebox{1}} \normalsize
In the first step we pick a field $X$
(elementary or composite) in the adjoint representation of the gauge
group\footnote{
A field in the fundamental representation would even be better, but in
QCD there are only the quarks which as fermions are more difficult to
treat.}.
Such a field is a $N\times N$-matrix which is hermitian,
\begin{eqnarray*}
X^\dagger=X
\end{eqnarray*}
and usually also traceless. It transforms as
\begin{eqnarray}
\label{Xtrans}
X\>^\Omega\!\!X=\Omega X \Omega^{-1}
\end{eqnarray}
The gauge field itself is ruled out because of the inhomogeneous
term. But there are other candidates containing the field strength
like
\begin{eqnarray*}
X^{ij}=G_{12}^{ij}\quad\mbox{or}\quad
X^{ij}=\left(\D_\mu G_{\alpha\beta}\D_\mu G_{\alpha\beta}\right)^{ij}
\qquad i,j=1\ldots N
\end{eqnarray*}
the latter having the advantage of being Lorentz invariant, but it is
more complicated.
Adding a scalar to the theory in order to fix the gauge would be the
easiest, but it would of course change the model.

As (\ref{Xtrans}) does not involve derivatives of $\Omega$ nor $\Omega$
at different points, it is a \textit{local} transformation, which will
be important when considering the ghosts.

\vspace{0.3cm}
\noindent\large\textbf{\framebox{2}} \normalsize
In the next step we use the field $X$ to fix the gauge (partially). We
choose the gauge $\Omega$ in which $X$ \textit{is diagonal}. Then $X$
is of the form
\begin{eqnarray*}
X=\left(\begin{array}{ccc}
\lambda_1&&0\\
&\ddots&\\
0&&\lambda_N
\end{array}\right)
\end{eqnarray*}
We further sort the eigenvalues,
\begin{eqnarray}
\label{sort}
\lambda_1\geq\lambda_2\geq\cdots\geq\lambda_N
\end{eqnarray} 
This can always be done, but it does not fix $\Omega$ entirely. In
technical terms we introduce Lagrange multipliers $\alpha$ for the
off-diagonal components,
\begin{eqnarray*}
\mathcal{L}^{\rm gauge}=\sum_{i<j}\alpha_{ij}X_{ij}
\end{eqnarray*}

\vspace{0.3cm}
\noindent\large\textbf{\framebox{3}} \normalsize
Observe that the previous step leaves a local $U(1)^N/U(1)=U(1)^{N-1}$
invariance. The gauge transformed $X$ equals the original one if and
only if $X$ and $\Omega$ commute,
\begin{eqnarray*}
X=\Omega X \Omega^{-1}\quad \mbox{iff}\:\:[X,\Omega]=0
\end{eqnarray*}
For diagonal $X$ this is true for (the group of) diagonal $\Omega$'s,
\begin{eqnarray}
\label{Omega}
\Omega&=&\left(\begin{array}{ccc}
\omega_1&&0\\
&\ddots&\\
0&&\omega_N
\end{array}\right)
\with  \prod_i \omega_i=1\\
&=&\exp i\left(\begin{array}{ccc}
\Lambda_1&&0\\
&\ddots&\\
0&&\Lambda_N
\end{array}\right)
\with  \sum_i \Lambda_i=0 
\end{eqnarray}
Every element of the diagonal stands for one $U(1)$ but the condition
\hbox{det $\Omega=1$} traces out an overall $U(1)$. For $SU(N)$ the number $N-1$ is
just the rank of the group, i.e. the dimension of the Cartan subalgebra. 

The stability group of $X$ is bigger if two eigenvalues coincide (see
step 7), but generically this is not the case.

The remainder of the gauge group will be fixed in the next step.

\vspace{0.3cm}
\noindent\large\textbf{\framebox{4}} \normalsize
The residual Abelian gauge might be fixed just as in QED, for instance
via the Lorentz gauge,
\begin{eqnarray*}
\mathcal{L}^{\rm gauge, Abelian}=\sum_{i=1}^{N-1}\beta_i\,\pa_\mu A^\mu_{ii}
\end{eqnarray*}
The total gauge fixing Lagrangian reads
\begin{eqnarray*}
\mathcal{L}^{\rm gauge}=\sum_{i<j}^N\alpha_{ij}X_{ij}+
\sum_{i=1}^{N-1}\beta_i\,\pa_\mu A^\mu_{ii}
\end{eqnarray*}
We have introduced $2\frac{N(N-1)}{2}+N-1$ real Lagrange multipliers
which together give the dimension of the group.

Renormalisation is still to be done, so we expect highly singular
Feynman rules.

\vspace{0.3cm}
\noindent\large\textbf{\framebox{5}} \normalsize
Fixing the gauge always includes a measure factor, the Faddeev Popov
determinant, which usuallly is exponentiated with the help of
\textit{ghosts}. As we will now show they do not interact in Abelian
Projections.

If the gauge is fixed by functions $C_k$ of the fields their change
under infinitesimal gauge transformations is
\begin{eqnarray*}
C_k\>C_k+m_{k\lambda}\Lambda^\lambda
\end{eqnarray*}
$\Lambda$ stands for the algebra element of the gauge transformations
under consideration $\Omega=\exp(i\Lambda)$. In our case $X_{ij}$ transforms
according to the adjoint action of the group, which infinitesimally
gives the commutator,
\begin{eqnarray*}
X_{ij}\>X_{ij}+i[\Lambda,X]_{ij}
\end{eqnarray*}
$\pa_\mu A^\mu$ transforms like the gauge field itself,
namely with the covariant derivative of the gauge parameter,
\begin{eqnarray*}
\pa_\mu A^\mu_{ii}\>\pa_\mu A^\mu_{ii}+\pa_\mu(\D^\mu \Lambda)_{ii}
\end{eqnarray*}

The Faddeev Popov determinant is included by adding $-\bar{\eta}_k
m_{k\lambda}\eta_\lambda$ to the Lagrangian. $\eta$ are
\textit{anti-commuting} variables, but as \textit{scalars} carry no
spin. Violating the spin statistics theorem they are not observable,
but integrating them out gives the right measure factor,
\begin{eqnarray*}
\int\D\eta\D\bar{\eta}\exp(-\int \d x \bar{\eta}_k
m_{k\lambda}\eta_\lambda) \propto \mbox{det}\: m
\end{eqnarray*}
For the gauge fixing above we get
\begin{eqnarray*}
\mathcal{L}^{\rm ghost}&=&i\sum_{i<j}\bar{\eta}_{ij}[\eta,X]_{ij}
+\sum_i\bar{\eta}_{ii}\pa_\mu (\D_\mu \eta)_{ii}\\
&=&i\sum_{i<j}\bar{\eta}_{ij}(\lambda_i-\lambda_j)\eta_{ij}+
\sum_i\bar{\eta}_{ii}\pa_\mu^2 \eta_{ii}-i g
\sum_i  \bar{\eta}_{ii}\pa_\mu(A_\mu^{ik}\eta_{ki}-
                                A_\mu^{ki}\eta_{ik})
\end{eqnarray*}
The third term transforms $\eta$ only one-way, hence it does not
contribute in loops. The second term corresponds to the free theory
for the diagonal $\eta$'s, while the first term is local in the
off-diagonal $\eta$'s, in other words these ghosts have infinite
mass. In total, there are \textbf{no (harmful) ghosts}.

\vspace{0.3cm}
\noindent\large\textbf{\framebox{6}} \normalsize
Observe that we have all charcteristics of a $U(1)^{N-1}$ Abelian
gauge theory.
The residual gauge transformations $\Omega$ (\ref{Omega}) transform
$A_\mu$ as
\begin{eqnarray*}
(A_\mu)_{ii}&\>&(A_\mu)_{ii}-\frac{1}{g}\pa_\mu \Lambda_{ii}\\
(A_\mu)_{ij}&\>&\exp(i(\Lambda_i-\Lambda_j))(A_\mu)_{ij}
\end{eqnarray*}
The diagonal gauge fields $(A_\mu)_{ii}$ are as photons in ($N-1$
independent) QED ('s). All other fields carry $U(1)^{N-1}$ charges
$Q_i$ with $\sum_i Q_i=0$. For the $(ij)$-component of $A_\mu$ we have
$Q_i=1$, $Q_j=-1$. In addition to the electrically charged quarks the
theory contains now electrically charged `off-diagonal photons'.

Moreover, these fields become massive. Since we removed the
off-diagonal gauge symmetry, their masses are \textit{not} protected
by gauge invariance anymore,
\begin{eqnarray*}
(\D_\mu X)_{ij}&=&\pa_\mu X_{ij}+ig(\lambda_i-\lambda_j)(A_\mu)_{ij}\\
\mbox{Tr}\:(\D_\mu X)^2 &\>& (\pa_\mu X)^2 +
g^2(\lambda_i-\lambda_j)^2 (A_\mu^{ij})^2
\end{eqnarray*}
But the theory is not exactly QED$^{N-1}$, something of the
non-Abelian character has to survive.

\vspace{0.3cm}
\noindent\large\textbf{\framebox{7}} \normalsize
The gauge fixing may lead to \textit{singularities} if
$\lambda_i=\lambda_j$, which we argued away so far by handwaving. Near
such a point the Higgs field looks like (from
the ordering (\ref{sort}) it is clear that $i$ and $j$ are
neighbours):
\begin{eqnarray*}
X=\left(\begin{array}{ccc}
\ddots&&0\\
&\framebox{$\begin{array}{cc}\lambda&0\\0&\lambda
                 \end{array}$}
&\\
0&&\ddots                 
\end{array}\right)
+\sum_{k=1}^3 a_k(x)
\left(\begin{array}{ccc}
\ddots&&\ldots\\
&\framebox{$\sigma_k$}
&\\
\ldots&&\ddots                 
\end{array}\right)
\end{eqnarray*}
We have used the parametrisation (\ref{para}) for the $SU(2)$
subset. The first part is gauge invariant and the second part shall
vanish when approaching some subspace,
\begin{eqnarray*}
x\>x_0: \: a_k(x)\>0 \quad k=1,2,3
\end{eqnarray*}
Generically these three conditions rule out three planes crossing at
$x_0$. In three-space this fixes a point, while in four-space it is a
(world) line.

Up to little deformations we can set $a_k(x)=(x-x_0)_k$ which is the
hedgehog from Chapter 3. At $x_0$ the residual gauge group is enlarged
from $U(1)^{N-1}$ to $U(1)^{N-3}\times U(2)$ and non-Abelian.
These are
\begin{eqnarray*}
\framebox{\rule[-1.5mm]{0mm}{5.5mm}\mbox{ magnetic monopoles }}
\end{eqnarray*}
w.r.t. that subgroup. Their magnetic charges are
$g_i=(0,\ldots,0,1,-1,0,\ldots,0)$ and sum up to zero.

We conclude that in the Abelian projection
we get electric and magnetic charges which
are all point-like. Note that in a $\theta$-vacuum (with instantons)
the magnetic monopoles receive electric charges $\frac{\theta}{2\pi}g$ 
(cf. Fig. \ref{diagrams}(b)).

For gauge group $SU(N)$ the Dirac condition reads
\begin{eqnarray*}
\sum_{i=1}^{N-1} g_i q_i=2\pi
\end{eqnarray*}

We expect complicated interactions among charged particles and  
magnetic monopoles. The latter will acquire a mass, since there is no
reason for them to be massless. Hence electro-magnetism provides the
only long-ranged fields, $N-1$ $U(1)$ photons. We can now ask what
happens to these objects in a Higgs mechanism?

\section{Phases of the Abelian Theory}

\begin{figure}[t]
\begin{minipage}{0.5\linewidth}
\begin{picture}(190,155)
\small
\put(15,40){\includegraphics{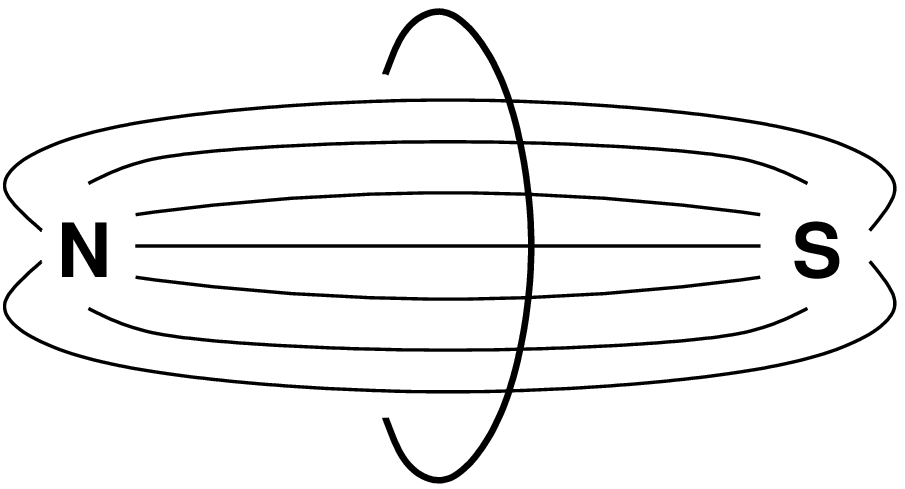}}
\put(90,30){$C$}
\put(90,0){(a)}
\normalsize
\end{picture}
\end{minipage}
\begin{minipage}{0.5\linewidth}
\begin{picture}(190,155)
\small
\put(20,20){\includegraphics{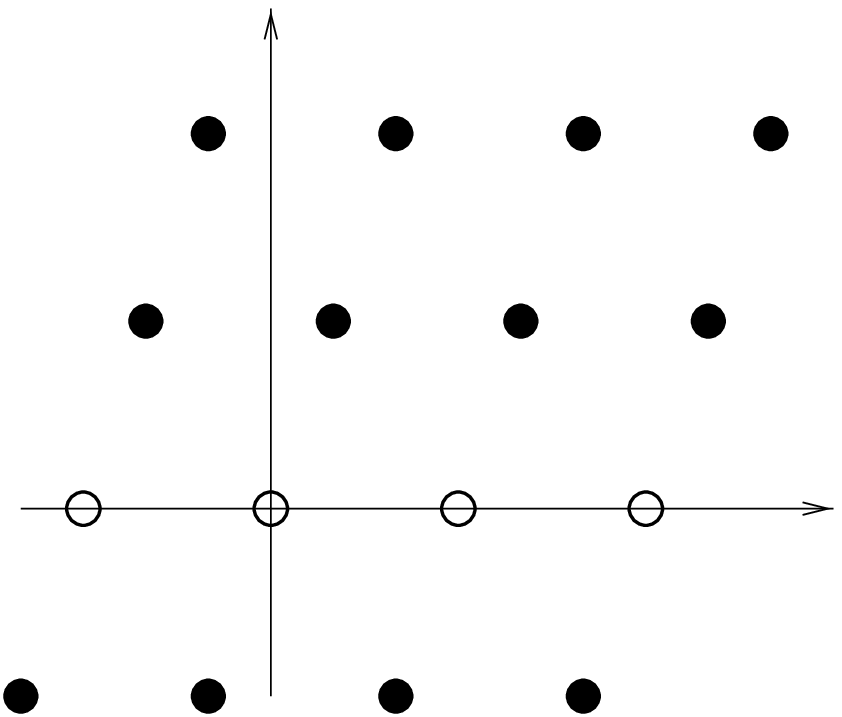}}
\put(62,148){$g_m$}
\put(168,55){$q_e$}
\put(90,0){(b)}
\normalsize      
\end{picture}
\end{minipage}
\caption{\small{In the Higgs phase magnetic charges are
connected by Meissner flux tubes. On the circle $C$ the Higgs field 
makes a full rotation (a). In this phase all magnetic charges
($\bullet$) are confined, while electric charges ($\circ$) are
free/screened (b). This and the following $gq$ diagrams refer to the
simplest case of $SU(2)_{\rm color}$.}}
\protect{\label{phaseH}}
\end{figure}

In the usual Higgs mechanism $\langle\phi\rangle\neq 0$
the Higgs field $\phi$ is an
`ordinary' elementary field with electric charges only.
This so-called \textbf{Higgs phase} is similar to the superconductor. All
magnetic charges will be confined by Meissner flux tubes, see Fig.
\ref{phaseH}(a). We do not see weak magnetic monopoles. Pure electric charges 
can move freely, see Fig. \ref{phaseH}(b).

\begin{figure}[b!]
\begin{minipage}{0.5\linewidth}
\begin{picture}(190,155)
\small
\put(20,20){\includegraphics{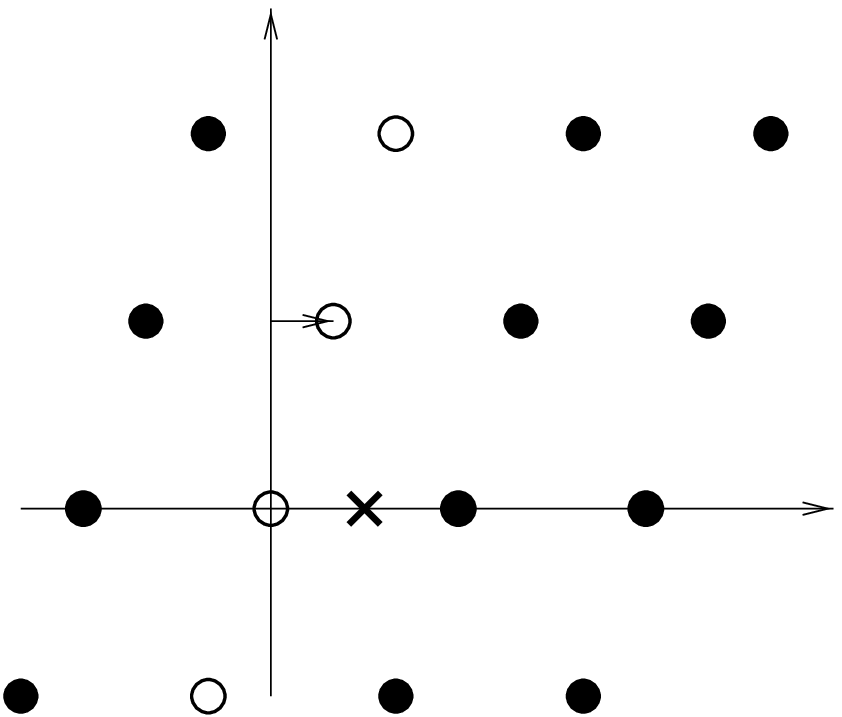}}
\put(62,148){$g_m$}
\put(168,55){$q_e$}
\put(69,95){$\theta/2\pi$}
\put(85,0){(a)}
\normalsize
\end{picture}
\end{minipage}
\begin{minipage}{0.5\linewidth}
\begin{picture}(190,155)
\small
\put(20,20){\includegraphics{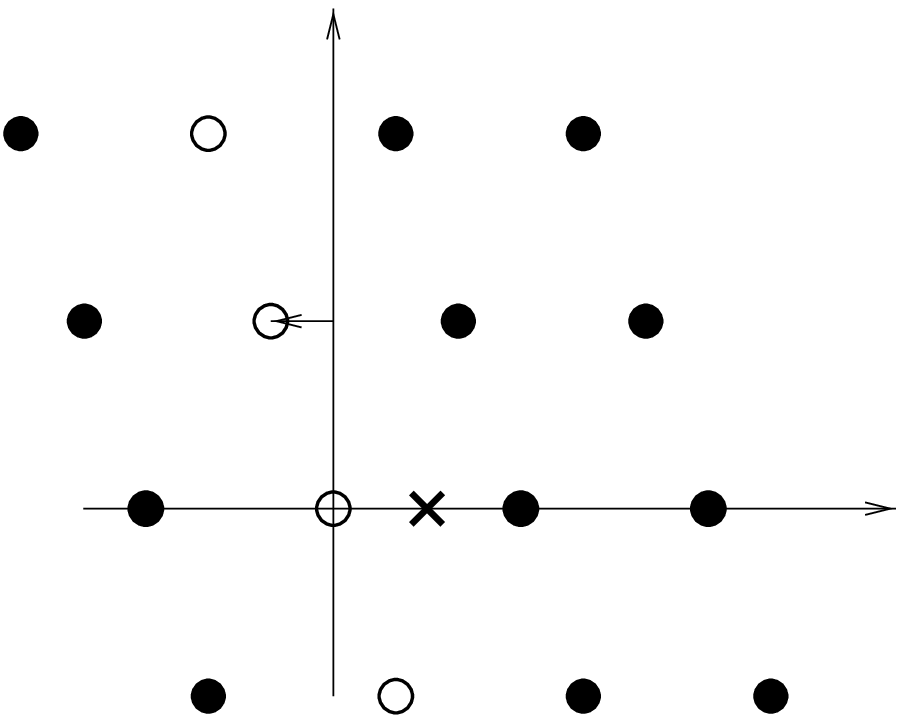}}
\put(72,148){$g_m$}
\put(178,55){$q_e$}
\put(34,95){$1-\theta/2\pi$}
\put(90,0){(b)}
\normalsize      
\end{picture}
\end{minipage}
\caption{\small{In the confinement phase all objects along a
particular line in the charge lattice are free ($\circ$). Now electric 
objects like gluons ($\bullet$)
and quarks ($\times$) are confined by vortices. The diagrams show the
confinement phase for small $\theta$ (a) and $\theta$ near $2\pi$
(b). The switch inbetween refers to a phase transition.}}
\protect{\label{phaseC}}
\end{figure}

The \textbf{confinement phase}
can be thought of as the \textit{dual transform} 
of the Higgs phase. The condensed  field $\phi$ is now a magnetic
object. Analogously all objects on the tilted line in Fig.
\ref{phaseC} are free, while gluons are confined, since they are
connected by $N$ vortices. Quarks in the fundamental representation
have electric charge $\frac{e}{N}$ and are connected by one vortex.

In the \textbf{Coulomb phase} no condensation takes place. Hence all charges
are free, but there are long-ranged electromagnetic fields. This
phase is \textit{self-dual}.\\

If we are in the confinement mode, then as $\theta$ runs from $0$ to
$2\pi$ there must be a phase transition: For small $\theta$ like in
Fig. \ref{phaseC}(a) the charges along $\theta$ are condensed, while
those along $2\pi-\theta$ are confined. For $\theta$ near $2\pi$ it is 
energetically favourable to have charges condensed along $2\pi-\theta$ 
and confined along $\theta$ like in Fig. \ref{phaseC}(b).

Note that physics is periodic in $\theta$, hence \textit{monopoles are 
not fundamentally distinguishable from dyons}.

The switch mentioned above refers to a phase transition, presumably at 
$\theta=\pi$. This phase transition is somewhat artificial, because in 
nature $\theta$ is a constant. It may be found in simulations, but the 
situation is hard to do on the lattice (complex action, instantons on
the lattice).

\begin{figure}[t]
\begin{picture}(380,120)
\small
\put(120,0){\includegraphics{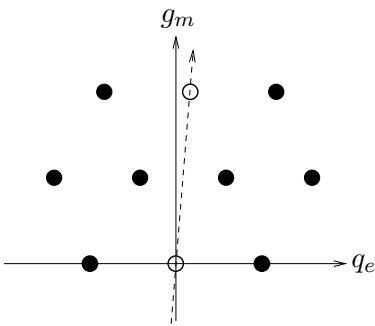}}
\put(180,115){$g_m$}
\put(252,22){$q_e$}
\normalsize
\end{picture}
\caption{\small{Oblique confinement removes the ambiguity for $\theta$ 
near $\pi$ by chosing a third charge ($\circ$) to be condensed.}}
\protect{\label{oblique}}
\end{figure}

We can imagine a more exotic condensation, \textit{oblique
confinement}. Let us have $\theta$ close to $\pi$. Now Buridan's
donkey cannot decide which of the haystacks to choose. Fortunately there is a
third choice in front of him (Fig. \ref{oblique}):
 Let the object inbetween (near
$g_m$-axis, small electric charge) be free and the residual charges be 
confined. This phase will not occur in QCD because there we know that
$\theta\simeq 0$, but it has peculiar features.\\

For some theories, the Higgs mode and the confinement mode are 
the same thing. As an example let us look at the electroweak theory in
the next section.

\section{A QCD-inspired Theory for the Electro\-weak Force}

The weak interaction is described by the Weinberg-Salam model with
gauge group
\begin{eqnarray*}
SU(2)_{\mbox{`color'}}\times U(1)_{\mbox{`em'}}
\end{eqnarray*}
Besides the gauge fields
\begin{eqnarray*}
U(1)\mbox{-photon: }\gamma,\quad SU(2)\mbox{-gluon: }g^a
\end{eqnarray*}
it has matter fields in the fundamental representation of $SU(2)$,
i.e.~color doublets. We call them `partons' (the physical particles
will come out later),
\hspace{0.5cm}\begin{tabbing}
leptonic partons ab \= $l_{i=1,2}$ ab\= (\mbox{spin} 1/2)\kill
leptonic partons \> $l_{i=1,2}$ \>
 (spin 1/2, charge $-1/2$)\\
quark partons    \> $q_{i=1,2}$ \>
 (spin 1/2, $SU(3)$-triplets, charge 1/6)\\
Higgs partons    \> $h_{i=1,2}$ \>
 (spin 0, charge $-1/2$)
\end{tabbing}
The theory is not fundamentally different from QCD, but the (scalar)
Higgs is really there.

This theory has `mesons' (parton-antiparton), `baryons' (two
partons) and the usual photon $\gamma$.
It comes out that the bound states without Higgs decay very
quickly. The scalar\footnote{w.r.t. the `color' gauge group}
part of the bound states are identified with 
physical particles as follows\\

\noindent \underline{\rule[-1.2mm]{0mm}{0mm}`Mesons'}
\hspace{0.5cm}\begin{tabbing}
ll,ll,ll abcd \= Higgs and $Z^0$ (orbital momentum)  ab\= \kill
$\bar{l}l, \bar{l}q, \bar{q}q$ \> unstable \> \\
$\bar{h}l$ \> neutrino \> (charge 0)\\
$\bar{h}q$ \> up-quark \> (charge $+2/3$)\\
$\bar{h}h$ \> Higgs and $Z^0$ (orbital momentum) \> (charge 0)\\
\end{tabbing}
\noindent \underline{`Baryons'}
\hspace{0.5cm}\begin{tabbing}
ll,ll,ll abcd \= down-quark  ab\= \kill
$ll, lq, qq$ \> unstable \> \\
$hl$ \> electron \> (charge $-1$)\\
$hq$ \> down-quark \> (charge $-1/3$)\\
$hh$ \> $W^\pm$ \> (charge $-1$)\\
\end{tabbing}

The only difference from QCD is that one can do a nice
\textit{pertubation expansion} and recovers everything from the
Standard Model. Like in Section \ref{construct} we fix the $SU(2)$
gauge first. In the unitary gauge we write the Higgs field $h$ as
\begin{eqnarray*}
h=\left(\begin{array}{c}F+h_1\\0
\end{array}\right)  
\end{eqnarray*}
Expanding in small fluctuations $h_1$ the bilinear mesonic field
combinations create the up-components,
\begin{eqnarray*}
\bar{h}l&=&Fl_1+\ldots\\
\bar{h}q&=&Fq_1+\ldots\\
\bar{h}h&=&F^2+2Fh_1+\ldots
\end{eqnarray*}
while the bilinear baryonic field combinations create the down-components,
\begin{eqnarray*}
\ep^{ij}h_il_j&=&Fl_2+\ldots\\
\ep^{ij}h_iq_j&=&Fq_2+\ldots
\end{eqnarray*}
By virtue of the covariant derivative we get the vector bosons,
\begin{eqnarray*}
\bar{h}\D_\mu h&=&FgW^3_\mu F+\ldots=gF^2 W^3_\mu\\
\ep^{ij}h_i\D_\mu h_j&=&Fg(W_\mu^1+iW_\mu^2)F+\ldots=gF^2W_\mu^- 
\end{eqnarray*}
We recovered the physical doublets,
\begin{eqnarray*}
\left(\begin{array}{c}l_1\\l_2
\end{array}\right)=
\left(\begin{array}{c}\nu\\e
\end{array}\right),\quad
\left(\begin{array}{c}q_1\\q_2
\end{array}\right)=
\left(\begin{array}{c}u\\d
\end{array}\right),\quad \mbox{etc.}
\end{eqnarray*} 
and all the rest is just the ordinary electroweak theory. In this
model physical particles are all confined, yet it trivially coincides
with the perturbative sector of the Standard Model,
 so confinement is no longer 
a mystery at all.

\section{Spontaneous Chiral Symmetry Breaking in QCD}

Let us take $N_f=2$ massless flavours (u and d quark). The composite
operator $\bar{\psi}^a_{\rm L}\psi^b_{\rm R}=\phi^{ab}$ with colour indices
$a,b$ will be \textit{coupled} to the agent of the Abelian
projection. An effective coupling $\bar{\psi}_{\rm L} X \psi_{\rm R}+$ 
h.c. will produce a \textit{constituent mass} (not algebraic mass) for 
the quarks.

But then each flavour has a Jackiw-Rebbi zero mode at the monopole
singularities. Hence the monopoles have `fractional' chiral flavour!
Although there are no explicit monopole solutions yet, we expect a
$2^2=4$-fold degeneracy for each monopole. This procedure would
suggest that these four states transform as
\begin{eqnarray*}
1(\mbox{no JR mode})+2 (\mbox{any flavour})+1 (\mbox{both flavours})
\end{eqnarray*}
under flavour $SU(2)$, see also Exercise (x).

But we have flavour $SU(2)_{\rm L}\times SU(2)_{\rm R}$. An
alternative possibility then is to attach one Jackiw-Rebbi mode to every 
\textit{left} quark,
\begin{eqnarray*}
\bar{2}_{\rm L}+2_{\rm R}=4
\end{eqnarray*}
Then we have $2^4=16$ states which transform as follows
\begin{eqnarray*}
\left.\begin{array}{ccccccccccc}
16&=&1&+&\bar{2}_{\rm L}+2_{\rm R}&+&1+\framebox{$\bar{2}_{\rm L}\times 2_{\rm
R}$}+1&+&\bar{2}_{\rm L}+2_{\rm R}&+&1\\
&&1&+&4&+&6&+&4&+&1
\end{array}\right.
\end{eqnarray*}
If the $\bar{2}_{\rm L}\times 2_{\rm R}$ monopole condenses, i.e. gets 
a vacuum expectation value, we have spontaneous chiral symmetry
breaking, as realised in nature.

Moreover, we know from lattice simulations, that the
confinement-de\-con\-fine\-ment phase transition  and the 
chiral symmetry breaking take place at the same temperature. We have
learned that different mechanisms are responsible for these effects,
but both refer to the same objects: magnetic monopoles.

It is interesting to study these mechanisms in supersymmetric theories 
like the Seiberg-Witten model.

\chapter{Effective Lagrangians for Theories with Confinement}

The effective mesonic fields $\phi_{ij}$ basically correspond to the
quark-antiquark composite operators,
\begin{eqnarray*}
\phi_{ij}=-\bar{q}^{\rm R}_j \, q^{\rm L}_i
\end{eqnarray*}
where $i,j=1..N_f$ are the indices of the chiral $U(N_f)^{\rm L}$ and 
$U(N_f)^{\rm R}$, respectively \cite{tHooftU1}. This symmetry acts as 
\begin{eqnarray*}
\phi^\prime_{ij}=U^{\rm L}_{ik} \, \phi_{kl} \, U^{{\rm R}\dagger}_{lj}
\end{eqnarray*}
Decomposing $\phi$ into its hermitean and anti-hermitean part we get
the scalars.  
Let us study the mesonic spectrum in an effective theory. Its
Lagrangian contains besides the kinetic term three potential terms,
\begin{eqnarray*}
\mathcal{L}_{\rm eff}&=&-\tr
\pa_\mu\phi\pa_\mu\phi^\dagger-V(\phi,\phi^\dagger)\\
V(\phi,\phi^\dagger)&=&V_0+V_m+V_{\rm inst.}
\end{eqnarray*}
which are the potential for spontaneous symmetry breaking, the
contribution of the quark masses and instanton contribution from
(\ref{det}), respectively,
\begin{eqnarray*}
V_0&=&-\mu^2 \, \tr \phi^\dagger\phi+A \, (\tr \phi^\dagger\phi)^2+
B \, \tr(\phi^\dagger\phi\phi^\dagger\phi)\\
V_m&=&-\sum_i m_i(\phi_{ii}+\phi_{ii}^\star)\\
V_{\rm inst.}&=&-2\kappa\mbox{ Re}(e^{i\theta}\det\phi) 
\end{eqnarray*}
where $A$, $B$ and $\mu^2$ are free parameters ($\mu^2>0$ for
spontaneous symmetry breaking), $m_1=m_u$, $m_2=m_d$, $m_3=m_s$, ...
are the quark masses and $\kappa$ contains the standard factor
$e^{-8\pi^2/g^2}$. $V_0$ preserves the $U(N_f)^{\rm L}\times
U(N_f)^{\rm R}$ symmetry, while $V_m$ and $V_{\rm inst.}$ break it
down to $U(N_f)$ and $SU(N_f)^{\rm L}\times
SU(N_f)^{\rm R}\times U(1)^{\rm V}$, respectively.

Let us study the
case of \textbf{3 flavours} with $m_1\simeq m_2\ll m_3$ and keep $\theta=0$ for
simplicity \cite{tHooftetaprime}. We expand the $3\times 3$ matrix
$\phi$ around its vacuum expectation values $F_i$,
\begin{eqnarray*}
\phi=\left(\begin{array}{ccc}
F_1&&\\&F_2&\\&&F_3
\end{array}\right)+\tilde{\phi}
\end{eqnarray*}
where the quark masses make the $F_i$ different. Scalar and
pseudoscalar particles can be identified by decomposing the
fluctuations $\tilde{\phi}$,
\begin{eqnarray*}
\tilde{\phi}_{ij}=S_{ij}+iP{ij}=-\ha\bar{q}_j(\id+\gamma_5)q_i,\quad
S=S^\dagger,\, P=P^\dagger
\end{eqnarray*}
By construction $\mathcal{L}$ is quadratic in $\tilde{\phi}$,
\begin{eqnarray*}
\mathcal{L}(F+\tilde{\phi})=\mathcal{L}(F)+0\cdot\tilde{\phi}+
\mathcal{L}_2(S)+\mathcal{L}_2^\prime(P)
\end{eqnarray*}
All scalars acquire masses via $V_0$ in $\mathcal{L}_2(S)$, the
pseudoscalar part becomes,
\begin{eqnarray*}
V_2^\prime(P)\propto\kappa F_1F_2F_3
\left(\frac{P_{11}}{F_1}+\frac{P_{22}}{F_2}+\frac{P_{33}}{F_3}\right)^2+
\sum_i \frac{m_i}{F_i}P_{ii}^2+\sum_{i\neq j}\ldots|P_{ij}|^2
\end{eqnarray*}
The instanton effect $\propto\kappa$ produces a mass in the
pseudoscalar sector: $\eta,\, \eta^\prime$. The other diagonal
pseudoscalars, $\pi$ and $K$, get $M^2\propto(m_1+m_2)$, the
off-diagonal ones carry masses anyway (through symmetry breaking).

Generalizing the model to \textbf{6 flavours}, we make a further
simplification,
\begin{eqnarray*}
V_0(\phi,\phi^\dagger)\rightarrow \delta|\phi\phi^\dagger-\id|^2,
\qquad \phi \mbox{ unitary}
\end{eqnarray*}
Now all $|F_i|^2$ equal $1$, and fluctuations of the modulus of $\phi$
cost infinite energy. Since the scalars seem not to play a role, their 
masses were sent to infinity. The chiral symmetry,
\begin{eqnarray*}
\phi^\prime=U^{\rm L}\,\phi\, U^{{\rm R}\dagger}
\end{eqnarray*}
is unchanged. For the other two potential terms we take the
straightforward generalisation of the $N_f=3$ model,
\begin{eqnarray*}
V_m&=&-\tr m_{ij}\phi_{ji}+\mbox{h.c.}\\
m_{ij}&=&\mbox{diag}\,(m_u,m_d,m_s,m_c,\ldots)\\
V_{\rm inst.}&=&-\kappa\det\phi+\mbox{h.c.}
\end{eqnarray*}
Note that the instanton angle may be absorbed in one of the masses,
\begin{eqnarray*}
m_u\> m_u e^{i\theta}
\end{eqnarray*}
Expanding $\phi$ now means
\begin{eqnarray*}
\phi=e^{iP},\qquad  P=P^\dagger
\end{eqnarray*}
The second order terms in $P$ are
\begin{eqnarray*}
V_2^\prime(P)&\propto&-\tr m(\id+iP-\ha P^2+\ldots)+\mbox{h.c.}-
\kappa \exp(i\tr P)+\mbox{h.c.}\\
&\>& -2\,\tr(m\cos P)-2\kappa\cos(\tr P)\\
&\>& \tr(m P^2)+\kappa(\tr P)^2\\
&=&\ha\sum_{ij}|P_{ij}|^2(m_i+m_j)+
\kappa (P_{11}+P_{22}+\ldots+P_{N_f N_f})^2
\end{eqnarray*}

These expressions refer to two main phenomenological observations. The 
first fact is that the meson masses squared are approximately linearly 
proportional to the quark masses,
\begin{eqnarray*}
M^2(\bar{q}_i\gamma_5q_j)=\mbox{const}\,(m_i+m_j)
\end{eqnarray*}
From the light pion one now concludes that the up and down quarks are
light, too: $m_u\simeq m_d\propto m^2_\pi\ll\Lambda_{\rm QCD}$.

The instanton produces a mass (only) for the $N_f$
pseudosinglet $\eta^\prime$. This explains the exception to the rule
above, namely the mystery of the $\eta$ and $\eta^\prime$ masses.
The Chern-Simons current $K_\mu$ (cf (\ref{CS})) is still conserved,
hence its Goldstone boson $\eta$ could not carry a mass. But $K_\mu$
is not gauge-invariant, it is rather like a ghost. Hence $K_\mu$ does
not protect $\eta^\prime$ from getting a mass. Indeed, instantons are
the only stable onjects with a non-vanishing value of the integral
(\ref{top/surf}) and they contribute to the $\eta^\prime$ mass.

\chapter{Exercises}

\begin{enumerate}

%*****************
\item[(i)]

Derive a Bogomol'nyi bound for the kink solution by writing
\begin{eqnarray*}
V(\phi)=\ha\left(\frac{\pa W(\phi)}{\pa \phi}\right)^2,\quad
\mathcal{H}=\ha\left(\pa_x\phi+\frac{\pa W(\phi)}{\pa \phi}\right)^2+\,\,
\mbox{total derivative.}\nonumber
\end{eqnarray*}
Discuss the function $W(\phi)$ and its extrema for the cases (a) and
(b). Can the bound always be saturated?

%*****************
\item[(ii)]

In case (a) a soliton and an anti-soliton approach each other. Assume an
approximate solution describing a stationary situation.

Why can this `solution' not be exact?

Show that the two Jackiw-Rebbi zero modes for fermions now mix, and
that their energies will no longer vanish. Estimate the amount to
which the energy levels split. Discuss the energy spectrum of the two
soliton system as a result of this effect.

%*****************
\item[(iii)]

Consider an $SU(2)$ gauge theory with an $I\!=\!1$ Higgs field
$\phi^{ab}\!=\!\phi^{ba}$, $\sum_a\phi^{aa}=0$. Assume the potential
such that
\begin{eqnarray*}
\langle\phi^{ab}\rangle_0=
F\left(\begin{array}{ccc}1&&\\&2&\\&&-3\end{array}\right)
\nonumber
\end{eqnarray*}
Find the (discrete) subgroup of $SU(2)$ that leaves this expectation
value invariant.

\textit{Hint}: First write these elements as $SO(3)$ matrices, then, by
exponentiation, as $SU(2)$ matrices.

Show that this is a non-Abelian group. Discuss the fusion rules for
the vortices that may occur in this system.

%*****************
\item[(iv)]

Consider the three-sphere $\sum_{\mu=1}^4 x_{\mu}^2=1$ and the $SU(2)$ matrices
\begin{eqnarray*}
U(\vec{x})=x_4\id+i\sum_{a=1}^3 x_a\sigma_a
\quad\quad x_\mu\,\,\mbox{real},\,\,
\sigma_a \mbox{ Pauli matrices}
\nonumber 
\end{eqnarray*}
Find a gauge transformation such that $^\Omega U(\vec{x})=\Omega U
\Omega^{-1}$ is diagonal,
$^\Omega U=\left(\begin{array}{cc}\omega_1&0\\0&\omega_2\end{array}\right)$.
Study the point $\vec{x}^\ast$ where your $\Omega(\vec{x})$ is
singular, and show that one cannot avoid that there is at least one
such point on $S^3$.

%*****************
\item[(v)]

In the 1+1 dimensional model case (a) with
$V(\phi)=\frac{\lambda}{4!}\left(\phi^2-F^2\right)^2$ 
construct an operator $\Omega(x)$ such that
\begin{eqnarray*}
\Omega(x_1)\phi(x_2)=\phi(x_2)\Omega(x_1)(-1)^{\theta(x_2-x_1)}
\quad \mbox{if}\,\,|x_2-x_1|>\epsilon>0
\nonumber
\end{eqnarray*}

Show that $\Omega(x)$ is the operator field that creates or
annihilates a soliton at $x$ (or at least some sort of kink).

Find an
approximate algorithm (or prescription) to compute the propagator
$\langle \mbox{T}\,\Omega(x_1,\,t_1)\Omega(x_2,\,t_2)\rangle_0$ in 
\textit{Euclidean} space-time. Find its behaviour at large
$|(x_2,\,t_2)-(x_1,\,t_1)|$.

\textit{Note}: We must assume that a local observer can observe
$|\phi(x)|$, but not the \textit{sign} of $\phi(x)$.

%*****************
\item[(vi)]

The magnetic monopole mass reads after the Bogomol'nyi trick
(\ref{monmass})
\begin{eqnarray*}
E=\int\!\!\d^3x\left[\ha(\vec{\D}\phi_a \pm \vec{B}_a)^2
+\frac{\lambda}{8}(\phi_a^2-F^2)^2\right]
+\frac{4\pi}{e}F
\end{eqnarray*}
Write down the usual ans\"atze (\ref{reg_mon_1}), (\ref{reg_mon_2})
for the fields,
\begin{eqnarray*}
\phi^a(x)=\hat{x}^a\phi(|\vec{x}|),\quad
A_i^a(x)=\ep_{iaj}\hat{x}_jA(|\vec{x}|),\quad
|\vec{x}|=\sqrt{\sum_{i=1}^3 x_i^2}
\end{eqnarray*}

Find the energy in terms of $\phi(|\vec{x}|)$, $A(|\vec{x}|)$, the field
equations and boundary conditions at $|\vec{x}|=0$, $|\vec{x}|\>\infty$
for these functions.

Give the Bogomol'nyi equations for $\phi(|\vec{x}|)$ and $A(|\vec{x}|)$.

%*****************
\item[(vii)]

Consider the candidate instanton solution $A_\mu^a(x)=\eta_{\mu\nu}^a
x^\nu A(|x|),\,\,
|x|=\sum_{\mu=1}^4 x_\mu^2$, where $A(|x|)$ is an arbitrary
function of $|x|$. Compute $G_{\mu\nu}^a$ and find the equation for
$A(|x|)$ corresponding to
\begin{eqnarray*}
(\mbox{a})\,\,G_{\mu\nu}=\tilde{G}_{\mu\nu}\quad\quad
(\mbox{b})\,\,G_{\mu\nu}=-\tilde{G}_{\mu\nu}
\nonumber
\end{eqnarray*}

Use
\begin{eqnarray*}
\epsilon_{abc}\eta_{\mu\nu}^b\eta_{\kappa\lambda}^c&=&
\delta_{\mu\kappa}\eta_{\nu\lambda}^a
-\delta_{\mu\lambda}\eta_{\nu\kappa}^a
-\delta_{\nu\kappa}\eta_{\mu\lambda}^a
+\delta_{\nu\lambda}\eta_{\mu\kappa}^a\\
\eta_{\mu\nu}^a\eta_{\mu\lambda}^b&=&
\delta^{ab}\delta_{\nu\kappa}+\epsilon_{abc}\eta^c_{\nu\kappa}
\end{eqnarray*}

\textit{Hint}: In case (b) one has $\eta_{\mu\nu}^a G_{\mu\nu}^b=0$. 

Solve these equations.

The two solutions look entirely different: (a) is an instanton, while
(b) is an anti-instanton. Explain the situation.

%*****************
\item[(viii)]

Consider the real scalar field $\phi(x)$ with
$\mathcal{L}=-\ha(\pa_\mu\phi)^2 +\frac{\lambda}{4!}\phi^4$, such that 
$\lambda$ has the `wrong sign'.

Show that there is an instanton solution in Euclidean four-space of the
form $\phi=\phi(|x|)$, $\phi(|x|\>\infty)\>0$.

Find the action and give a physical interpretation of this event and
the quantity $e^{S_{\rm inst.}}$.

%*****************
\item[(ix)]

Consider a theory with $N$ kinds of Maxwell fields: $G=U(1)^N$. Let
there exist objects $p$ with electric charges $(e_1,\ldots,e_N)_p$ and   
magnetic charges $(g_1^m,\ldots,g_N^m)_p$.

Write down the Dirac condition for two such objects $p$ and $q$.

%*****************
\item[(x)]

Let there be a monopole coupled to three fermion species $\psi_1$,
$\psi_2$, $\psi_3$ which are 3-representations of a \textit{global}
$SU(3)$-symmetry. Each of them has \textit{one} Jackiw-Rebbi zero mode 
solution with the monopole. Suppose that the `completely empty'
monopole is an $SU(3)$-singlet. There is an $2^3=8\,$-fold degeneracy.

How do the other states transform under $SU(3)$?

\end{enumerate}

%\printindex

\end {document}